\documentclass[1p]{elsarticle}
\pdfoutput=1 
\usepackage{amssymb,amstext,amsmath}
\usepackage{graphicx}
\usepackage{epstopdf}
\usepackage{color}
\usepackage{multirow}
\usepackage{bm}
\usepackage{url}

\newlength{\dhatheight}

\def\s{\bm{s}}
\def\n{\bm{n}}
\def\x{{\bm{x}}}
\def\y{{\bm{y}}}
\def\xo{\y}
\def\L{\bm{L}}
\def\R{\mathbb{R}}

\def\G{\mathcal{G}}

\def\A{\mathbf{A}}
\def\I{\mathbf{I}}
\def\U{\mathbf{U}}
\def\V{\mathbf{V}}
\def\W{\mathbf{W}}
\def\a{\mathbf{a}}
\def\b{\mathbf{b}}
\def\nn{\mathbf{n}}

\def\pa{{\partial\Omega}}

\journal{Journal of Computational Physics}

\begin{document}

\begin{frontmatter}

\title{Semi-analytical computation of Laplacian Green functions in three-dimensional domains with disconnected spherical boundaries}

\author[pmc,pon]{Denis S. Grebenkov}  
 \ead{denis.grebenkov@polytechnique.edu}

\author[sem]{Sergey D. Traytak}
 \ead{sergtray@mail.ru}

\address[pmc]{Laboratoire de Physique de la Mati\`{e}re Condens\'{e}e (UMR 7643), \\ 
CNRS -- Ecole Polytechnique, University Paris-Saclay, 91128 Palaiseau, France}

\address[pon]{Interdisciplinary Scientific Center Poncelet \\
(UMI 2615 CNRS/ IUM/ IITP RAS/ Steklov MI RAS/ Skoltech/ HSE)  \\
Bolshoy Vlasyevskiy Pereulok 11, 119002 Moscow, Russia}

\address[sem]{Semenov Institute of Chemical Physics of the Russian Academy of Sciences, \\ 4 Kosygina St., 117977 Moscow, Russia}


\date{Received: \today / Revised version: }

\begin{abstract}
We apply the generalized method of separation of variables (GMSV) to
solve boundary value problems for the Laplace operator in
three-dimensional domains with disconnected spherical boundaries
(i.e., an arbitrary configuration of non-overlapping partially
reactive spherical sinks or obstacles).  We consider both exterior and
interior problems and all most common boundary conditions: Dirichlet,
Neumann, Robin, and conjugate one.  Using the translational addition
theorems for solid harmonics to switch between the local spherical
coordinates, we obtain a semi-analytical expression of the Green
function as a linear combination of partial solutions whose
coefficients are fixed by boundary conditions.  Although the numerical
computation of the coefficients involves series truncation and matrix
inversion, the use of the solid harmonics as basis functions naturally
adapted to the intrinsic symmetries of the problem makes the GMSV
particularly efficient, especially for exterior problems.  The
obtained Green function is the key ingredient to solve boundary value
problems and to determine various characteristics of stationary
diffusion such as reaction rate, escape probability, harmonic measure,
residence time, and mean first passage time, to name but a few.  The
relevant aspects of the numerical implementation and potential
applications in chemical physics, heat transfer, electrostatics, and
hydrodynamics are discussed.
\end{abstract}



\end{frontmatter}


\section{Introduction}

Diffusion-reaction processes in porous materials and biological media
play an important role in various fields, from physics to chemistry,
biology and ecology \cite{Rice,Calef,Weiss,Zhou2010}.  The geometric
structure of these media is often modeled by packs of spheres.  Some
spheres can be just inert reflecting obstacles to diffusing particles,
the others can fully or partially absorb the particles, while the
third ones allow for diffusive exchange between interior and exterior
compartments.  In the stationary regime, the local concentration of
diffusing particles, $n(\x)$, obeys the Laplace equation, $\nabla^2
n(\x) = 0$, subject to appropriate boundary conditions.  Similar
boundary value problems arise in various sciences such as heat
transfer \cite{Carslaw,Samarskii}, electrostatics \cite{Jackson},
hydrodynamics \cite{Milne-Thomson},
geophysics \cite{James69}, and probability theory
\cite{Gardiner,Keilson}.  Although the Laplace equation is probably
the most well studied partial differential equation (PDE), its
explicit analytical solutions are available only for a very limited
number of three-dimensional domains \cite{Carslaw,Crank}.  Among them
one usually distinguishes solutions obtained by separation of
variables in separable curvilinear coordinate systems that are
determined by the Euclidean symmetry group of the Laplace equation
\cite{Miller}.  The most common examples are a sphere and a circular
cylinder.  In more complicated but also more practically relevant
cases, one has to resort to numerical methods.  Except for Monte Carlo
simulations, essentially all numerical methods aim to reduce the PDE
to an infinite system of linear algebraic equations (ISLAE) that is
then solved numerically.  The efficiency of a numerical method depends
thus on the chosen reduction scheme.  For instance, in a finite
element method (FEM), the PDE is projected onto basic functions which
are piecewise polynomials on each element of a meshed computational
domain.  The unknown coefficients in front of these functions are then
determined by solving a system of linear equations.  Without relying
on specific geometrical properties of the domain, the FEM is a
powerful tool to solve general PDEs in general bounded domains
\cite{Reddy,Eun13}.  In turn, a (much) higher computational efficiency
is expected for a method that is specifically adapted to the
geometrical structure of the domain.  In particular, when the domain
has disconnected spherical boundaries, one can profit from the
underlying local spherical symmetries to build up more efficient but
less generic methods.

In this paper, we revisit the so-called generalized method of
separation of variables (GMSV) that goes back to Rayleigh's seminal
paper on the conductivity of heat and electricity in a medium with
cylindrical or spherical obstacles arranged in a rectangular array
\cite{Rayleigh}.  In its modern form, the GMSV was thoroughly
developed for studying diffraction of electromagnetic waves on
surfaces of several bodies \cite{Ivanov} and then applied in various
fields.  It is striking how many names were given to the method under
consideration by different authors: ``the method of addition
theorems'' \cite{Erofeenko}, ``the method of reduction to the ISLEA''
\cite{Alexandrov}, ``the method of irreducible Cartesian tensors''
\cite{Tray90}, ``the method based on the theory of multipole
expansions'' \cite{Clercx93}, ``the generalized Fourier method''
\cite{Nikolaev}, ``the Rayleigh multipole method''
\cite{Yardley99}, ``the method of twin multipole expansions''
\cite{Tsao02}, ``the direct method of re-expansion'' \cite{Traytak03},
``a twin spherical expansions method'' (or just ``twin expansions
technique'') \cite{McDonald03}, ``the method of a bispherical
expansion'' \cite{McDonald04}, ``the multipole re-expansion method''
\cite{Gumerov05}, ``the multipole expansion method'' \cite{Kushch13},
and a particular case of ``the method of series'' \cite{Guz}.  The
GMSV has been successfully applied in the elasticity theory
\cite{Miyamoto}, heat transfer \cite{Jefferey}, diffraction theory
\cite{Guz,Linton} and other branches of mathematical physics
\cite{Erofeenko}.  In the diffusion context, Mitra and later Goodrich
were the first to apply the GMSV to find the steady-state diffusion
field in the vicinity of two identical ideal spherical sinks (drops)
\cite{Mitra,Goodrich}.  Felderhof investigated diffusion-controlled
reactions in regular and random arrays of static ideal spherical sinks
\cite{Felderhof,Mattern87}.  Later Venema and Bedeaux applied the GMSV
to study similar problems for a periodic array of penetrable spherical
sinks \cite{Venema}.  Traytak employed the GMSV in the form of
expansions with respect to irreducible Cartesian tensors
\cite{Tray90,Traytak03,Tray0,Tray01}.  Traytak and Tachiya investigated
diffusive interaction between two spherical sinks in an electric field
by means of the GMSV \cite{Traytak97}.  Tsao, Strieder {\it et al.}
and Traytak {\it et al.} used the GMSV to calculate rigorously the
electric field effects and to study reactions on two different
spherical sinks and on spherical source and sink
\cite{Tsao02,McDonald03,McDonald04,Seki,TrBarTa07,Traytak08,Strieder16}.
A more general form of the GMSV was elaborated to compute the
steady-state reaction rate for an irreversible bulk
diffusion-influenced chemical reaction between a mobile point-like
particle and static finite three-dimensional configurations of
spherical active particles
\cite{Traytak03,Traytak05,Galanti16a,Galanti16b}.  Since diffusion-reaction
processes among spherical sinks is a long-standing problem, many other
theoretical methods (such as variational estimates and perturbative
analysis) have been employed
\cite{Samson77,Kayser83,Kayser84,Berezhkovskii90,Berezhkovskii92,Berezhkovskii92b,Berezhkovskii93,Makhnovskii99,Makhnovskii02,Oshanin98,Yang98,Kipr5,Barzykin,Torquato02,Nguyen10}.

In a nutshell, the GMSV for many spheres consists in representing the
solution of a boundary value problem as a linear combination of
partial solutions written in local spherical coordinates of each
sphere.  The coefficients in front of the underlying solid harmonics
are fixed to respect boundary conditions by using so-called addition
theorems for solid harmonics to switch between local spherical
coordinates.  As in other numerical methods, the original PDE is
reduced to an ISLAE that in general has to be solved numerically.
However, the natural choice of the solid harmonics in local spherical
coordinates as basis functions preserves the intrinsic symmetries of
the domain and provides superior computational efficiency.  In
particular, the resulting ISLEA can be truncated at smaller sizes,
yielding faster and more accurate solutions \cite{Greengard97}.
Moreover, there is no need for meshing the domain: once the
coefficients in front of partial solutions are computed, the
concentration field can be easily and very rapidly evaluated at any
point due to the explicit analytical dependence on the coordinates.
For exterior problems, this method does not require imposing a distant
artificial outer boundary that is needed for many other numerical
methods (such as FEM) to deal with a bounded domain.  It is important
to note that the GMSV is not limited to spherical shapes and can be
applied to other canonical domains such as spheroids, cylinders,
cones, etc. \cite{Traytak18}.

To our knowledge, the GMSV has not been applied to compute Green
functions for Laplacian boundary value problems in three-dimensional
domains with disconnected spherical boundaries (however, see
\cite{Chen09} for the planar case).  The Green function is harmonic
everywhere in a given domain except for a fixed singularity point, and
satisfies imposed homogeneous boundary conditions.  The corresponding
boundary value problems are known to be well-posed in simply-connected
three-dimensional domains bounded by piecewise smooth boundaries
(i.e., as in our setting) \cite{Courant,Egorov,Melnikov,Bogolyubov}.
We compute the Green functions for both exterior and interior domains
and for all most common boundary conditions: Dirichlet, Neumann,
Robin, and conjugate one (also known as the fourth boundary condition,
transmission condition and exchange condition).  We describe all the
steps of the method, from analytical derivations to numerical
implementations.  We deduce the semi-analytical formula for the Green
function, which is the key ingredient to solve general boundary value
problems for Laplace and Poisson equations and to determine various
characteristics of stationary diffusion such as reaction rate, escape
probability, harmonic measure, residence time, and mean first passage
time, to name but a few.  An implementation of this method as a Matlab
package is released and made freely accessible.

The paper is organized as follows.  Section \ref{sec:solution}
presents the main results and their derivation, for both interior and
exterior boundary value problems.  With increasing complexity, we
treat the exterior Dirichlet problem (Sec. \ref{sec:ExtD}), the
exterior Robin problem (Sec. \ref{sec:ExtR}), the interior Robin
problem (Sec. \ref{sec:Int}), and the conjugate (or exchange) problem
(Sec. \ref{sec:trans}).  To illustrate the general scheme, we
summarize in Sec. \ref{sec:examples} several examples for which the
solution is fully explicit.  In Sec. \ref{sec:applications}, we
discuss various applications of the derived semi-analytical formula
for the Green function.  Section \ref{sec:numerics} is devoted to a
practical implementation of the proposed method and some numerical
results.  Section \ref{sec:conclusion} concludes the paper, whereas
some technical points are moved to \ref{sec:derivations}.

\section{Semi-analytical solution}
\label{sec:solution}

In this section, we present the detailed derivation of the Green
function for the exterior Dirichlet problem (Sec. \ref{sec:ExtD}), the
exterior Robin problem (Sec. \ref{sec:ExtR}), the interior Robin
problem (Sec. \ref{sec:Int}), and the conjugate problem
(Sec. \ref{sec:trans}).

\begin{figure}
\begin{center}
\includegraphics[width=63mm]{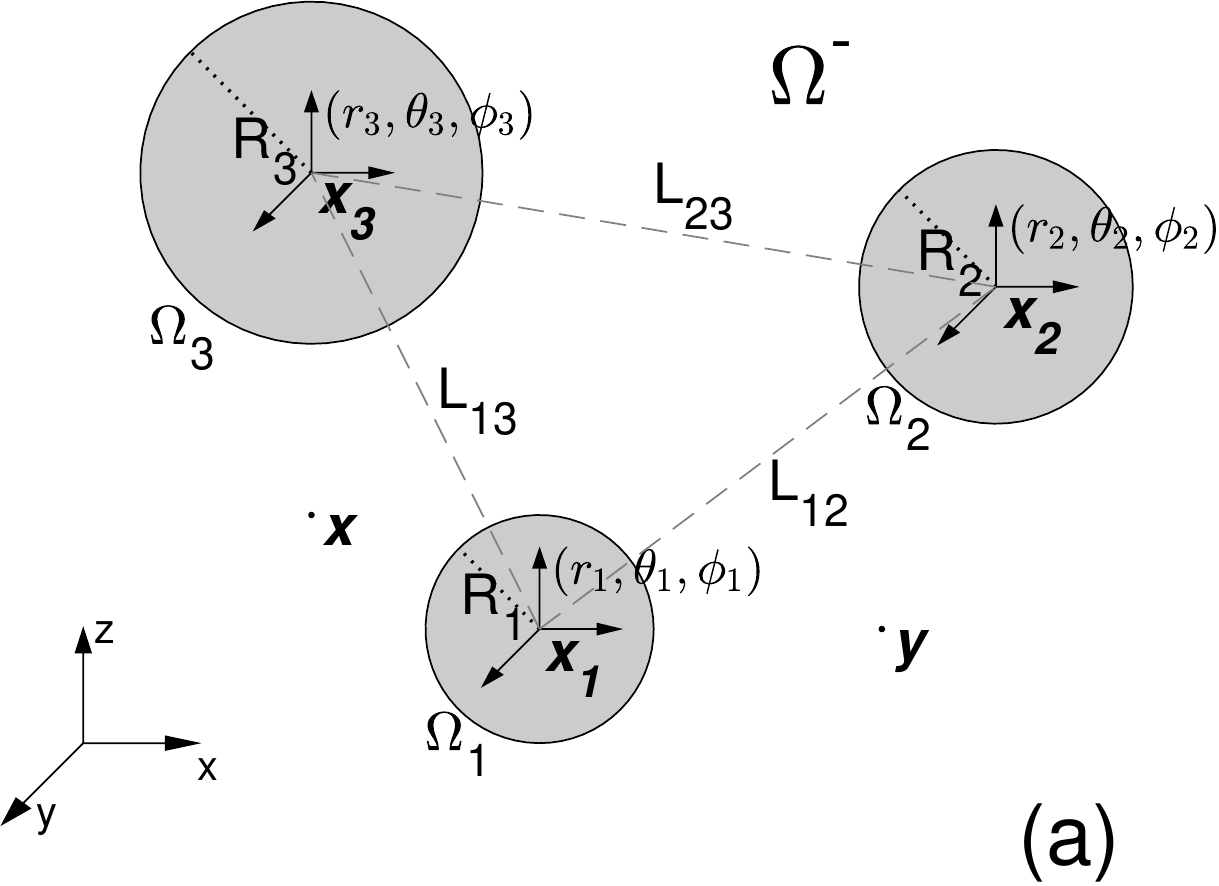} 
\includegraphics[width=57mm]{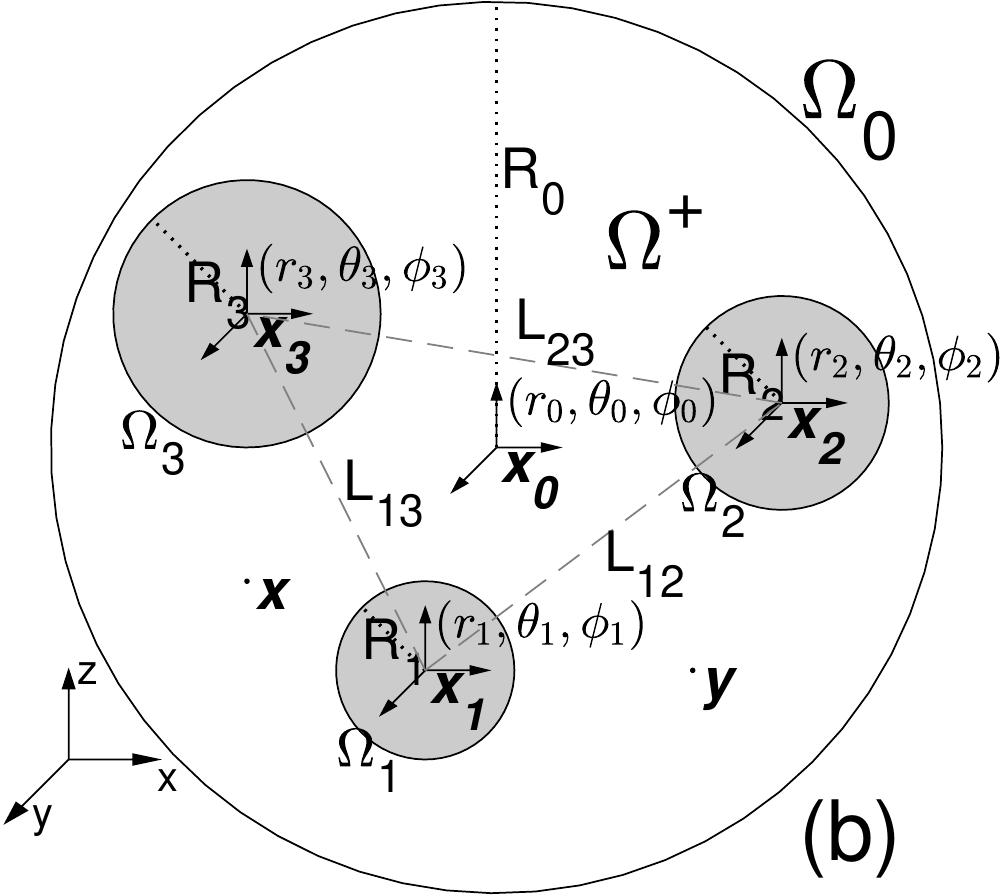} 
\end{center}
\caption{
{\bf (a)} Illustration of an unbounded exterior domain $\Omega^{-} =
\Bbb{R}^3 \backslash \bigcup\limits_{i=1}^N\overline{\Omega }_i$ with three
balls ($N=3$).  The Cartesian coordinates $\x_1$, $\x_2$, $\x_3$ of
the centers of these balls are given in some fixed global coordinate
system.  In turn, a local spherical coordinate system,
$(r_i,\theta_i,\phi_i)$, is associated with each ball.  The Green
function $G(\x,\xo)$ is computed at any pair of points $\x$ and $\xo$
of $\Omega^{-}$.  {\bf (b)} Illustration of a bounded interior domain
$\Omega^{+} = \Omega_0 \backslash
\bigcup\limits_{i=1}^N\overline{\Omega }_i$ with three balls ($N=3$).}
\label{fig:schema}
\end{figure}

For exterior problems, we consider an unbounded domain $\Omega^{-}$
outside $N$ non-overlap\-ping balls $\Omega_i = \{ \x \in \R^3~:~ \|
\x - \x_i\| < R_i\}$ of radii $R_i$, centered at $\x_i$, with
$i=\overline{1,N}$ (see Fig. \ref{fig:schema}(a)):
\begin{equation}
\Omega^{-} := \Bbb{R}^3\backslash \bigcup\limits_{i=1}^N\overline{\Omega }_i,
\quad \overline{\Omega }_i:= \Omega_i \cup \partial\Omega_i ,  \label{GMSV1}
\end{equation}
where $\partial\Omega_i $ is the surface of the $i$-th ball, and $\|
\cdot \|$ is the Euclidean distance.  The non-overlapping condition
reads 
\begin{equation}  \label{eq:nonoverlapping}
{\overline{\Omega }_i}\cap {\overline{\Omega }_j}=\emptyset \quad (i\neq j).  
\end{equation}
The $ N $-connected boundary of the domain $\Omega^{-}$ is
\begin{equation}
{\partial \Omega^{-}}=\bigcup\limits_{i=1}^N\partial{\Omega }_i , \label{GMSV2}
\end{equation}
i.e., partial surfaces $ \partial{\Omega }_i $ are the connected
components of the boundary of the exterior domain $ \Omega^{-} $.  In the
literature, domains like $\Omega^{-}$ are called ``periphractic
domains'' \cite{Maxwell}, ``perforated domains''
\cite{Marchenko,Hofer16} and ``domains with disconnected boundary''
\cite{Traytak03}.

For interior problems, we consider that $N$ formerly introduced
non-overlapping balls $\Omega_i$ are englobed by a larger spherical
domain $\Omega_{0} = \{ \x\in \R^3~:~ \| \x - \x_0\| < R_0\}$ of
radius $R_0$, centered at $\x_0$.  The interior boundary value
problems are posed in a bounded interior domain $\Omega^{+}$ (see
Fig. \ref{fig:schema}(b))
\begin{equation}
\Omega^{+} := \Omega_{0}\backslash \bigcup\limits_{i=1}^N\overline{\Omega }_i, \qquad  \overline{\Omega}_i \subset \Omega_{0} .
 \label{GMSV1b}
\end{equation}
The $ (N+1) $-connected boundary of the domain $\Omega^{+}$ is
\begin{equation}
{\partial \Omega^{+}}=\bigcup\limits_{i=0}^N\partial{\Omega }_i \,, \label{GMSV2a}
\end{equation}
that includes the outer boundary $\pa_0$.

\subsection{Exterior Dirichlet problem}
\label{sec:ExtD}

We first consider a general exterior Dirichlet boundary value problem
for the Poisson equation in the unbounded domain $\Omega^{-}\subset
\Bbb{R}^3$:
\begin{subequations} \label{green3a_all}
\begin{eqnarray}
-\nabla ^2u &=&F\quad (\x\in \Omega^{-}),  \label{green3a} \\
\label{green3a2}
\left. u\right| _{{\partial \Omega }_i} &=&f_i\quad (i=\overline{1,N}), \\
\label{green3a3}
\left. u\right| _{\Vert \x\Vert \rightarrow \infty } &\rightarrow &0,
\end{eqnarray}
\end{subequations}
where $F\in L_2 (\Omega^{-})\cap C^1(\Omega^{-})$ is a given function
of ``sources'' and $f_i \in C(\partial\Omega_i)$ are given continuous
functions.  The last relation (\ref{green3a3}) is the regularity
condition at infinity.  This problem is well posed and has a unique
classical solution
\cite{Courant,Egorov}.

There are at least two standard ways to get the classical solution of
this problem.  

(i) One can use the fundamental solution,
\begin{equation}  \label{eq:Gfund}
\G(\x,\xo)=\frac{1}{4\pi \Vert \x-\xo\Vert} \,,
\end{equation}
which is the Green function of the Laplace operator in ${\mathbb R}^3
\backslash\lbrace\xo\rbrace$:
\begin{equation}  \label{eq:fundamental}
-\Delta_\x \G(\x,\xo) = \delta(\x - \xo)   \quad (\x \in {\mathbb R}^3 \backslash\lbrace\xo\rbrace),
\end{equation}
where $\delta$ is the Dirac distribution, and $\xo$ is a fixed
point-like ``source''.  Multiplying Eqs. (\ref{green3a},
\ref{eq:fundamental}) by $\G(\x,\xo)$ and $u(\x)$ respectively,
subtracting them, integrating over $\x\in\Omega^{-}$ and using the
Green formula, one gets
\begin{equation}
u(\xo) = \int\limits_{\Omega^{-}} d\x \, F(\x) \, \G(\x,\xo) + \int\limits_{{\partial \Omega }^{-}}d\s\, 
\left. \left(\G(\x,\xo) \frac{\partial u(\x)}{\partial \n_{\x}}  - u(\x)\frac{\partial \G(\x,\xo)}{\partial \n_{\x}} \right) \right| _{\x=\s} ,
\label{eq:uDirichlet0}
\end{equation}
where $\partial/\partial \n_\x$ is the normal derivative at the
surface point $\x$, directed outward the domain $\Omega^{-}$.
Combined with the boundary condition (\ref{green3a2}), this
representation yields a boundary integral equation on $\partial
u(\x)/\partial \n_{\x}$, whose solution then determines $u(\xo)$
according to Eq. (\ref{eq:uDirichlet0}).  The solution of this
integral equation for domains with spherical and prolate spheroidal
boundaries has been recently proposed by Chang {\it et al.}
\cite{Chang16}.  Although the proposed method is conceptually close to
the GMSV that we describe here, the use of translational addition
theorems for solid harmonics significantly facilitates and speeds up
the computation (see below).  Most importantly, the dependence on the
``source'' point $\xo$ will appear explicitly in our analysis.

(ii) The solution of the problem (\ref{green3a_all}) can alternatively
be written as
\begin{equation}
\begin{split}
u(\xo)& = \int\limits_{\Omega ^{-}}d\x\,F(\x)\,G(\x,\xo) 
 +\sum\limits_{i=1}^N\int\limits_{{\partial \Omega }_i}d\s\,f_i(\s)
\left. \left( -\frac{\partial G(\x,\xo)}{\partial \n_\x}\right) \right| _{\x=\s}, \\
\end{split}
\label{eq:uDirichlet}
\end{equation}
where $G(\x,\xo)$ is the Dirichlet Green function in $\Omega^{-}$
satisfying for any $\xo\in \Omega^{-}$ the boundary value problem
\begin{subequations}
\begin{eqnarray}
-\nabla _{\x}^2G(\x,\xo) &=&
\delta \left( \x -\xo\right) \quad (\x\in \Omega ^{-}),  \label{green3a1} \\
\left. G\right| _{{\partial \Omega }_i} &=&0   \quad (i=\overline{1,N}), \\
\left. G\right| _{\Vert \x\Vert \rightarrow \infty } &\rightarrow &0.
\end{eqnarray}
\end{subequations}
The representation (\ref{eq:uDirichlet}) is obtained by multiplying
Eqs. (\ref{green3a}, \ref{green3a1}) by $G(\x,\xo)$ and $u(\x)$
respectively, subtracting them, integrating over $\x\in\Omega^{-}$ and
using the Green formula.  In spite of apparent similarity between
Eqs. (\ref{eq:uDirichlet0}, \ref{eq:uDirichlet}), the major difference
is that Eq. (\ref{eq:uDirichlet}) is an explicit solution in terms of
yet unknown Green function $G(\x,\xo)$, whereas
Eq. (\ref{eq:uDirichlet0}) is an integral equation on $\partial
u(\x)/\partial \n_\x$ involving the known fundamental solution
$\G(\x,\xo)$.  We recall that the Green function $G(\x,\xo)$ can be
physically interpreted as the electric potential at $\x$ created by a
charge at $\xo$ with grounded balls \cite{Duffy}.

To compute the Green function, one can represent it as
\begin{equation}
G(\x,\xo)= \G(\x,\xo)-g(\x;\xo), \label{green4}
\end{equation}
with an auxiliary function $g(\x;\xo)$ satisfying for any point
$\xo\in \Omega^{-}$:
\begin{subequations}  \label{eq:g_extD}
\begin{eqnarray}
-\nabla _{\x}^2g(\x;\xo) &=&0\quad (\x\in \Omega^{-} ),  \label{green3b} \\  
\label{green2b}
\left. g(\x;\xo)\right| _{\x\in \partial \Omega_i} &=&\left. \G(\x,\xo)\right| _{\x\in \partial \Omega_i}   \quad (i=\overline{1,N}), \\
\left. g(\x;\xo)\right| _{\Vert \x\Vert \rightarrow
\infty } &\rightarrow &0.
\end{eqnarray}
\end{subequations}
In other words, one can separate the universal singular part
$\G(\x,\xo)$ (yielding the Dirac $\delta$ distribution) and the
remaining regular part $g(\x,\xo)$ (ensuring the boundary conditions).
In spite of the known reciprocity,
\begin{equation*}
G(\x,\xo)=G(\xo,\x),
\end{equation*}
we will treat the point $\xo$ as a fixed parameter.  In the remaining
part of this subsection, we focus on the particular problem
(\ref{eq:g_extD}), bearing in mind that its solution gives access via
Eq. (\ref{eq:uDirichlet}) to a solution of any exterior Dirichlet
problem (\ref{green3a_all}).

We search for the solution of Eqs. (\ref{eq:g_extD}) in the form of
superposition
\begin{equation}
g(\x;\xo)=\sum\limits_{i=1}^N g_i(r_i,\theta_i,\phi_i ~;~ \xo),
\label{green4a}
\end{equation}
where $g_i$ is the partial solution in the local spherical coordinates
of the ball $\Omega _i$, with $(r_i,\theta _i,\phi _i)$ being the
local spherical coordinates of point $\x$, i.e., the spherical
coordinates of $\x-\x_i$.  The above expression follows immediately
from the representation (\ref{eq:uDirichlet0}) of the function
$g(\x;\xo)$ (with $F \equiv 0$) and the additivity of the Riemann
integral over the disconnected boundary $\partial\Omega^{-}$.  Each
partial solution can be expanded onto a complete basis of functions
$\{\psi _{mn}^{-}\}$ outside the $i$-th ball:
\begin{equation}
g_i(r_i,\theta _i,\phi _i~;~\xo)=\sum\limits_{n=0}^\infty
\sum\limits_{m=-n}^n A_{mn}^i\psi _{mn}^{-}(r_i,\theta _i,\phi _i),
\label{green4b}
\end{equation}
where $A_{mn}^i$ are the unknown coefficients (depending
parametrically on $\xo$).  Basis functions $\{\psi _{mn}^{-}\}$ are
the irregular (also called singular with respect to the origin) solid
harmonics:
\begin{equation}
\psi_{mn}^{-}(r,\theta,\phi) := \frac{1}{r^{n+1}}  Y_{mn}(\theta,\phi ) , \label{green6a}
\end{equation}
where 
\begin{equation}
Y_{mn}(\theta,\phi ):=P_n^m(\cos \theta) \, e^{im\phi}\quad 
\end{equation}
are the (non-normalized) spherical harmonics.  The (non-normalized)
associated Legendre functions $P_n^m(z)$ of degree $n$ ($n =
0,1,2,\ldots$) and order $m$ ($m = -n,-n+1,\ldots,n-1,n$) are
\begin{equation}
\begin{split}
P_n^m(z) &= (-1)^m (1-z^2)^{m/2}  \frac{d^m}{dz^m} P_n(z)  \quad (m = \overline{0,n}) , \\
P_n^{-m}(z) &= (-1)^m \frac{(n-m)!}{(n+m)!} \, P_n^m(z)  \hskip 12.5mm (m = \overline{1,n}),  \\
\end{split}
\end{equation}
where $P_n(z)$ are Legendre polynomials of degree $n$.  Note that
irregular solid harmonics $ \psi _{mn}^{-} $ and the coefficients
$A_{mn}^i$ are sometimes called the ``multipoles'' and the ``moments
of the expansion'', respectively \cite{Greengard97}.  The irregular
solid harmonics are well defined in the exterior of a ball.  In turn,
the regular solid harmonics,
\begin{equation}
\psi_{mn}^{+}(r,\theta,\phi) := r^{n} \, Y_{mn}(\theta,\phi ) , \label{green6aa}
\end{equation}
are well defined in the interior of a ball.

In order to satisfy boundary conditions, each partial solution $g_i$,
written in the local spherical coordinates of the ball $\Omega_i$,
should be represented in local spherical coordinates of other balls.
Such representations can be efficiently performed by so-called
translational addition theorems (TATs) \cite{Miller,Epton} which
express a basic of solid harmonics $\{\psi _{mn}^{\pm}(\x_i)\}$ in
local coordinates $(O;\x_i)$ via a new basis of solid harmonics
$\{\psi _{mn}^{\pm}(\x_j)\}$ in translated coordinates $\x_j = \x_i +
\L_{ij}$.  There are three TATs that ``translate'' regular to regular
(R$\to$R), irregular to regular (I$\to$R), and irregular to irregular
(I$\to$I) solid harmonics (noting that regular harmonics cannot be
expanded onto irregular ones).  We have thus
\begin{subequations}
\begin{eqnarray}    \label{ad2_RR}
\textrm{R$\to$R:} && \psi_{mn}^{+}(r_j, \theta_j, \phi_j) = \sum\limits_{l=0}^n\sum\limits_{k=-l}^l U_{mnkl}^{(+j,+i)} \, \psi_{kl}^{+}(r_i, \theta_i, \phi_i) , \\
\label{green7a}
\textrm{I$\to$R:} && \psi_{mn}^{-}(r_j, \theta_j, \phi_j) = \sum\limits_{l=0}^\infty\sum\limits_{k=-l}^l 
U_{mnkl}^{(-j,+i)} \, \psi_{kl}^{+}(r_i, \theta_i, \phi_i)  \qquad (r_i < L_{ij}), \\
\label{ad2_II}
\textrm{I$\to$I:} && \psi_{mn}^{-}(r_j, \theta_j, \phi_j) = \sum\limits_{l=0}^\infty \sum\limits_{k=-l}^l 
U_{mnkl}^{(-j,-i)} \,  \psi_{kl}^{-}(r_i, \theta_i, \phi_i)  \qquad (r_i>L_{ij}), 
\end{eqnarray}
\end{subequations}
where $\L_{ij} = \x_j - \x_i$ is the vector connecting the centers of
balls $j$ and $i$, and $(L_{ij}, \Theta_{ij}, \Phi_{ij})$ are the
spherical coordinates of the vector $\L_{ij}$:
\begin{equation}
\begin{split}
x_j & = x_i + L_{ij}\sin \Theta _{ij} \cos \Phi _{ij},  \\
y_j & = y_i + L_{ij}\sin \Theta _{ij} \sin \Phi _{ij},  \\
z_j & = z_i + L_{ij}\cos \Theta _{ij} . \\
\end{split}
\end{equation}
The coefficients $U_{mnkl}^{(\pm j,\pm i)}$ are the matrix elements of
the translation operator \cite{Epton}, which are also known as
mixed-basis matrix elements \cite{Miller}.  For $i\ne j$, we have
\begin{subequations}
\begin{eqnarray}
\label{ad3_RR}
U_{mnkl}^{(+j,+i)} &=& \frac{(n+m)!}{(n-l+m-k)! \, (k+l)!} \, \psi_{(m-k)(n-l)}^{+}(L_{ij},\Theta _{ij},\Phi _{ij}) , \\
\label{green8a}
U_{mnkl}^{(-j,+i)} &=& (-1)^{k+l} \, \frac{(n+l-m+k)!}{(n-m)! \, (l+k)!} \, \psi_{(m-k)(n+l)}^{-}(L_{ij}, \Theta_{ij}, \Phi_{ij})  ,  \\
\label{ad3_II}
U_{mnkl}^{(-j,-i)} &=& \frac{(-1)^{m-n+l-k} (l-k)!}{(n-m)! \, (m-n+l-k)!} \, \psi_{(m-k)(l-n)}^{+}(L_{ij}, \Theta_{ij}, \Phi_{ij})  . 
\end{eqnarray}
\end{subequations}
Note that we use the convention that $\psi^{\pm}_{mn}$ is zero for $n
< 0$ or $|m| > n$.  For instance, the elements of the matrix
$U_{mnkl}^{(-j,-i)}$ are zero when the inequalities $l \geq n$ and
$n+m-l\leq k\leq m-n+l$ are not fulfilled.  Similarly, there is no
contribution of terms with the index $k$ such that $|m-k| > n-l$,
i.e., the second sum in Eq. (\ref{ad2_RR}) runs over $k$ from
$-\min(l,n-l-m)$ to $\min(l,n-l+m)$.  In particular, the first sum in
Eq. (\ref{ad2_RR}) can be formally extended to $+\infty$, as in other
expansions.  Since the relation (\ref{ad2_RR}) involves polynomials on
both sides, it is applicable for any $r_i$.  In turn, two other TATs
are applicable for $r_i < L_{ij}$ and $r_i > L_{ij}$ respectively.


Here and throughout the text, we use the triple indices $(i,m,n)$ and
$(j,k,l)$ to encode the elements of the involved vectors and matrices.
These indices facilitate the visual interpretation of these elements,
the superscript $i$ (or $j$) always referring to the ball number,
while the subscript $mn$ (or $kl$) to the order $m$ and the degree $n$
of the solid harmonics (see also Sec. \ref{sec:numerics} for details
on the numerical implementation).  Note that the sign in front of $i$
(or $j$) refers to regular (plus) and irregular (minus) solid
harmonics.  We will also employ the shortcut summation notations:
\begin{equation}
\sum\limits_{n,m} = \sum\limits_{n=0}^\infty \sum\limits_{m=-n}^n  \qquad  \textrm{and} \qquad 
\sum\limits_{l,k} = \sum\limits_{l=0}^\infty \sum\limits_{k=-l}^l
\end{equation}

The unknown coefficients $A_{mn}^i$ are fixed by the boundary
condition (\ref{green2b}).  To fulfill this condition at the surface
$\pa_i$, one needs to represent $g(\x;\xo)$ in the local coordinates
of the $i$-th ball.  Combining Eqs. (\ref{green4a}, \ref{green4b},
\ref{green7a}), one gets any $i=\overline{1,N}$
\begin{equation}
g(\x;\xo) =\sum\limits_{n,m}\Biggl(\frac{A_{mn}^i}{r_i^{2n+1}} 
+ \sum\limits_{j(\neq i)=1}^N \sum\limits_{l,k} A_{kl}^j U_{klmn}^{(-j,+i)}\Biggr) 
 \psi _{mn}^{+}(r_i,\theta _i,\phi _i),
\label{eq:v_auxil2}
\end{equation}
which is valid for $\x$ in a close vicinity of the $i$-th ball.  We
also use the Laplace expansion for the Newton's potential (see
\ref{sec:ANewton}) to expand the right-hand side of Eq. (\ref{green2b})
on the regular solid harmonics in the local coordinates of the $i$-th
ball:
\begin{equation}  \label{green10b}
\G(\x,\xo) = \sum\limits_{n,m} V_{mn}^i \,  \psi_{mn}^{+}(r_i, \theta_i,\phi_i) \qquad (r_i < L_{i}),
\end{equation}
with 
\begin{equation}  \label{eq:V}
V_{mn}^i = \frac{(-1)^m}{4\pi} \, \psi_{(-m)n}^{-}(L_{i}, \Theta_{i}, \Phi_{i}),
\end{equation}
where $\L_{i} = \xo - \x_i$ and $(L_{i}, \Theta_{i}, \Phi_{i})$ are
the spherical coordinates of $\L_{i}$.
Equating Eqs. (\ref{eq:v_auxil2}) and (\ref{green10b}) at $r_i = R_i$
and using (\ref{green2b}), one gets the equality that must be
fulfilled for all points on $\partial\Omega_i$ (i.e., all $\theta_i$
and $\phi_i$), implying that the coefficients in front of
$\psi_{mn}^{+}(R_i,\theta_i,\phi_i)$ must be identical:
\begin{equation}  \label{green11a}
\frac{A_{mn}^i}{R_i^{2n+1}} + \sum\limits_{j(\ne i)=1}^N
\sum\limits_{l,k} A_{kl}^j \, U_{klmn}^{(-j,+i)} = V_{mn}^i .
\end{equation}
Multiplying by $R_i^{2n+1}$ and denoting
\begin{eqnarray}  \label{eq:Uhat}
\hat{U}^{ij}_{mnkl} &=& \begin{cases} \displaystyle R_i^{2n+1} \,
U^{(-j,+i)}_{klmn} \quad (i\ne j), \cr \delta_{mk} \, \delta_{nl} \hskip 15.5mm
(i=j), \end{cases} \\
\label{eq:Vhat}
\hat{V}^i_{mn} &=& R_i^{2n+1}\, V^i_{mn} ,
\end{eqnarray}
one rewrites the above relation as an ISLAE:
\begin{equation}  \label{green11b}
\sum\limits_{j=1}^N \sum\limits_{l,k} \hat{U}_{mnkl}^{ij} A_{kl}^j = 
\hat{V}_{mn}^i  \qquad (i=\overline{1,N}, ~ n = \overline{0,\infty}, ~ m = \overline{-n,n}).
\end{equation}
Writing this system in a matrix form, one gets the vector $\A$ of
coefficients $A_{mn}^i$ by inverting the matrix $\hat{\U}$:
\begin{equation}  \label{eq:UA}
\A = \W \hat{\V}, \qquad \W = \hat{\U}^{-1}.
\end{equation}
The Green function follows with the aid of Eqs. (\ref{green4},
\ref{green4a}, \ref{green4b}):
\begin{equation}  \label{eq:G}
G(\x,\xo) = \G(\x,\xo) -
\sum\limits_{i=1}^N \sum\limits_{n,m} A_{mn}^i \,
\psi_{mn}^{-}(r_i,\theta_i,\phi_i) .
\end{equation}
In the second term, the dependence on $\x$ is captured explicitly via
$(r_i,\theta_i,\phi_i)$, whereas the dependence on $\xo$ is also
explicit but hidden in the coefficients $A_{mn}^i$ via $V_{mn}^i$.  In
turn, the mixed-basis matrix $\U$ (and thus $\hat{\U})$ depends on the
positions and radii of the balls but is independent of the point
$\xo$.  In practice, one first inverts numerically a truncated matrix
$\hat{\U}$ and then, for each point $\xo$, rapidly computes the vector
$\hat{\V}$, from which the coefficients $A_{mn}^i$ are found (see
Sec. \ref{sec:numerics}).

As said earlier, the Green function allows one to solve any exterior
Dirichlet problem (\ref{green3a_all}).  The general representation
(\ref{eq:uDirichlet}) includes both the Green function and its normal
derivative at spheres ${\partial\Omega}_i$, which is also known as the
harmonic measure density \cite{Garnett}
\begin{equation}
\omega_{\xo}(\s) := - \left(\frac{\partial G(\x,\xo)}
{\partial {\n_\x}} \right)_{\x = \s} \quad (\s
\in {\partial\Omega^{-}}) .
\end{equation}
In \ref{sec:HM_derivation}, we deduce the decomposition of this
density onto irregular solid harmonics:
\begin{equation}  \label{eq:HM}
\omega_{\xo}^i(\s) := \left . \omega_{\xo}(\s) \right|_{{\partial\Omega}_i} = 
\frac{1}{R_i} \sum\limits_{n,m} (2n+1)A_{mn}^i \,
\psi_{mn}^{-}(R_i,\theta_i,\phi_i) .
\end{equation}

\subsection{Exterior Robin problem}
\label{sec:ExtR}

The above technique can be extended to finding the Green function with
Robin boundary conditions
\begin{subequations}
\begin{eqnarray}
-\nabla _{\x}^2G(\x,\xo) &=&\delta \left( \x-\xo\right) \quad (\x\in\Omega^{-}),   \\
\label{eq:GRobin_BC}
\left. \left( a_iG+b_iR_i\frac{\partial G}{\partial \n_\x}\right) 
\right| _{\x\in {\partial \Omega }_i} &=&0\quad (i=\overline{1,N}), \\
\label{eq:GRobin_inf}
\left. G\right| _{\Vert \x\Vert \rightarrow \infty } &\rightarrow &0,
\end{eqnarray}
\end{subequations}
with a fixed source point $\xo\in\Omega^{-}$ and nonnegative constants
$a_i$ and $b_i$ such that $a_i+b_i>0$ for each $i$.  Using again
Eq. (\ref{green4}), one gets the Robin boundary conditions for another
auxiliary function $g(\x;\xo)$:
\begin{equation}
\left. \left( a_ig+b_iR_i\frac{\partial g}{\partial \n_\x}\right) 
\right| _{\x\in \partial \Omega _i}=\left. \left( a_i \G +b_iR_i\frac{\partial
\G}{\partial \n_\x}\right) \right| _{\x\in \partial \Omega _i}.
\label{eq:RBC}
\end{equation}
Re-writing Eqs. (\ref{eq:v_auxil2}, \ref{eq:dg}) as 
\begin{subequations}
\label{eq:g_dg}
\begin{eqnarray}
\left. g(\x;\xo)\right| _{\x\in {\partial \Omega }_i}
&=&\sum\limits_{n,m}\bigl(\hat{\U}\A\bigr)_{mn}^i\psi _{mn}^{-}(R_i,\theta_i,\phi _i), \\
\left. \left( \frac{\partial g(\x;\xo)}{\partial \n_\x}
\right) \right| _{\x\in {\partial \Omega }_i}
&=& \frac1{R_i}\sum\limits_{n,m}\bigl((2n+1)A_{mn}^i -n(\hat{\U}\A)_{mn}^i \bigr)
\psi _{mn}^{-}(R_i,\theta _i,\phi _i), 
\end{eqnarray}
\end{subequations}
we represent the left-hand side of Eq. (\ref{eq:RBC}) as 
\begin{equation}
\sum\limits_{n,m}\bigl( (a_i-nb_i)(\hat{\U}\A)_{mn}^i + (2n+1)b_iA_{mn}^i\bigr)
\psi_{mn}^{-}(R_i,\theta _i,\phi _i).  \label{eq:RBC_auxil1}
\end{equation}
Using Eq. (\ref{green10b}, \ref{eq:dn_G0}), we represent the
right-hand side of Eq. (\ref{eq:RBC}) as
\begin{equation}
\sum\limits_{n,m}\bigl(a_i-nb_i\bigr) \hat V_{mn}^i\psi _{mn}^{-}(R_i,\theta
_i,\phi _i).  \label{eq:RBC_auxil2}
\end{equation}
Equating Eqs. (\ref{eq:RBC_auxil1}, \ref{eq:RBC_auxil2}), one finally
gets the equalities on the coefficients $A_{mn}^i$ in the Robin case
for $i=\overline{1,N}, ~ n = \overline{0,\infty}, ~ m =
\overline{-n,n}$:
\begin{equation}
(2n+1)b_iA_{mn}^i+(a_i-nb_i)\bigl(\hat{\U}\A)_{mn}^i=(a_i-nb_i)\hat V_{mn}^i .
\label{eq:Robin_A}
\end{equation}
This ISLAE generalizes Eq. (\ref{green11b}) to the Robin boundary
condition.  Representing the multiplication by $a_i$, $b_i$, and $n$
in a matrix form by diagonal matrices $\hat{\a}$, $\hat{\b}$, and
$\hat{\nn}$, the ISLAE can be written in a matrix form as:
\begin{equation}
\label{eq:UA_cond}
\bigl[(2\hat{\nn} + \I) \hat{\b} + (\hat{\a}- \hat{\nn} \hat{\b})
\hat{\U}\bigr] \A =(\hat{\a}-\hat{\nn}\hat{\b})\hat{\V},
\end{equation}
where $\I$ stands for the identity matrix.  Inverting the matrix in
front of $\A$, one represents the vector of coefficients $A_{mn}^i$ as
\begin{equation}
\A = \W\hat{\V},\qquad   \W=\bigl[(2\hat{\nn}+\I)\hat{\b} + 
(\hat{\a}- \hat{\nn}\hat{\b})\hat{\U}\bigr]^{-1}(\hat{\a}-\hat{\nn}\hat{\b}).  
\label{eq:UA_Robin}
\end{equation}
In the Dirichlet case ($b_i=0$ and $a_i=1$), this expression is
reduced to Eq. (\ref{eq:UA}).  The coefficients $A_{mn}^i$ fully
determine the Robin Green function:
\begin{equation}  \label{eq:G_Rob}
G(\x,\xo) = \G(\x,\xo) -
\sum\limits_{i=1}^N \sum\limits_{n,m} A_{mn}^i \,
\psi_{mn}^{-}(r_i,\theta_i,\phi_i) .
\end{equation}

With this Green function, the solution of a general exterior Robin
boundary value problem,
\begin{subequations}  \label{eq:Robin_all}
\begin{eqnarray}
-\nabla ^2u &=&F\quad (\x\in \Omega ^{-}),  \label{Robin1} \\
\left. \left( a_iu+b_iR_i\frac{\partial u}{\partial \n_\x}\right) 
\right| _{{\partial \Omega }_i} &=&f_i\quad (i=\overline{1,N}), \label{Robin2} \\
\left. u\right| _{\Vert \x\Vert \rightarrow \infty } &\rightarrow &0,
\label{Robin3}
\end{eqnarray}
\end{subequations}
can be represented as 
\begin{equation}
u(\xo)= \int\limits_{\Omega ^{-}}d\x\,F(\x)\,G(\x,\xo)+\sum\limits_{i=1}^N\int\limits_{{\partial \Omega }%
_i}d\s\,f_i(\s)\,\omega _{\xo}^i(\s),
\label{eq:u_general}
\end{equation}
where 
\begin{equation}
\omega _{\xo}^i(\s)=\frac{\left. G(\x,\xo)\right| _{\x\in {\partial \Omega }_i}}{b_iR_i}=\left. \left( -%
\frac{\partial G(\x,\xo)}{a_i\,\partial \n_\x}\right) \right|
_{\x\in {\partial \Omega }_i}  \label{eq:HMspread}
\end{equation}
is the spread harmonic measure density on the sphere ${\partial \Omega
}_i$ \cite{Grebenkov06,Grebenkov15}.  This is a natural extension of
the harmonic measure density to partially absorbing boundaries with
Robin boundary condition.  When both $a_i$ and $b_i$ are nonzero, two
representations in Eq.  (\ref{eq:HMspread}) are equivalent due to
Eq. (\ref{eq:GRobin_BC}).  In turn, one uses the first representation
for the Neumann case ($a_i=0$) and the second representation for the
Dirichlet case ($b_i=0$).  For $b_i\ne 0$, we use
Eqs. (\ref{green10b}, \ref{eq:G}) to get
\begin{equation} \label{eq:HMspread_R}
\omega _{\xo}^i(\s)=\frac 1{b_iR_i}\sum\limits_{n,m}\bigl(%
\hat V_{mn}^i-\bigl(\hat{\U}\A\bigr)_{mn}^i\bigr) \psi _{mn}^{-}(R_i,\theta
_i,\phi _i),
\end{equation}
whereas Eq. (\ref{eq:HM}) is used for the Dirichlet case ($b_i=0$, $a_i=1$).

\subsection{Interior Robin problem}
\label{sec:Int}

In many applications, a domain is limited by an outer boundary which
can significantly affect the diffusion characteristics.  A prominent
example is the mean first passage time which is infinite for unbounded
domains.  In order to deal with such problems, one needs to
incorporate an outer boundary, transforming the exterior problem to
the interior problem in a bounded domain $\Omega^+$ from
Eq. (\ref{GMSV1b}), with $N$ non-overlapping balls $\Omega_i$ ($i =
\overline{1,N}$), englobed by a larger ball $\Omega_0$ of radius $R_0$
and centered at $\x_0$.  The Robin Green function for the interior
problem in $\Omega^{+}$ satisfies for any $\xo\in\Omega^{+}$:
\begin{subequations}
\begin{eqnarray}
-\nabla _{\x}^2G(\x,\xo) &=&\delta \left( \x -\xo\right) \quad (\x\in\Omega^{+}),   \\
\label{eq:RBC_Green_Int}
\left. \left( a_iG+b_iR_i\frac{\partial G}{\partial \n_\x}\right) 
\right| _{\x\in {\partial \Omega }_i} &=&0\quad (i=\overline{0,N}), 
\end{eqnarray}
\end{subequations}
with nonnegative parameters $a_i$ and $b_i$ such that $a_i + b_i > 0$
for each $i$, and $a_0 + \cdots + a_N > 0$.  The last inequality
excludes the case with Neumann conditions at all boundaries, for which
the Green function of an interior problem does not exist.  Since the
Green function is now defined in a bounded domain, there is no
regularity condition (\ref{eq:GRobin_inf}) at infinity.  As
previously, one represents the Green function as in Eq. (\ref{green4})
and then searches for an auxiliary function $g(\x;\xo)$ in the form
\begin{equation}
g(\x;\xo) = \sum\limits_{i=0}^N g_i(r_i,\theta_i,\phi_i ~;~ \xo),
\end{equation}
to which a new function $g_0$ is added
\begin{equation}
\label{eq:gN}
g_0(r_0,\theta_0,\phi_0 ~;~ \xo) = \sum\limits_{n,m} A_{mn}^{0} \, \psi_{mn}^{+}(r_{0},\theta_{0},\phi_{0}) ,
\end{equation}
where $A_{mn}^{0}$ are the unknown coefficients, and
$(r_0,\theta_0,\phi_0)$ are the spherical coordinates of $\x - \x_0$.
As this function describes the behavior inside the ball $\Omega_{0}$,
one uses regular harmonics $\psi_{mn}^{+}$ instead of irregular ones
for other functions $g_i$.  The remaining derivation is similar to the
exterior case, i.e., one needs to find the coefficients $A_{mn}^i$
from the boundary conditions.  

At the boundary of an inner ball $\Omega_i$, one re-expand
$\psi_{mn}^{-}(r_j,\theta_j,\phi_j)$ for $j=\overline{1,N}$ (with
$j\ne i$) as previously.  In turn, one needs the R$\to$R addition
theorem (\ref{ad2_RR}) to re-expand the function $g_0$ in the local
coordinates of the $i$-th ball:
\begin{equation}
g_{0}(r_0,\theta_0,\phi_0 ~;~ \xo) = \sum\limits_{n,m} A_{mn}^{0} \sum\limits_{l,k} U_{mnkl}^{(+0,+i)} \, \psi_{kl}^{+}(r_i,\theta_i,\phi_i) .
\end{equation}
At the boundary $\pa_i$, one finds then
\begin{subequations}
\begin{eqnarray}
g_{0}|_{\pa_i} &=& \sum\limits_{n,m} \psi_{mn}^{+}(R_i,\theta_i,\phi_i) \sum\limits_{l,k} U_{klmn}^{(+0,+i)} A_{kl}^{0} , \\
R_i \left.\left(\frac{\partial g_{0}}{\partial \n_\x}\right)\right|_{\pa_i} 
& =& - \sum\limits_{n,m} \, n \, \psi_{mn}^{+}(R_i,\theta_i,\phi_i) \sum\limits_{l,k} U_{klmn}^{(+0,+i)} A_{kl}^{0}  . 
\end{eqnarray}
\end{subequations}
Combining these contributions with other $g_i$, we retrieve
Eqs. (\ref{eq:g_dg}), in which the matrix $\hat{\U}$ from
Eq. (\ref{eq:Uhat}) is modified by adding a new column $j = 0$ (with
$i>0$):
\begin{equation}
\label{eq:U_extend1}
\hat{U}^{i0}_{mnkl} = R_i^{2n+1} \, U_{klmn}^{(+0,+i)} .
\end{equation}
As a consequence, the Robin boundary condition (\ref{eq:GRobin_BC}) at
each $\pa_i$ implies the ISLAE (\ref{eq:Robin_A}), as for the exterior
problem.  Here, the effect of the outer boundary is captured through
the additional elements of the matrix $\hat{\U}$ in
Eq. (\ref{eq:U_extend1}).

Moreover, one has to fulfill the Robin boundary condition
(\ref{eq:RBC_Green_Int}) at the outer boundary $\pa_{0}$.  For this
purpose, each $g_i$ is re-expanded by using the I$\to$I addition
theorem (\ref{ad2_II}) as
\begin{equation}
g_i(r_i,\theta_i,\phi_i ~;~ \xo) = \sum\limits_{n,m} A_{mn}^{i} \sum\limits_{l,k} U_{mnkl}^{(-i,-0)} \, 
\psi_{kl}^{-}(r_{0},\theta_{0},\phi_{0})    \qquad (r_0 > L_{0i}),
\end{equation}
from which
\begin{subequations}
\begin{eqnarray}
\left. (g_i)\right|_{\pa_{0}} & =& \sum\limits_{n,m} \psi_{mn}^{-}(R_{0},\theta_{0},\phi_{0}) 
 \sum\limits_{l,k} U_{klmn}^{(-i,-0)} \, A_{kl}^{i} , \\
R_{0} \left.\left(\frac{\partial g_i}{\partial \n_\x}\right)\right|_{\pa_{0}} 
& =& - \sum\limits_{n,m} (n+1) \psi_{mn}^{-}(R_{0},\theta_{0},\phi_{0}) 
\sum\limits_{l,k} U_{klmn}^{(-i,-0)} \, A_{kl}^{i}  , 
\end{eqnarray}
\end{subequations}
where $\L_{0i} = \x_i - \x_0$.

In addition, Eq. (\ref{eq:gN}) implies
\begin{subequations}
\begin{eqnarray}
\left. (g_{0})\right|_{\pa_{0}} & =& \sum\limits_{n,m} R_{0}^{2n+1} A_{mn}^{0} \psi_{mn}^{-}(R_{0},\theta_{0},\phi_{0}) , \\
R_{0} \left.\left(\frac{\partial g_{0}}{\partial \n_\x}\right)\right|_{\pa_{0}} 
& =& \sum\limits_{n,m} n R_{0}^{2n+1} A_{mn}^{0} \psi_{mn}^{-}(R_{0},\theta_{0},\phi_{0}) . 
\end{eqnarray}
\end{subequations}
Combining these relations, one finds
\begin{subequations}
\begin{eqnarray}
\left. g\right|_{\pa_{0}} & =& \sum\limits_{n,m} \psi_{mn}^{+}(R_{0},\theta_{0},\phi_{0}) \bigl(\hat{\U} \A\bigr)_{mn}^{0} ,\\
R_{0} \left.\left(\frac{\partial g}{\partial \n_\x}\right)\right|_{\pa_{0}} 
& =& \sum\limits_{n,m} \psi_{mn}^{+}(R_{0},\theta_{0},\phi_{0})  
\bigl((2n+1)A_{mn}^0 - (n+1)(\hat{\U}\A)_{mn}^{0}\bigr) , 
\end{eqnarray}
\end{subequations}
where the matrix $\hat{\U}$ is modified by adding a new row at $i=0$
(with $j>0$) as
\begin{equation}
\hat{U}_{mnkl}^{0j} = R_{0}^{-(2n+1)} \, U_{klmn}^{(-j,-0)}  .
\end{equation}

On the other hand, using again the Laplace expansion for the Newton's
potential (see \ref{sec:ANewton}), one can write the fundamental
solution $\G(\x,\xo)$ as
\begin{equation}  \label{eq:Gfund_int}
\G(\x,\xo) = \sum\limits_{n,m} \tilde V^{0}_{mn} \, \psi_{mn}^{-}(r_{0},\theta_{0},\phi_{0}) \qquad (r_{0} > L_{0}),
\end{equation}
with
\begin{equation}
\tilde V_{mn}^{0} = \frac{(-1)^m}{4\pi} \, \psi_{(-m)n}^{+}(L_{0}, \Theta_{0}, \Phi_{0}) ,
\end{equation}
where $\L_{0} = \xo - \x_0$.  One also gets
\begin{equation}
R_{0} \left. \left(\frac{\partial \G}{\partial \n_\x}\right)\right|_{\pa_{0}} 
= - \sum\limits_{n,m} (n+1) \tilde V^{0}_{mn} \psi_{mn}^{-}(R_{0},\theta_{0},\phi_{0}) .
\end{equation}

Combining the above expressions for $g$ and $\G$ and their normal
derivatives according to the Robin boundary condition
(\ref{eq:RBC_Green_Int}) at the outer boundary $\pa_{0}$, one gets the
ISLAE for all $n=\overline{0,\infty}$ and $m = \overline{-n,n}$:
\begin{equation}
\label{eq:Robin_A2}
(2n+1) b_{0} A_{mn}^{0} + (a_{0} - (n+1)b_{0}) (\hat{\U} \A)_{mn}^{0} 
 = (a_{0}-(n+1)b_{0}) \hat{V}_{mn}^{0} , 
\end{equation}
where the vector $\hat{\V}$ is modified at $i=0$ as
\begin{equation}
\hat{V}_{mn}^{0} = \tilde V_{mn}^{0} \, R_{0}^{-2n-1} .
\end{equation}
Combining Eqs. (\ref{eq:Robin_A}, \ref{eq:Robin_A2}), one gets a
complete ISLEA that fully determines all the coefficients $A_{mn}^i$.
As previously, the solution can be written in a matrix form as
\begin{equation}
\label{eq:UA_cond_int}
\A = \W \hat{\V}, \qquad \W = \bigl[(2\hat{\nn} + \I)\hat{\b} + 
(\hat{\a} - \hat{\nn}' \hat{\b}) \hat{\U}\bigr]^{-1} (\hat{\a} - \hat{\nn}' \hat{\b}) ,
\end{equation}
where the new matrix $\hat{\nn}'$ includes the change of $n$ to $n+1$
in front of $b_0$ in Eq. (\ref{eq:Robin_A2}) for the outer boundary:
\begin{equation}
\bigl(\hat{\nn}'\bigr)_{mnkl}^{ij} = \delta_{ij} \delta_{mk} \delta_{nl} (n + \delta_{i0}) .
\end{equation}
The Green function reads
\begin{equation}
\label{eq:GRobin_int}
\begin{split}
G(\x,\xo) & = \G(\x,\xo) - \sum\limits_{i=1}^{N} \sum\limits_{n,m} A_{mn}^i \psi_{mn}^{-}(r_i,\theta_i,\phi_i) 
 - \sum\limits_{n,m} A_{mn}^{0} \psi_{mn}^{+}(r_{0},\theta_{0},\phi_{0}) . \\
\end{split}
\end{equation}
Note that when $R_{0}$ goes to infinity, the elements
$\hat{V}_{mn}^{0}$, as well as the nondiagonal elements of the matrix
$\hat{\U}$ corresponding to $A_{mn}^{0}$, vanish, so that $A_{mn}^{0}
= 0$ and one retrieves the solution for the exterior problem.

With this Green function, the solution of a general interior Robin
boundary value problem,
\begin{subequations}
\begin{eqnarray}
-\nabla ^2u &=&F\quad (\x\in \Omega ^{+}),  \label{Robin1_int} \\
\left. \left( a_iu+b_iR_i\frac{\partial u}{\partial \n_\x}\right) 
\right| _{{\partial \Omega }_i} &=&f_i\quad (i=\overline{0,N}), \label{Robin2_int} 
\end{eqnarray}
\end{subequations}
can be represented as 
\begin{equation}
u(\xo)= \int\limits_{\Omega ^{+}}d\x\,F(\x)\,G(\x,\xo)+\sum\limits_{i=0}^N
\int\limits_{{\partial \Omega }_i}d\s\,f_i(\s)\,\omega _{\xo}^i(\s),
\label{eq:u_general_int}
\end{equation}
where the spread harmonic measure density $\omega _{\xo}^i(\s)$ is
expressed through $G(\x,\xo)$ by Eq. (\ref{eq:HMspread}).  For $i >
0$, $\omega _{\xo}^i(\s)$ is given by Eq. (\ref{eq:HMspread_R}) for
$b_i > 0$, and by Eq. (\ref{eq:HM}) for $b_i = 0$.  In turn, for $i =
0$, one finds
\begin{equation}
\label{eq:omegaN}
\omega _{\xo}^0(\s) = \begin{cases} 
\displaystyle \frac{1}{b_0 R_0} \sum\limits_{n,m} \bigl(\hat{V}_{mn}^0 
- \bigl(\hat{\U}\A\bigr)_{mn}^0 \bigr) \psi_{mn}^{+}(R_0,\theta_0,\phi_0)  \qquad (b_0 > 0) ,\cr
\displaystyle \frac{1}{a_0 R_0} \sum\limits_{n,m} (2n+1) A_{mn}^0 \, \psi_{mn}^{+}(R_0,\theta_0,\phi_0) 
\hskip 14.5mm (b_0 = 0) .\end{cases}
\end{equation}

\subsection{Conjugate problems}
\label{sec:trans}

In many biological applications, the diffusive transport occurs in
heterogeneous media, with distinct diffusion coefficients in different
regions.  A pack of balls is a basic model of a tissue that is formed
by individual cells located in the extracellular space
\cite{Grebenkov10}.  A diffusing molecule can cross cell membranes and
move from a cell to the extracellular space and back.  Such diffusion
processes are often described with conjugate boundary conditions on
the surface between any two adjacent ``compartments'' of the medium
(also known as the fourth boundary condition, transmission condition,
and exchange condition).  When the surface is fully permeable, the
concentration $u$ of diffusing molecules obeys two conditions: (i) the
continuity of the concentration,
\begin{equation}  \label{eq:BCT_1a}
\left. u \right|_{\pa^{-}} = \left. u\right|_{\pa^{+}} ; 
\end{equation}
and (ii) the continuity of the diffusive flux at the surface,
\begin{equation}  \label{eq:BCT_2}
\left. -\left(D^{-} \frac{\partial u}{\partial_{\n_\x}}\right)\right|_{\pa^{-}} 
= \left. \left(D^{+} \frac{\partial u}{\partial_{\n_\x}}\right)\right|_{\pa^{+}}, 
\end{equation}
where $D^{\pm}$ are diffusion coefficients on both sides of the
surface (denoted by $\pa^{\pm}$).  Note that the normal derivatives
$\partial/\partial_{\n_\x}$ on both sides are directed outwards the
corresponding compartment and thus opposite.  When the membrane
presents some ``resistance'' to exchange between compartments, the
first condition is replaced by
\begin{equation}  \label{eq:BCT_1b}
\left. -\left(D^{-} \frac{\partial u}{\partial_{\n_\x}}\right)\right|_{\pa^{-}} 
= \kappa \bigl( \left. u\right|_{\pa^{-}} - \left. u\right|_{\pa^{+}}\bigr) ,
\end{equation}
which states that the diffusive flux is proportional to the drop of
concentrations on both sides.  Here $\kappa \geq 0$ is the
permeability of the surface quantifying how difficult is to cross it:
the limit $\kappa = 0$ corresponds to a fully impermeable boundary (in
which case one recovers two uncoupled Neumann conditions on both
sides), whereas the limit $\kappa \to\infty$ describes the former
situation of a fully permeable surface (in which case
Eq. (\ref{eq:BCT_1b}) is reduced to Eq. (\ref{eq:BCT_1a})).  Only in
the case of a fully impermeable surface, one can treat the boundary
value problem separately in two compartments, in particular, one can
use solutions from previous subsections to describe separately
intracellular and extracellular diffusions.  In contrast, whenever
$\kappa > 0$, the two problems are coupled and should thus be treated
simultaneously.  As a consequence, the conjugate problems are more
difficult to solve.  Moreover, even a general formulation of conjugate
problems is more challenging because one can imagine a large
compartment (e.g., the extracellular space) filled with smaller
compartments, each of them is filled with even smaller compartments,
and so on (like Russian nested dolls).  Although the GMSV can still be
applied to such complicated cases (when all compartments are
spherical), we do not consider this general setting.

For illustration purposes, we limit ourselves to the practically
relevant situation of $N$ non-overlapping balls $\Omega_i$ and the
extracellular space $\Omega^{-} = \R^3\backslash \bigl(\cup_{i=1}^N
\overline{\Omega}_i\bigr)$.  We search for the Green function
$G(\x,\xo)$ that satisfies general conjugate boundary conditions
\begin{subequations}  \label{eq:BCT}
\begin{eqnarray}  \label{eq:BCTa}
\left. \biggl(a_i G + b_i R_i \frac{\partial G}{\partial \n_\x} \biggr)\right|_{\x\in \pa_i^{-}} &=&
\left. \biggl(\bar{a}_i G + \bar{b}_i R_i \frac{\partial G}{\partial \n_\x} \biggr)\right|_{\x\in \pa_i^{+}} , \\
\label{eq:BCTb}
\left. \biggl(c_i G + d_i R_i \frac{\partial G}{\partial \n_\x} \biggr)\right|_{\x\in \pa_i^{-}} &=&
\left. \biggl(\bar{c}_i G + \bar{d}_i R_i \frac{\partial G}{\partial \n_\x} \biggr)\right|_{\x\in \pa_i^{+}} ,
\end{eqnarray}
\end{subequations}
where the parameters $a_i,\, b_i,\, c_i,\, d_i$ characterize the
exterior compartment $\Omega^{-}$ (near the surface $\pa_i$), while
the parameters $\bar{a}_i,\, \bar{b}_i,\, \bar{c}_i,\, \bar{d}_i$
characterize the interior spherical compartment $\Omega_i$.  As one
needs to relate the Green function in the exterior compartment to that
in the interior compartment, there are {\it two} conjugate relations
at $\pa_i$, in contrast to former Robin boundary conditions with a
single relation.  These two relations should be linearly independent,
i.e. one relation should not be reduced to the other (e.g., as
Eqs. (\ref{eq:BCT_2}, \ref{eq:BCT_1b})).

For convenience, we denote the restrictions of $G(\x,\xo)$ to
$\Omega^{-}$ and to $\Omega_i$ as $G^{-}$ and $G^{+i}$, respectively.
We consider separately two cases: $\xo \in \Omega^{-}$ and $\xo \in
\Omega_i$.

(i) When $\xo \in \Omega^{-}$, each function $G^{+i}(\x,\xo)$
satisfies the Laplace equation in $\Omega_i$, $\nabla^2_\x
G^{+i}(\x,\xo) = 0$, and it is thus naturally decomposed onto the
regular solid harmonics in the local spherical coordinates of the ball
$\Omega_i$:
\begin{equation}  \label{eq:G+i}
G^{+i}(\x,\xo) = \sum\limits_{n,m} \bar{A}_{mn}^i \, \psi_{mn}^{+}(r_i,\theta_i,\phi_i) ,
\end{equation}
with unknown coefficients $\bar{A}_{mn}^i$.  In turn, the function
$G^{-}(\x,\xo)$ can be represented as $\G(\x,\xo) - g(\x;\xo)$, with
an auxiliary function $g$ satisfying the Laplace equation in
$\Omega^{-}$, and $\G$ given by Eq. (\ref{eq:Gfund}).  The conjugate
boundary conditions (\ref{eq:BCT}) read
\begin{subequations}  \label{eq:BCT2}
\begin{align}
& \left. \biggl(a_i g + b_i R_i \frac{\partial g}{\partial \n_\x} +
\bar{a}_i G^{+i} + \bar{b}_i R_i \frac{\partial G^{+i}}{\partial \n_\x} \biggr)\right|_{\x\in \pa_i} =
\left. \biggl(a_i \G + b_i R_i \frac{\partial \G}{\partial \n_\x} \biggr)\right|_{\x\in \pa_i}  , \\
& \left. \biggl(c_i g + d_i R_i \frac{\partial g}{\partial \n_\x} +
\bar{c}_i G^{+i} + \bar{d}_i R_i \frac{\partial G^{+i}}{\partial \n_\x} \biggr)\right|_{\x\in \pa_i} =
\left. \biggl(c_i \G + d_i R_i \frac{\partial \G}{\partial \n_\x} \biggr)\right|_{\x\in \pa_i} ,
\end{align}
\end{subequations}
where $\pa_i^\pm$ were replaced by $\pa_i$, as the appropriate side of
the surface is now clear from notations.  The function $g(\x;\xo)$ is
represented again as the sum (\ref{green4a}) of partial solutions.
With the aid of addition theorems, one can express $g$ in the local
coordinates of the ball $\Omega_i$, whereas the left-hand side of
Eqs. (\ref{eq:BCT2}) is an explicit function, which can be decomposed
over the regular solid harmonics.  Repeating the steps of
Sec. \ref{sec:ExtR}, we get the ISLAE for $i=\overline{1,N}, ~ n =
\overline{0,\infty}, ~ m = \overline{-n,n}$:
\begin{subequations} \label{eq:trans_A}
\begin{align}
(2n+1)b_iA_{mn}^i+(a_i-nb_i)\bigl(\hat{\U}\A)_{mn}^i + R_i^{2n+1}(\bar{a}_i + n \bar{b}_i) \bar{A}_{mn}^i &=& (a_i-nb_i)\hat V_{mn}^i ,\\
(2n+1)d_iA_{mn}^i+(c_i-nd_i)\bigl(\hat{\U}\A)_{mn}^i + R_i^{2n+1}(\bar{c}_i + n \bar{d}_i) \bar{A}_{mn}^i &=& (c_i-nd_i)\hat V_{mn}^i ,
\end{align}
\end{subequations}
where the elements of $\hat{\U}$ and $\hat{\V}$ were defined by
Eqs. (\ref{eq:Uhat}, \ref{eq:Vhat}).  These relations generalize
Eqs. (\ref{eq:Robin_A}) by the inclusion of the terms with
$\bar{A}_{mn}^i$ that account for coupling between exterior and
interior compartments.  Although the number of unknowns is doubled
($A_{mn}^i$ and $\bar{A}_{mn}^i$), the number of equations is also
doubled.  Writing these equations in a matrix form, one can solve the
truncated system to determine the unknown coefficients and thus the
Green function.

(ii) When $\xo \in \Omega_k$ for some $k$, each function
$G^{+i}(\x,\xo)$ with $i\ne k$ satisfies the Laplace equation in
$\Omega_i$ and can thus be searched in the form (\ref{eq:G+i}).
Moreover, $G^{-}$ satisfies the Laplace equation in $\Omega^{-}$ so
that one can set $G^{-} = g$ and search it as the sum (\ref{green4a})
of partial solutions.  In turn, the function $G^{+k}(\x,\xo)$ can be
represented as $\G(\x,\xo) - g^{+k}(\x;\xo)$, with
\begin{equation}
g^{+k}(\x;\xo) = \sum\limits_{n,m} \bar{A}_{mn}^k \, \psi_{mn}^{+}(r_k,\theta_k,\phi_k) .
\end{equation}
The conjugate boundary conditions (\ref{eq:BCT}) read
\begin{subequations}  \label{eq:BCT3a}
\begin{align}
& \left. \biggl(a_k g + b_k R_k \frac{\partial g}{\partial \n_\x} +
\bar{a}_k g^{+k} + \bar{b}_k R_k \frac{\partial g^{+k}}{\partial \n_\x} \biggr)\right|_{\x\in \pa_k} =
\left. \biggl(a_k \G + b_k R_k \frac{\partial \G}{\partial \n_\x} \biggr)\right|_{\x\in \pa_k}  , \\
& \left. \biggl(c_k g + d_k R_k \frac{\partial g}{\partial \n_\x} +
\bar{c}_k g^{+k} + \bar{d}_k R_k \frac{\partial g^{+k}}{\partial \n_\x} \biggr)\right|_{\x\in \pa_k} =
\left. \biggl(c_k \G + d_k R_k \frac{\partial \G}{\partial \n_\x} \biggr)\right|_{\x\in \pa_k} 
\end{align}
\end{subequations}
for $i = k$, and 
\begin{subequations}  \label{eq:BCT3b}
\begin{align}
& \left. \biggl(a_i g + b_i R_i \frac{\partial g}{\partial \n_\x} - 
\bar{a}_i G^{+i} - \bar{b}_i R_i \frac{\partial G^{+i}}{\partial \n_\x} \biggr)\right|_{\x\in \pa_i} = 0  , \\
& \left. \biggl(c_i g + d_i R_i \frac{\partial g}{\partial \n_\x}  -
\bar{c}_i G^{+i} - \bar{d}_i R_i \frac{\partial G^{+i}}{\partial \n_\x} \biggr)\right|_{\x\in \pa_i} = 0
\end{align}
\end{subequations}
for $i \ne k$.  One can repeat again the steps of Sec. \ref{sec:ExtR}
to derive linear equations on the unknown coefficients.  The only
difference is that one needs to employ another representation of the
fundamental solution inside $\Omega_k$ (similar to
Eq. (\ref{eq:Gfund_int}) for the interior problem):
\begin{equation}
\G(\x,\xo) = \sum\limits_{n,m} \tilde{V}_{mn}^k \, \psi_{mn}^{-}(r_k,\theta_k,\phi_k) \qquad (r_k > L_k),
\end{equation}
where 
\begin{equation}
\tilde{V}_{mn}^k = \frac{(-1)^m}{4\pi} \, \psi_{(-m)n}^{+}(L_k,\Theta_k,\Phi_k),
\end{equation}
with $\L_k = \xo - \x_k$.  We get thus
\begin{subequations} \label{eq:trans_A2a}
\begin{align}
& (2n+1)b_kA_{mn}^k+(a_k-nb_k)\bigl(\hat{\U}\A)_{mn}^k + R_k^{2n+1}(\bar{a}_k + n \bar{b}_k) \bar{A}_{mn}^k = (a_k-(n+1)b_k)\hat V_{mn}^k ,\\
& (2n+1)d_kA_{mn}^k+(c_k-nd_k)\bigl(\hat{\U}\A)_{mn}^k + R_k^{2n+1}(\bar{c}_k + n \bar{d}_k) \bar{A}_{mn}^k = (c_k-(n+1)d_k)\hat V_{mn}^k 
\end{align}
\end{subequations}
for $i = k$, and
\begin{subequations} \label{eq:trans_A2b}
\begin{align}
& (2n+1)b_iA_{mn}^i+(a_i-nb_i)\bigl(\hat{\U}\A)_{mn}^i - R_i^{2n+1}(\bar{a}_i + n \bar{b}_i) \bar{A}_{mn}^i = 0 ,\\
& (2n+1)d_iA_{mn}^i+(c_i-nd_i)\bigl(\hat{\U}\A)_{mn}^i - R_i^{2n+1}(\bar{c}_i + n \bar{d}_i) \bar{A}_{mn}^i = 0 
\end{align}
\end{subequations}
for $i \ne k$, with $n = \overline{0,\infty}$ and $m =
\overline{-n,n}$.

The above computation can be straightforwardly extended to the case
when the extracellular space is bounded by a large ball $\Omega_0$.
In this case, the analysis of the exterior part (i.e., the evaluation
of the function $g$) should follow Sec. \ref{sec:Int} instead of
Sec. \ref{sec:ExtR}.  Finally, one can consider more general problems,
in which some balls are partially absorbing sinks or impermeable
obstacles (with Robin or Neumann boundary conditions), whereas the
other ball allow for interior diffusion (with conjugate boundary
conditions).  One just combines the corresponding conjugate conditions
with Robin boundary conditions.  Moreover, it is worth noting that the
conjugate problem naturally includes the Robin boundary value problem
as a particular case.  In fact, setting $G^{+i} \equiv 0$ in
Eq. (\ref{eq:BCTa}) and removing Eq. (\ref{eq:BCTb}), one recovers the
Robin boundary condition (\ref{eq:GRobin_BC}).

\section{Basic examples}
\label{sec:examples}

\subsection{Interior problem for two concentric spheres}

As an example of an interior Robin problem, we determine the Green
function in a bounded domain $\Omega^{+}$ between two concentric
spheres, centered at $\x_0 = \x_1 = 0$ and of radii $R_0 > R_1$.
Although this problem could be solived via the spectral decomposition
over the known Laplacian eigenfunctions, our derivation yields a more
explicit formula and serves as an illustration for the proposed
method.

One can check that
\begin{equation}
\hat{U}_{mnkl}^{10} = R_1^{2n+1} \, \delta_{ln} \, \delta_{km} , \qquad
\hat{U}_{mnkl}^{01} = R_0^{-2n-1} \, \delta_{ln} \, \delta_{km} , 
\end{equation}
i.e., the matrix $\hat{\U}$ is formed by four diagonal matrices.  This
structure is preserved by multiplication by diagonal matrices
$\hat{\a}$, $\hat{\b}$, $\hat{\nn}$, and $\hat{\nn}'$ in the matrix
relation (\ref{eq:UA_cond_int}).  The resulting matrix $\W$ has the
block three-diagonal structure:
\begin{equation}
\begin{split}
W_{mnkl}^{11} &=   \delta_{nl} \, \delta_{km} \, w_n^{11}, \qquad
W_{mnkl}^{10}  = - \delta_{nl} \, \delta_{km} \, w_n^{10} R_1^{2n+1}, \\
W_{mnkl}^{01} &= - \delta_{nl} \, \delta_{km} \, w_n^{01} R_0^{-2n-1}, \qquad
W_{mnkl}^{00}  =   \delta_{nl} \, \delta_{km} \, w_n^{00}, \\
\end{split}
\end{equation}
where
\begin{equation}
\begin{split}
w_n^{11} & = w_n (a_1-nb_1)(a_0+nb_0) , \qquad
w_n^{10} = w_n (a_1-nb_1)^2  , \\
w_n^{01} & = w_n (a_0-(n+1)b_0)^2 ,\qquad
w_n^{00}  = w_n (a_1+(n+1)b_1)(a_0-(n+1)b_0) , \\
\end{split}
\end{equation}
and
\begin{equation}
w_n = \frac{1}{(a_1+(n+1)b_1)(a_0+nb_0) - (a_1-nb_1)(a_0-(n+1)b_0) (R_1/R_0)^{2n+1}} \,.
\end{equation}
Since $\L_1 = \xo - \x_1 = \xo = \xo - \x_0 = \L_0$, one has
\begin{subequations}
\begin{align}
& \hat{V}_{mn}^1 = R_1^{2n+1} \frac{(-1)^m}{4\pi} \psi_{(-m)n}^{-}(L_{1},\Theta_{1},\Phi_{1}) 
= \frac{(-1)^m}{4\pi} \frac{R_1^{2n+1}}{r_\xo^{n+1}} Y_{(-m)n}(\theta_\xo,\phi_\xo) , \\
& \hat{V}_{mn}^2 = R_0^{-2n-1} \frac{(-1)^m}{4\pi} \psi_{(-m)n}^{+}(L_{0},\Theta_{0},\Phi_{0}) 
= \frac{(-1)^m}{4\pi} \frac{r_\xo^n}{R_0^{2n+1}} Y_{(-m)n}(\theta_\xo,\phi_\xo) , 
\end{align}
\end{subequations}
where $(r_\xo,\theta_\xo,\phi_\xo)$ are the spherical coordinates of
$\y$.  One gets thus
\begin{subequations}
\begin{align}
& A_{mn}^1 = \frac{(-1)^m}{4\pi} Y_{(-m)n}(\theta_\xo,\phi_\xo) \frac{R_1^{2n+1}}{r_\xo^{n+1}} \bigl[w_n^{11} - w_n^{10} (r_\xo/R_0)^{2n+1}\bigr] , \\
& A_{mn}^2 = \frac{(-1)^m}{4\pi} Y_{(-m)n}(\theta_\xo,\phi_\xo) \frac{r_\xo^n}{R_0^{2n+1}} \bigl[w_n^{00} - w_n^{01} (R_1/r_\xo)^{2n+1}\bigr], 
\end{align}
\end{subequations}
from which
\begin{eqnarray} \nonumber
G(\x,\xo) &=& \G(\x,\xo) - \frac{1}{4\pi} \sum\limits_{n=0}^\infty P_n\left(\frac{(\x\cdot \xo)}{\|\x\|\, \|\xo\|}\right) 
\biggl\{\frac{R_1^{2n+1}}{(\|\x\| \, \|\xo\|)^{n+1}} \bigl(w_n^{11} - w_n^{10} (\|\xo\|/R_0)^{2n+1}\bigr)  \\
\label{eq:GR_twospheres}
&+& \frac{(\|\x\| \, \|\xo\|)^n}{R_0^{2n+1}} \bigl(w_n^{00}  - w_n^{01} (R_1/\|\xo\|)^{2n+1}\bigr)  \biggr\} , 
\end{eqnarray}
where the sum over $m$ was calculated by using the classical addition
theorem for two unit vectors $\mathbf{e}$ and $\mathbf{e}^{\prime}$:
\begin{equation}  \label{eq:Ymn_addition}
\sum\limits_{m=-n}^n (-1)^m Y_{mn}(\theta,\phi)
Y_{(-m)n}(\theta^{\prime},\phi^{\prime}) 
= P_n(\mathbf{e}\cdot \mathbf{e}^{\prime}) .
\end{equation}
The spread harmonic measure density $\omega_{\xo}$ follows
from its definition (\ref{eq:HMspread}):
\begin{align}
& \omega_{\xo}^1(\s) = \left. \left(- \frac{\partial G}{a_1 \partial \n_\x}\right)\right|_{\x=\s\in \pa_1} = 
\frac{1}{4\pi a_1 R_1^2} \sum\limits_{n=0}^\infty P_n\left(\frac{(\s\cdot \xo)}{R_1\, \|\xo\|}\right) \\
\nonumber
& \times \biggl\{ \frac{R_1^{n+1}}{\|\xo\|^{n+1}} \bigl(n + (n+1)w_n^{11} + n(R_1/R_0)^{2n+1} w_n^{01}\bigr)
- \frac{\|\xo\|^n \, R_1^{n+1}}{R_0^{2n+1}} \bigl(n w_n^{00} + (n+1) w_n^{10}\bigr)   \biggr\} , 
\end{align}
from which the absorption probability (see also
Sec. \ref{sec:applications}) reads
\begin{equation}
p_1(\xo) = a_1 \int\limits_{\pa_1} d\s \, \omega_{\xo}^1(\s) = w_0^{11} R_1/\|\xo\| - w_0^{10} R_1/R_0.
\end{equation}
Similarly, we get
\begin{equation}
\begin{split}
& \omega_{\xo}^0(\s) = \left. \left(- \frac{\partial G}{a_0 \partial \n_\x}\right)\right|_{\x=\s\in \pa_0} = 
\frac{1}{4\pi a_0 R_0^2} \sum\limits_{n=0}^\infty P_n\left(\frac{(\s\cdot \xo)}{R_0\, \|\xo\|}\right) \frac{1}{R_0^n} \\
& \times \biggl\{ \|\xo\|^n \bigl((n+1) + (n+1)w_n^{10}(R_1/R_0)^{2n+1} + n w_n^{00}\bigr)
- \frac{R_1^{2n+1}}{\|\xo\|^{n+1}} \bigl( (n+1)w_n^{11} + n w_n^{01}\bigr) \biggr\} , \\
\end{split}
\end{equation}
from which
\begin{equation}
p_0(\xo) = a_0 \int\limits_{\pa_0} d\s \, \omega_{\xo}^0(\s) = 1 + w_0^{10} R_1/R_0 - w_0^{11} R_1/\|\xo\| .
\end{equation}
Note that $p_1 + p_0 = 1$ as expected.  

We are not aware of earlier derivations of the Robin Green function
and the spread harmonic measure for two concentric spheres in such
simple forms.  The Green functions in four limiting cases (i.e.,
Dirichlet-Dirichlet, Dirichlet-Neumann, Neumann-Dirichlet, and
Neumann-Neumann conditions on the inner and outer spheres) was
provided in \cite{Chang16}.  These solutions can be easily deduced
from our general formula (\ref{eq:GR_twospheres}).  For instance, in
the Dirichlet case ($b_1 = b_0 =0$), one has $w_n^{ij} = w_n$, and the
formula (\ref{eq:GR_twospheres}) reads
\begin{eqnarray}
\label{eq:GD_twospheres}
G(\x,\xo) & = & \G(\x,\xo) - \sum\limits_{n=0}^\infty  P_n\biggl(\frac{(\x\cdot \xo)}{\|\x\| \, \|\xo\|}\biggr) \\
\nonumber
& \times &\frac{R_1^{2n+1} (R_0^{2n+1} - \|\x\|^{2n+1}) + \|\xo\|^{2n+1} (\|\x\|^{2n+1} - R_1^{2n+1})}
{4\pi \, \|\x\|^{n+1}\, \|\xo\|^{n+1} (R_0^{2n+1} - R_1^{2n+1})} . 
\end{eqnarray}
Note also that an explicit form of the Dirichlet Green function for
two nonconcentric spheres was derived by using bispherical coordinates
in \cite{Chen13}.

\subsection{Interior Robin problem for one sphere}

In the limit $R_1\to 0$, Eq. (\ref{eq:GR_twospheres}) is reduced to
\begin{equation}
\label{eq:GR_onephere_int}
G(\x,\xo) = \G(\x,\xo) - \frac{1}{4\pi} \sum\limits_{n=0}^\infty P_n\left(\frac{(\x\cdot \xo)}{\|\x\|\, \|\xo\|}\right) 
\frac{\|\x\|^n\, \|\xo\|^n}{R_0^{2n+1}} \, \frac{a_0 -(n+1)b_0}{a_0 + n b_0} \,,
\end{equation}
i.e., we get the Green function for the interior Robin problem in a
ball of radius $R_0$.  In the Dirichlet case, setting $b_2 = 0$ and
using the identity
\begin{equation}  \label{eq:auxil12}
\frac{1}{\sqrt{1-2qz + z^2}} = \sum\limits_{n=0}^\infty P_n(q) \, z^n ,
\end{equation}
one retrieves the classical result
\begin{equation}  \label{eq:Gonesphere}
G(\x,\xo) = \frac{1}{4\pi \| \x - \xo\|} - \frac{R_0/\|\xo\|}{4\pi \left\|\x - \xo R_0^2/\|\xo\|^2\right\|} \,,
\end{equation}
which is usually deduced by the image method \cite{Courant}.  From the
identity (\ref{eq:auxil12}), one also gets
\begin{equation}  \label{eq:auxil12b}
\frac{qz-z^2}{(1-2qz + z^2)^{3/2}} = \sum\limits_{n=0}^\infty n P_n(q)\, z^n ,
\end{equation}
that helps to deduce the harmonic measure density
\begin{equation}  \label{eq:HM_onesphere}
\omega_{\xo}(\s) = \frac{1}{4\pi R_0} \, \frac{R_0^2 - \|\xo\|^2}{\|\xo - \s\|^3} \,.
\end{equation}
Note that the Green function does not exist for the interior Neumann
problem.  This can be seen directly from
Eq. (\ref{eq:GR_onephere_int}) which diverges as $a_0 \to 0$.

\subsection{Exterior Robin problem for one sphere}

In the limit $R_0\to\infty$, Eq. (\ref{eq:GR_twospheres}) is reduced
to
\begin{equation}
\label{eq:GR_onephere_ext}
G(\x,\xo) = \G(\x,\xo) - \frac{1}{4\pi} \sum\limits_{n=0}^\infty 
P_n\left(\frac{(\x\cdot \xo)}{\|\x\|\, \|\xo\|}\right) \frac{R_1^{2n+1}}{\|\x\|^{n+1}\, \|\xo\|^{n+1}}
\, \frac{a_1 - n b_1}{a_1 + (n+1)b_1} \,,
\end{equation}
i.e., we get the Green function for the exterior Robin problem outside
the ball of radius $R_1$.  The integral of the spread harmonic measure
over the sphere $\pa_1$ yields the absorption probability on the
partially absorbing sink of radius $R_1$:
\begin{equation}
p_1(\xo) = \frac{a_1}{a_1+b_1} \, \frac{R_1}{\|\xo\|} \,.
\end{equation}

In the Dirichlet case, the sum in Eq. (\ref{eq:GR_onephere_ext}) is
again reduced to (\ref{eq:Gonesphere}) (in which $R_0$ is replaced by
$R_1$), whereas the (non-normalized) harmonic measure density becomes
\begin{equation}  \label{eq:HM_sphereone_ext}
\omega_{\xo}(\s) = \frac{1}{4\pi R_1} \, 
\frac{\|\xo\|^2 - R_1^2}{\|\xo - \s\|^3} \,.
\end{equation}
Its integral over the sphere yields the classical result
\begin{equation}  \label{eq:p_onesphere}
p_1(\xo) = \frac{R_1}{\|\xo\|} \,.
\end{equation}

In the Neumann case, the sum in Eq. (\ref{eq:GR_onephere_ext}) becomes
\begin{equation}
G(\x,\xo) = \G(\x,\xo) + \frac{1}{4\pi} \sum\limits_{n=0}^\infty P_n\left(\frac{(\x\cdot \xo)}{\|\x\|\, \|\xo\|}\right) 
\frac{R_1^{2n+1}}{\|\x\|^{n+1}\, \|\xo\|^{n+1}} \biggl(1 - \frac{1}{n+1}\biggr) .
\end{equation}
The first sum was already computed in the Dirichlet case, whereas the
second sum can be evaluated by taking the integral of
Eq. (\ref{eq:auxil12}) from $0$ to $t$, yielding
\begin{eqnarray}
\label{eq:GN_onephere_ext}
G(\x,\xo) & =& \frac{2}{4\pi \| \x - \xo\|} - 
\frac{R_1/\|\xo\|}{4\pi \left\|\x - \xo R_1^2/\| \xo\|^2\right\|} \\
\nonumber
& +& \frac{1}{4\pi R_1} \ln\left( \frac{R_1^2 - (\x\cdot \xo) + \|\xo\|\, \| \x - \xo R_1^2/\|\xo\|^2\|}{\|\x\| \, \|\xo\| - (\x\cdot \xo)}\right). 
\end{eqnarray}
Note that the Dirichlet Green function for exterior and interior
problems for a prolate spheroid was recently analyzed in \cite{Xue17}.

\section{Applications}
\label{sec:applications}

The knowledge of the Green function $G(\x,\xo)$ provides the solution
of any boundary value problem associated to the Laplace or Poisson
equation.  In this section, we just mention several quantities that
often appear in various applications and can be directly deduced by
using our solution.

\subsection{Hitting and splitting probabilities}

As mentioned earlier, the normal derivative of the Green function
yields the harmonic measure density $\omega_\xo^i(\s)$, which
characterizes the likelihood for Brownian motion started from $\xo$ to
arrive at the absorbing boundary for the first time in a vicinity of
the boundary point $\s \in \Omega_i$ \cite{Garnett}.  Integrating
Eq. (\ref{eq:HM}) over the sphere ${\partial\Omega}_i$, one gets the
probability of the first arrival onto the ball $\Omega_i$:
\begin{equation}  \label{eq:pi}
p_i(\xo) = \int\limits_{{\partial\Omega}_i} d\s \left . \omega_{\xo}^i(\s) \right|_{{\partial\Omega}_i} = 4\pi A_{00}^i ,
\end{equation}
where only the rotation-invariant term with $n=m=0$ survived.  This is
also known as the hitting probability or the splitting probability,
i.e., the probability of arrival at the ball $\Omega_i$ before
arriving onto other balls or escaping at infinity.  From this
relation, the escape probability $P_\infty (\xo)$ from a fixed
starting point $\xo$ to infinity is
\begin{equation}
P_\infty(\xo) =1-\sum\limits_{i=1}^N p_i(\xo) = 1- 4\pi \sum\limits_{i=1}^N A_{00}^i.  \label{eq:Pinf}
\end{equation}
In other words, $P_\infty $ is the probability that the particle does
not hit any absorbing sink.  This is a nontrivial quantity in three
dimensions because of the transient character of Brownian motion (in
contrast, $P_\infty $ is always zero in two dimensions).  We recall
that the dependence of $P_\infty$ on $\xo$ enters through the
coefficients $A_{00}^i$ that are expressed as linear combinations of
$\hat V_{mn}^i$.

When the balls are only partially absorbing with Robin boundary
conditions, the first arrival onto the ball does not necessarily imply
absorption or chemical reaction, as the particle can be reflected.
The absorption can thus be realized after numerous returns to the
ball.  The probability density of such absorption events is called the
spread harmonic measure density $\omega_\xo^i(\s)$ and given by Eq.
(\ref{eq:HMspread_R}).  Integrating again this density over the
boundary $\pa_i$, one gets the probability of absorption on the
partially absorbing sphere ${\partial \Omega }_i$ as
\begin{equation}  \label{eq:HMs}
p_i(\xo) := \int\limits_{{\partial \Omega }_i}d\s\, \left. \left( -%
\frac{\partial G(\x,\xo)}{\partial \n_\x}\right) \right|_{\x=\s\in {\partial \Omega }_i} 
 = a_i \int\limits_{{\partial \Omega }_i}d\s\,
\omega _{\xo}^i(\s)= 4\pi A_{00}^i \qquad (a_i > 0), 
\end{equation}
the last equality coming from Eq. (\ref{eq:Robin_A}).  Note that
$p_i(\xo) =0$ if $a_i = 0$ that corresponds to the Neumann boundary
condition.  The escape probability is still given by
Eq. (\ref{eq:Pinf}).

Finally, when the balls $\Omega_i$ are englobed by a larger ball
$\Omega_0$, the diffusing particle cannot escape to infinity but can
be absorbed by the outer boundary $\pa_0$.  The corresponding spread
harmonic measure density $\omega_\xo^0(\s)$ is given by
Eq. (\ref{eq:omegaN}).  Integrating this density over the sphere
$\pa_0$, one gets
\begin{equation}
\label{eq:pN_int}
p_0(\xo) := a_0 \int\limits_{\pa_0} d\s \, \omega_{\xo}^0(\s) = 4\pi R_0 \times
\begin{cases}
 \bigl(\hat{V}_{00}^0 - \bigl(\hat{\U}\A\bigr)_{00}^0 \bigr)a_0/b_0 \quad (b_0 > 0) , \cr
 A_{00}^0   \hskip 31mm (b_0 = 0) .\end{cases}
\end{equation}
Note also that if at least one $a_i$ is nonzero, then the
probabilities $p_i$ in a bounded domain satisfies
\begin{equation}
\sum\limits_{i=0}^N p_i(\xo) = 1 .
\end{equation}
This relation can be obtained by integrating Eq. (\ref{green3a1}) over
$\x\in \Omega^{+}$, applying the Green formula and using
Eq. (\ref{eq:HMspread}).  In probabilistic terms, it simply means that
a particle released at $\xo$ unavoidably arrives at some sink in a
bounded domain.

\subsection{Diffusive flux and reaction rate}

In chemical kinetics, the escape probability $P_\infty(\xo)$ from
Eq. (\ref{eq:Pinf}) can be interpreted as a concentration $n(\xo)$ of
species B diffusing from infinity towards partially absorbing sinks
(species A) \cite{Tachiya}.  Although this particular problem was
thoroughly investigated by using the GMSV in \cite{Galanti16a}, one
can easily re-derive the former results from our more general
semi-analytical solution for the Green function.  In fact, the latter
problem is conventionally formulated as an exterior boundary value
problem
\begin{subequations}
\begin{eqnarray}
-\nabla ^2 n_B &=&0\quad (\x\in \Omega ^{-}),  \label{green3c} \\
\left. \left( a_i n_B +b_iR_i\frac{\partial n_B}{\partial \n_{\x}}%
\right) \right| _{{\partial \Omega }_i} &=&0\quad (i=\overline{1,N}), \\
\label{green3c3}
\left. n_B\right| _{\Vert \x\Vert \rightarrow \infty } &\rightarrow
&n_0,
\end{eqnarray}
\end{subequations}
i.e., the field of concentration with a constant $n_0$ at infinity and
partially absorbing sinks.  Setting $n(\x) = n_0[1-u(\x)]$, one easily
shows that this is a specific case of the general problem
(\ref{eq:Robin_all}) so that the solution is
\begin{equation}
u(\xo)=\sum\limits_{i=1}^N\int\limits_{{\partial \Omega }_i} d\s\,\omega _{\xo}^i(\s) = \sum\limits_{i=1}^N p_i(\xo).
\end{equation}
As a consequence, 
\begin{equation}
n_B(\xo)=n_0\,P_\infty (\xo),  \label{eq:cPinf}
\end{equation}
i.e., the concentration field is proportional to the escape
probability.  

The flux onto the sink $\Omega _i$ can be computed as (see
\ref{sec:Aflux})
\begin{equation}
J_i:=\int\limits_{{\partial \Omega }_i}d\s\left. \left( 
-D\frac{\partial n_B}{\partial \n_{\xo}}\right) \right| _{\xo
=\s}=\pi n_0 D R_i\sum\limits_{j=1}^N W_{0000}^{ji},  \label{eq:Ji}
\end{equation}
where $D$ is the diffusion coefficient and the matrix $\W$ is defined
by Eq. (\ref{eq:UA_Robin}).  The total flux is just the sum of $J_i$:
\begin{equation}
J:=\sum\limits_{i=1}^NJ_i=4\pi n_0D\sum\limits_{i,j=1}^NW_{0000}^{ji}\,R_i.
\end{equation}
In the case of a single spherical sink, this formula yields the
classical Collins-Kimball relation \cite{Rice,Collins49} 
\begin{equation}
J = \frac{4\pi n_0 D R_1}{1+b_1/a_1} \,,
\end{equation}
which for $b_1 = 0$ is reduced to the famous Smoluchowski formula.

In some applications, the source of particles cannot be treated as
infinitely distant.  To account for a finite distance to the source,
one assumes that the particles are constantly released from an outer
spherical boundary $\pa_0$, in which case the concentration of
particles, $n_B(\x)$, satisfies
\begin{subequations}  \label{eq:n_int}
\begin{eqnarray}
-\nabla ^2 n_B &=&0\quad (\x\in \Omega ^{+}),  \label{green3d} \\
\left. \left( a_i n_B + b_iR_i\frac{\partial n_B}{\partial \n_{\x}}%
\right) \right| _{{\partial \Omega }_i} &=&0\quad (i=\overline{1,N}), \\
\left. n_B\right|_{\pa_0} & = & n_0.
\end{eqnarray}
\end{subequations}
The solution of this interior problem is simply 
\begin{equation} \label{eq:n_pN}
n_B(\x) = n_0\, p_0(\x), 
\end{equation}
where $p_0$ is given by Eq. (\ref{eq:pN_int}).  This problem has found
numerous applications in physics, electrochemistry, and biology
\cite{Sapoval94,Sapoval02,Grebenkov06b,Gill11}.  In particular, the
diffusive flux can be expressed by using the Dirichlet-to-Neumann
operator \cite{Grebenkov06b}.  Our general solution allows one to
investigate the spectral properties of this pseudo-differential
operator in various configurations of spherical sinks.

\subsection{Residence time and other functionals of Brownian motion}

The Dirichlet Green function is related to the expectation of
functionals of Brownian motion $B_t$ started from a point $\xo$,
according to the formula \cite{Morters}
\begin{equation}
\mathbb{E}_{\xo}\left\{ \int\limits_0^\tau dt\,f(B_t)\right\}
=\int\limits_{\Omega^{\pm}} d\x\,f(\x)\,G(\x,\xo), \label{green3ab} 
\end{equation}
where $\mathbb{E}_{\xo}$ is the expectation, $f$ is a measurable
function, and $\tau $ is the first passage time to the boundary of
$\Omega^{\pm}$ : $\tau =\inf \{t>0~:~B_t\in {\partial \Omega^{\pm}
}\}$ (this is valid for both exterior and interior cases).  In
particular, if
\begin{equation*}
f(\x)= \frac{1}{D} \, {\mathbb I}_C(\x),
\end{equation*}
where $D$ is the diffusion coefficient and ${\mathbb I}_C(\x)$ is the
indicator function of a subset $C\subset \Omega^{\pm}$, i.e.
\begin{equation*}
{\mathbb I}_C(\x):=\left\{ 
\begin{array}{c}
1 \quad \text{ if  }\x\in C ,\\ 
0 \quad \text{ if  }\x\notin C,
\end{array}
\right. 
\end{equation*}
then the functional (\ref{green3ab}) is the residence (or occupation)
time in $C$, i.e., the time that Brownian motion spends in $C$ until
the first arrival onto the boundary $\partial \Omega^{\pm}$, or escape
at infinity \cite{Morters,Grebenkov07b}.

When $C$ is a ball $\Omega_I$ of radius $R_I$ that is centered at
$\x_I$ and does not intersect the sinks, the addition theorem
(\ref{green7a}) allows one to compute the residence time as (see
\ref{sec:Aresidence})
\begin{equation}  \label{eq:Tresidence}
{\mathcal T}(\xo) = \frac{1}{D} \int\limits_{\Omega_I} d\x \, G(\x,\xo) 
 = \frac{4\pi R_I^3}{3D} \biggl\{ \frac{1}{4\pi L_{I}} -
\sum\limits_{j=1}^N \sum\limits_{n,m} A_{mn}^j
\psi_{mn}^{-}(L_{Ij},\Theta_{Ij},\Phi_{Ij}) \biggr\}, 
\end{equation}
where $\L_{Ij} = \x_j - \x_I$, $(L_{Ij},\Theta_{Ij},\Phi_{Ij})$ are
the spherical coordinates of $\L_{Ij}$, and $L_{I} = \| \xo - \x_I\|$.
Note that this result can also be extended to the case when $C$ is an
arbitrary union of non-overlapping balls.

\subsection{Mean first passage time}

For the interior problem, an immediate application of the
semi-analytical form of the Green function is related to the mean
first passage time (MFPT), $T_i(\xo)$, to the sink $\Omega_i$ when a
particle is started from $\xo$ and reflected from all other sinks.
The MFPT satisfies
\begin{subequations}
\begin{eqnarray}
-\nabla^2 T_i &=& 1/D \quad (\xo \in \Omega^{+}), \\
T_i|_{\pa_i} &=& 0 , \\
\left.\frac{\partial T_i}{\partial \n_{\xo}}\right|_{\pa_j} &=& 0  \quad (j\ne i) .
\end{eqnarray}
\end{subequations}
By definition of the Green function, one has
\begin{equation}
\label{eq:Ti}
T_i(\xo) = \frac{1}{D} \int\limits_{\Omega^{+}} d\x \, G(\x,\xo) ,
\end{equation}
where the Green function satisfies the boundary conditions
(\ref{eq:RBC_Green_Int}), with $a_j = \delta_{ij}$ and $b_j = 1 -
\delta_{ij}$.  Note that the integral in Eq. (\ref{eq:Ti}) can be
computed explicitly (see \ref{sec:Aint}).  Moreover, Eq. (\ref{eq:Ti})
for the MFPT resembles Eq. (\ref{eq:Tresidence}) for the residence
time, the main difference between two quantities lying in the boundary
conditions and thus in the coefficients $A_{mn}^i$.

More generally, one can find the MFPT to any combination of
absorbing/reflecting sinks or with more general partial reflections.
In addition, one can consider the space-dependent diffusion
coefficient, in which case the factor $1/D(\x)$ would remain under the
integral.

\section{Numerical aspects}
\label{sec:numerics}

\subsection{Implementation}

Our semi-analytical solution for the Green function in both exterior
and interior problems is exact and valid for any configuration of
non-overlapping balls (with or without an outer spherical boundary).
An approximation is only involved at the implementation step of this
solution that requires truncation of infinite-dimensional matrices,
vectors and series.  By setting the maximal degree $n_{\mathrm{max}}$
of solid harmonics, one truncates all the series for $n >
n_{\mathrm{max}}$ or $l > n_{\mathrm{max}}$.  We replace the triple
index $(i,m,n)$ of $A_{mn}^i$ by a single index of a vector $\A$ of
size $M = N({n_{\mathrm{max}}}+1)^2$ for exterior problems, and of
size $M = (N+1)({n_{\mathrm{max}}}+1)^2$ for interior problems (note
that the size is doubled for conjugate problems).  Similarly, the
truncated matrix $\hat{\U}$ and the truncated vector $\hat{\V}$ are of
sizes $M \times M$ and $M$, respectively.  For implementing Robin
boundary conditions, one also constructs the truncated diagonal
matrices $\hat{\a}$, $\hat{\b}$, $\hat{\nn}$, and $\hat{\nn}'$ as
illustrated in Table \ref{tab:indices}.  This table also shows one
possible ordering of the coefficients $A_{mn}^i$ as elements of the
truncated vector $\A$.  For a given configuration of balls, the
truncated matrix $\hat{\U}$ has to be computed only once.  If the
parameters $a_i$ and $b_i$ of Robin boundary conditions are fixed, the
truncated matrix $\W$ in Eq. (\ref{eq:UA_Robin}) or
Eq. (\ref{eq:UA_cond_int}) has to be computed only once by a numerical
inversion.  When $M$ is large, this is the most time-consuming
operation.  Once the truncated matrix $\W$ is found, the coefficients
$A_{mn}^i$ and the resulting Green function are computed rapidly.  In
particular, the Green function $G(\x,\xo)$ can be evaluated at any
spatial points $\x$ and $\xo$ with a low computational cost.  Note
also that many other diffusion characteristics such as the escape
probability, the residence time, and the mean first passage time (see
Sec. \ref{sec:applications}) are immediately accessible from the
computed $A_{mn}^i$ and $\hat{\U}$.

We implemented the computation of the Green function in
three-dimensional domains with disconnected spherical boundaries as a
Matlab package ``GreenBallsL'' that can be freely downloaded at
\begin{center}
\url{https://pmc.polytechnique.fr/pagesperso/dg/GBL/gbl.html}
\end{center}
In this package, one needs to specify the radii, positions and surface
properties (coefficients $a_i$ and $b_i$) of the non-overlapping
balls, as well as sets of points $\x$ and $\xo$, at which the Green
function should be calculated.  In spite of the mathematical condition
(\ref{eq:nonoverlapping}) needed to formally prove the convergence of
the solution, the package allows one to consider touching balls as
well.

\begin{table}
\begin{center}
\begin{tabular}{l l}
$A_{mn}^i$ & \small $\overbrace{\underbrace{\scriptstyle 00}_{1}  \underbrace{\scriptstyle (-1)1 ~ 01 ~ 11}_{3} 
 \ldots  \underbrace{\scriptstyle (-n)n ~ (-n+1)n ~ \ldots ~ nn}_{2n+1}  \ldots\,}^{(n_{\rm max}+1)^2~\textrm{elements for ball~}1} ~\ldots~
\overbrace{\underbrace{\scriptstyle 00}_{1} \underbrace{\scriptstyle (-1)1 ~ 01 ~ 11}_{3} 
\ldots \underbrace{\scriptstyle (-n)n ~ (-n+1)n ~ \ldots ~ nn}_{2n+1} \ldots}^{(n_{\rm max}+1)^2~\textrm{elements for ball~}N}$ \\
$\hat{\nn}$ & \small $\overbrace{\underbrace{\scriptstyle 0}_{1} \underbrace{\scriptstyle ~~~1 ~~~ 1 ~~~ 1}_{3} 
 \ldots  \underbrace{\scriptstyle ~~~~ n ~~~~~~~ n ~~~~~ \ldots ~~ n}_{2n+1} \ldots\,}^{(n_{\rm max}+1)^2~\textrm{elements for ball~}1} ~ \ldots~
\overbrace{\underbrace{\scriptstyle 0}_{1} \underbrace{\scriptstyle ~~~1 ~~~ 1 ~~~ 1}_{3} 
\ldots \underbrace{\scriptstyle ~~~~ n ~~~~~~~ n ~~~~~ \ldots ~~ n}_{2n+1} \ldots}^{(n_{\rm max}+1)^2~\textrm{elements for ball~}N}$ \\
$\hat{\a}$ & \small $\overbrace{\underbrace{\scriptstyle a_1}_{1} \underbrace{\scriptstyle ~~a_1 ~~ a_1 ~~ a_1}_{3} 
 \ldots  \underbrace{\scriptstyle ~~~ a_1 ~~~~~~ a_1 ~~~~~ \ldots \, a_1}_{2n+1} \ldots}^{(n_{\rm max}+1)^2~\textrm{elements for ball~}1} ~ \ldots~ 
 \overbrace{\underbrace{\scriptstyle a_N}_{1} \underbrace{\scriptstyle ~ a_N ~ a_N ~ a_N}_{3} 
 \ldots  \underbrace{\scriptstyle ~~ a_N ~~~~~ a_N ~~~ \ldots ~\, a_N}_{2n+1} \ldots}^{(n_{\rm max}+1)^2~\textrm{elements for ball~}N}$  \\
$\hat{\b}$ & \small $\overbrace{\underbrace{\scriptstyle b_1}_{1} \underbrace{\scriptstyle ~~b_1 ~~~ b_1 ~~ b_1}_{3} 
 \ldots \underbrace{\scriptstyle ~~ b_1 ~~~~~~ b_1 ~~~~~ \ldots ~ b_1}_{2n+1} \ldots}^{(n_{\rm max}+1)^2~\textrm{elements for ball~}1} ~ \ldots~ 
 \overbrace{\underbrace{\scriptstyle b_N}_{1} \underbrace{\scriptstyle ~~ b_N ~~ b_N ~ b_N}_{3} 
 \ldots  \underbrace{\scriptstyle ~~ b_N ~~~~~ b_N ~~~ \ldots ~~ b_N}_{2n+1} \ldots}^{(n_{\rm max}+1)^2~\textrm{elements for ball~}N}$  \\
\end{tabular}
\end{center}
\caption{
Ordering the coefficients $A_{mn}^i$ as elements of the truncated
vector $\A$ of size $M = N({n_{\mathrm{max}}}+1)^2$ (for an exterior
problem), where $N$ is the number of balls and $n_{\mathrm{max}}$ is
the maximal degree of spherical harmonics.  The diagonal elements of
the matrices $\hat{\nn}$, $\hat{\a}$ and $\hat{\b}$, involved in the
boundary conditions (\ref{eq:UA_cond}), are also shown.}
\label{tab:indices}
\end{table}

\subsection{Monopole approximation}
\label{sec:escape_monopole}

When the absorbing sinks are small, one can resort to the monopole
approximation (MOA) which consists in truncating all expansions to the
zeroth degree: $n_{\rm max} = 0$.  For diffusion problems, this
approximation was first proposed by Borzilov and Stepanov to study the
growth of $N$ drops immersed in an unbounded gas medium
\cite{Borzilov} and by Deutch {\it et al.} to get approximate
solutions of the trap problem in regular arrays with $N$ ideal sinks
\cite{Deutch}.  Later on, this approximation was often employed by
many authors (e.g., see \cite{Tray0,Galanti16a,Biello15} and
references therein).  For the exterior problem, one only needs the
elements
\begin{equation}
\hat U_{0000}^{ij}=\frac{R_i}{L_{ij}}\quad (i\ne j),\qquad 
\hat V_{00}^i=\frac{R_i}{4\pi L_{i}} \,,
\end{equation}
while Eq. (\ref{eq:Robin_A}) is reduced to the set of $N$ linear
equations on $A_{00}^i$:
\begin{equation}
(a_i + b_i) \frac{A_{00}^i}{R_i} + a_i 
\sum\limits_{j(\neq i)=1}^N \frac{A_{00}^j}{L_{ij}} =\frac{a_i}{4\pi L_{i}} \,.
\end{equation}
The monopole approximation accounts for the inter-sink distances
$L_{ij}$ but fully ignores the angular part.  The monopole
approximation for the interior problem is summarized in
\ref{sec:MOA_int}.

\subsection{Numerical analysis and validation}

In order to illustrate the efficiency of the proposed method, we
consider two basic configurations of sinks.

\subsubsection{Two concentric spheres}

\begin{figure}
\begin{center}
\includegraphics[width=65mm]{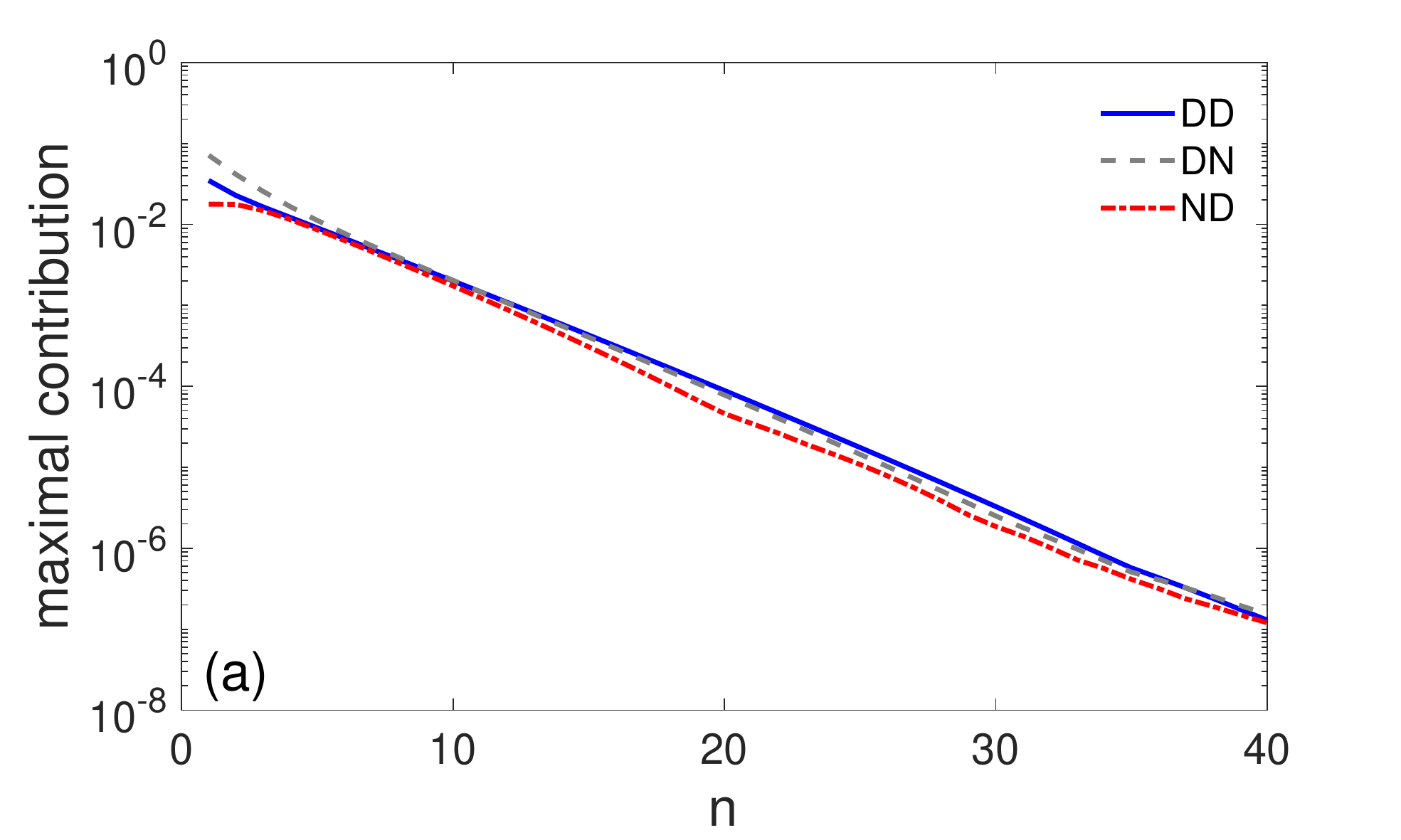} 
\includegraphics[width=65mm]{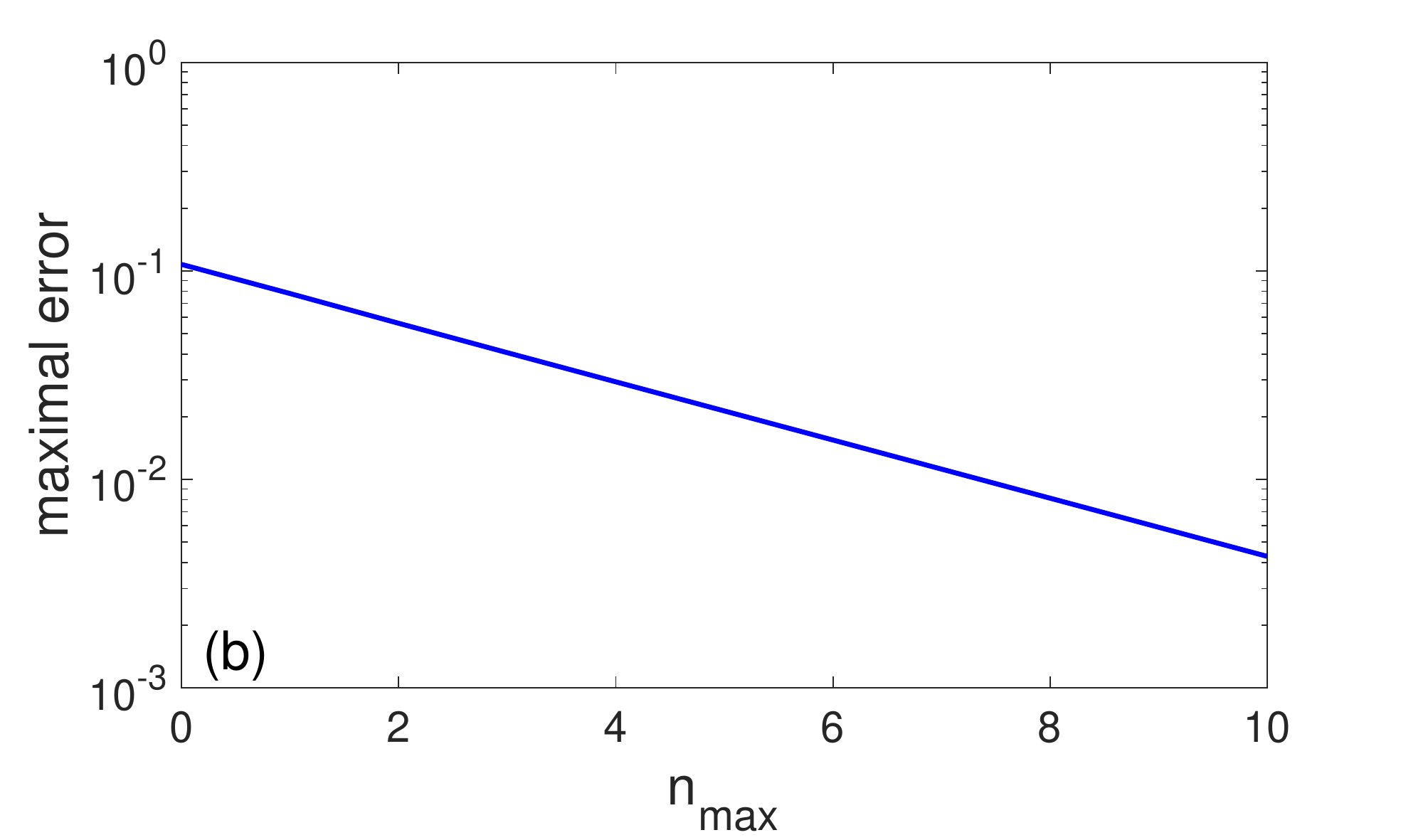} 
\end{center}
\caption{
{\bf (a)} The maximal contribution of the $n$-th term in
Eq. (\ref{eq:GR_twospheres}) computed numerically for two concentric
spheres of radii $R_1 = 1$ and $R_0 = 2$, with the starting point $\xo
= (0,0,1.5)$ and three combinations of boundary conditions at
inner/outer spheres: Dirichlet-Dirichlet, Dirichlet-Neumann, and
Neumann-Dirichlet.  {\bf (b)} The maximal error ($L_{\infty}$-norm) of
the Dirichlet Green function obtained via our numerical implementation
of the GMSV for the same configuration, as a function of the
truncation degree $n_{\rm max}$.  The numerical solution is compared
to the exact formula (\ref{eq:GR_twospheres}) truncated at $n = 40$.}
\label{fig:accuracy}
\end{figure}

We start with the case of two concentric spheres for which an explicit
solution in Eq. (\ref{eq:GR_twospheres}) was derived.  First, we
evaluate how the contribution of the $n$-th term in
Eq. (\ref{eq:GR_twospheres}) decreases with $n$.  This analysis
assesses the accuracy of the truncated explicit solution that will
serve as a reference point to check the accuracy of our numerical
implementation of the GMSV.  Figure \ref{fig:accuracy}(a) shows that
the contribution of the $n$-th term decreases exponentially fast with
$n$, regardless of the type of boundary condition used.  In
particular, the truncation size of $n = 40$ provides the accuracy of
the order of $10^{-7}$ which is enough for our illustrative purposes.

Now, we use the analytical solution as a reference to check the
accuracy of our implementation of the GMSV for the same geometric
configuration.  For this purpose, we compute the Dirichlet Green
function $G(\x,\xo)$ inside the domain $\Omega^{+}$ between two
concentric spheres of radii $R_1 = 1$ and $R_0 = 2$ in two ways:
analytically via Eq. (\ref{eq:GR_twospheres}) and numerically
according to Eq. (\ref{eq:GRobin_int}) truncated at $n_{\rm max}$.
The starting point $\xo$ is fixed at $(0,0,1.5)$.  The Green function
is computed on a set of $10\,000$ points $\x$ uniformly distributed in
the domain.  The maximal error, i.e., the $L_\infty$-norm of the
difference between analytical and numerical solutions, is then
evaluated.  Figure \ref{fig:accuracy}(b) shows the maximal error as a
function of the truncation degree $n_{\rm max}$.  One can see that the
error decreases exponentially fast.

\subsubsection{Co-axial configurations of spheres}

Now we switch to a co-axial configuration of balls that are englobed
by a larger ball.  This type of configurations is particularly
suitable because the axial symmetry facilitates both a visualization
of the obtained results and a numerical solution by a finite element
method that we use as an independent verification scheme.  In fact,
one can use cylindrical coordinates, $(z,\rho,\phi)$, to reduce the
original three-dimensional problem to an effectively two-dimensional
problem if the boundary conditions do not depend on the angular
coordinate $\phi$.  For illustrative purposes, we investigate the
stationary concentration of particles, $n_B(\xo)$, that are constantly
released from a source at the outer sphere $\pa_0$ and diffuse towards
partially reactive inner sinks.  According to Eq. (\ref{eq:n_pN}),
this concentration is proportional to the absorption probability
$p_0(\xo)$.  Setting $n_0 = 1$, we focus on the latter quantity.  On
one hand, we solve the boundary value problem (\ref{eq:n_int}) by a
FEM implemented in Matlab PDE toolbox.  As a numerical method, the FEM
provides an approximate solution whose accuracy depends on the maximal
mesh size $h_{\rm max}$ used.  To control the accuracy of the FEM, we
compute $p_0(\xo)$ with two values of $h_{\rm max}$: $0.05$ and
$0.02$.  On the other hand, we calculate $p_0(\xo)$ from
Eq. (\ref{eq:pN_int}) by using the GMSV and computing the underlying
matrices.  In the following, we analyze how the accuracy of the GMSV
depends on the truncation degree $n_{\rm max}$ and on the reactivity
of the spherical sinks.

\begin{figure}
\begin{center}
\includegraphics[width=42mm]{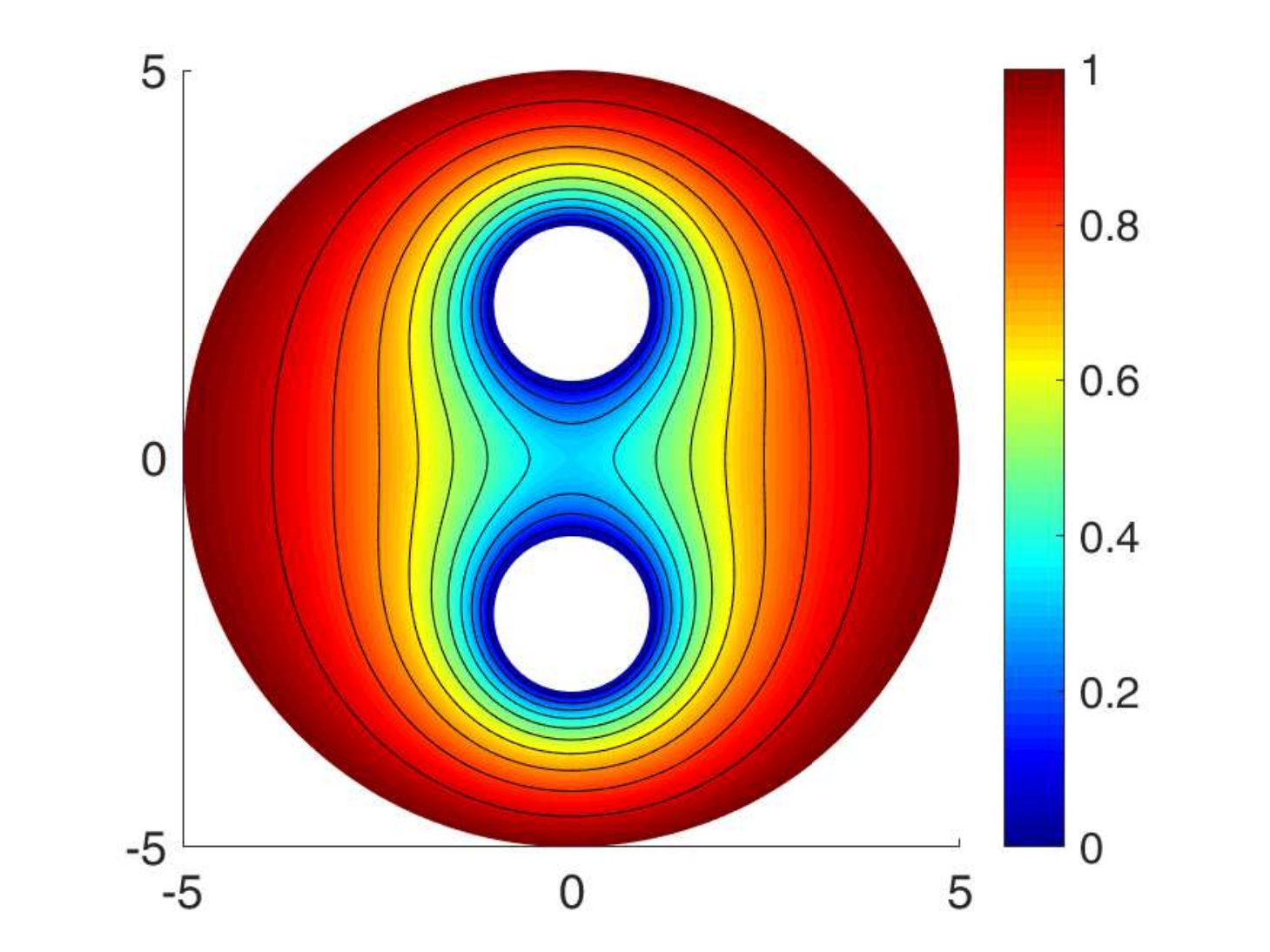} 
\includegraphics[width=42mm]{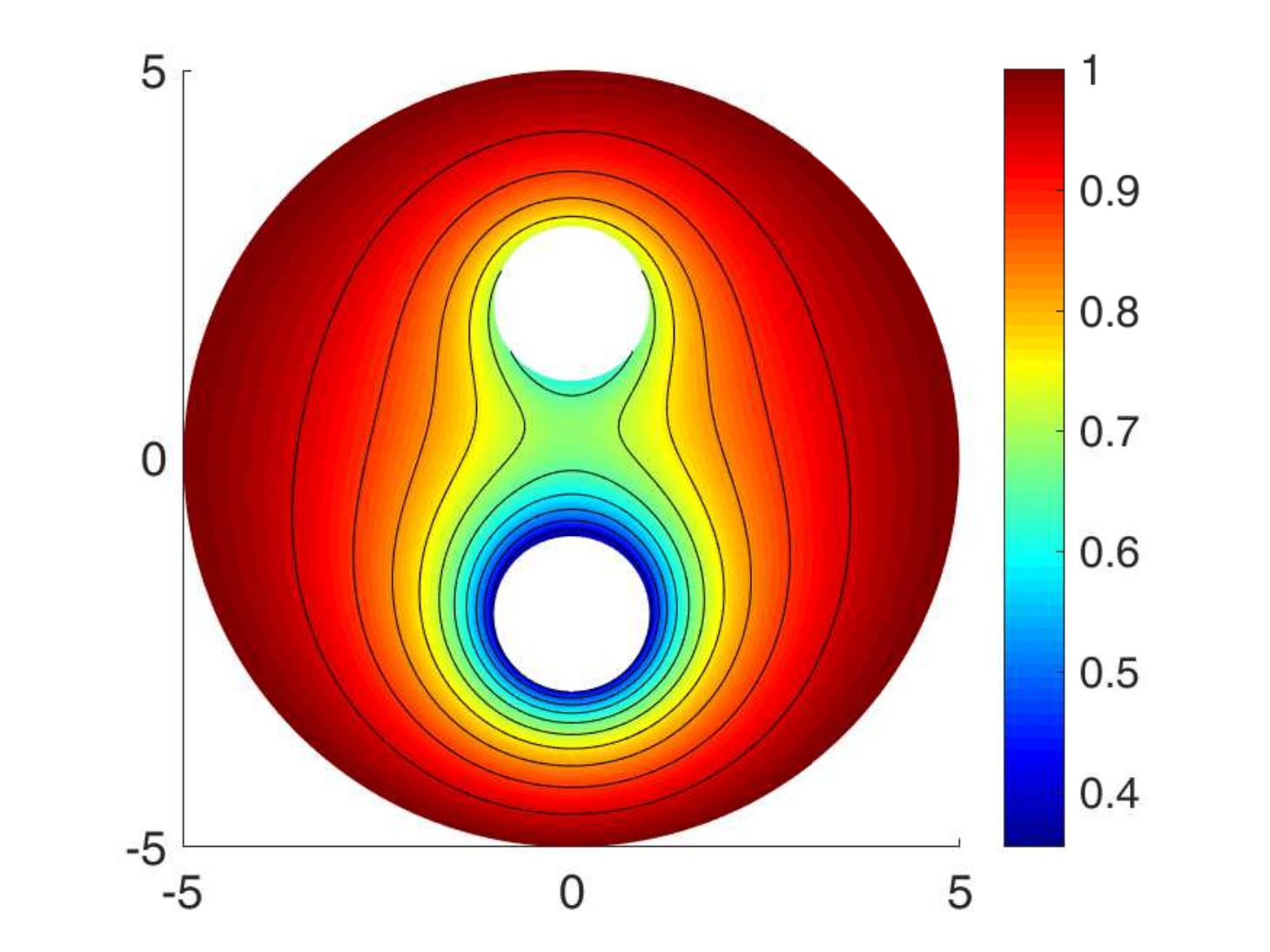} 
\includegraphics[width=42mm]{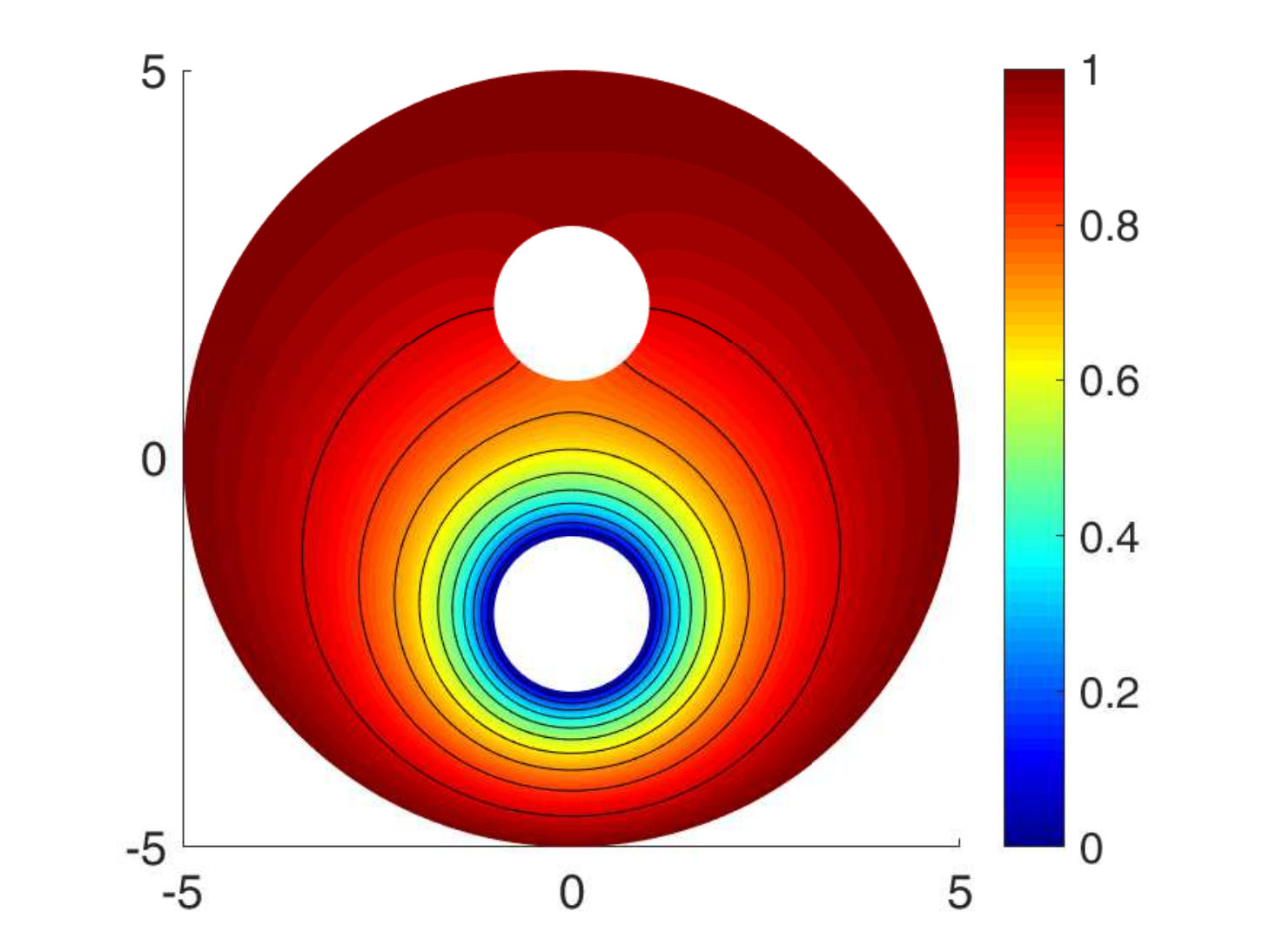} 
\end{center}
\caption{
The absorption probability $p_0(\xo)$ for the domain composed of two
inner spherical sinks of radii $R_1 = R_2 = 1$ centered at $(0,0,\pm
2)$, englobed by the outer source of radius $R_0 = 5$ centered at
$(0,0,0)$, with $a_0 = 1$ and $b_0 = 0$.  On two inner sinks, we set
either Dirichlet conditions ($a_1 = a_2 = 1$, $b_1 = b_2 = 0$, left),
or Robin conditions ($a_1 = a_2 = 1$, $b_1 = 0.5$, $b_2 = 2$, middle),
or Dirichlet-Neumann conditions ($a_1 = b_2 =0$, $a_2= b_1 = 0$,
right).}
\label{fig:coaxial_distrib}
\end{figure}

Figure \ref{fig:coaxial_distrib} shows the absorption probability
$p_0(\xo)$ for the domain $\Omega^{+}$ composed of two inner spherical
sinks of radii $R_1 = R_2 = 1$ centered at $(0,0,\pm 2)$, englobed by
the outer source of radius $R_0 = 5$ centered at $(0,0,0)$.  Setting
the Dirichlet boundary condition ($a_0 = 1$ and $b_0 = 0$) at the
outer source, we compare several combinations of boundary conditions
at the inner sinks: two fully absorbing sinks ($a_1 = a_2 = 1$ and
$b_1 = b_2 = 0$), two partially reflecting sinks ($a_1 = a_2 = 1$,
$b_1 = 0.5$, $b_2 = 2$), and one absorbing sink with one reflecting
obstacle ($a_1 = b_2 = 1$ and $a_2 = b_1 = 0$).  These solutions are
obtained by the GMSV with the truncation degree $n_{\rm max} = 7$.

\begin{figure}
\begin{center}
\includegraphics[width=32mm]{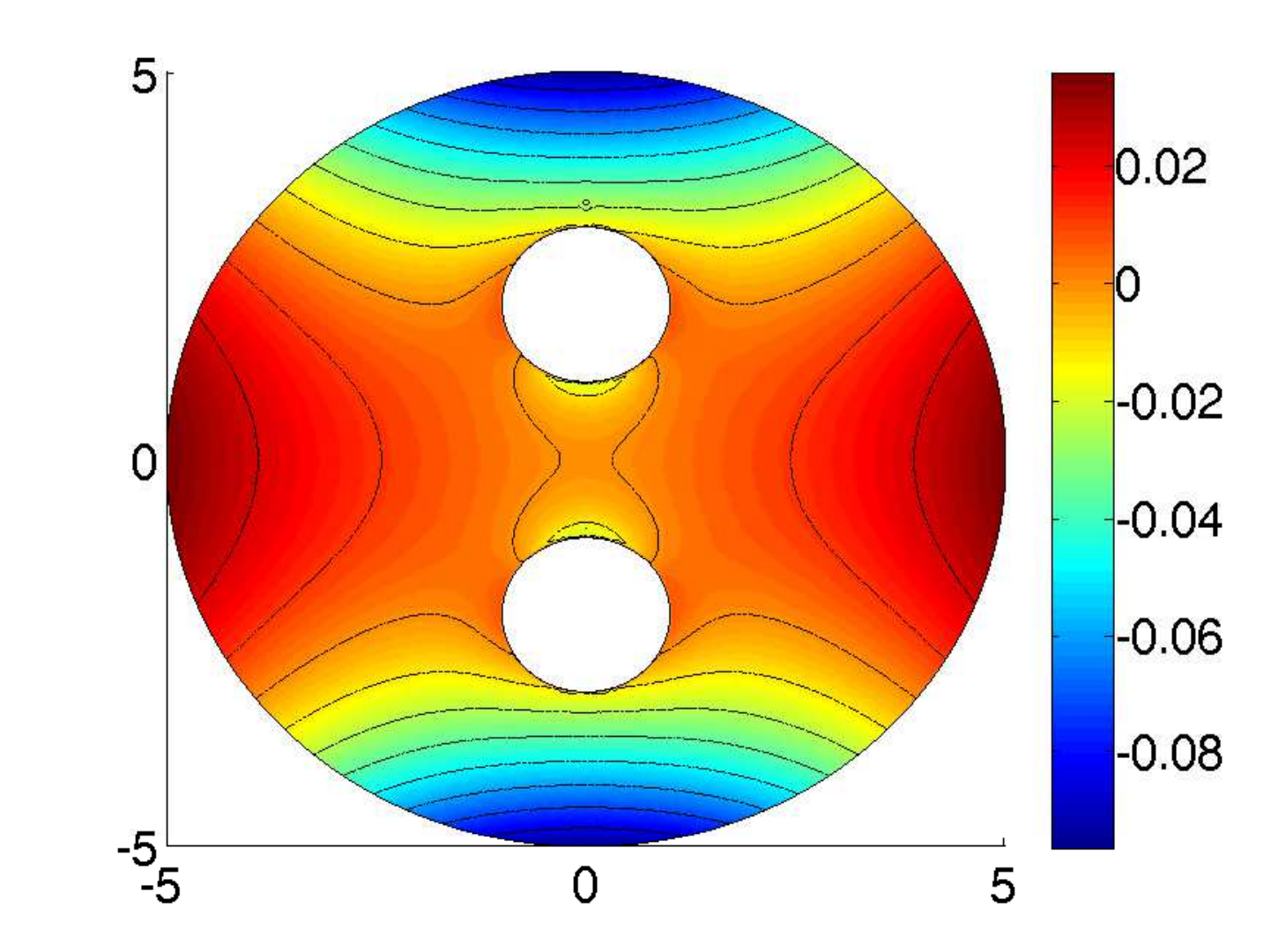} 
\includegraphics[width=32mm]{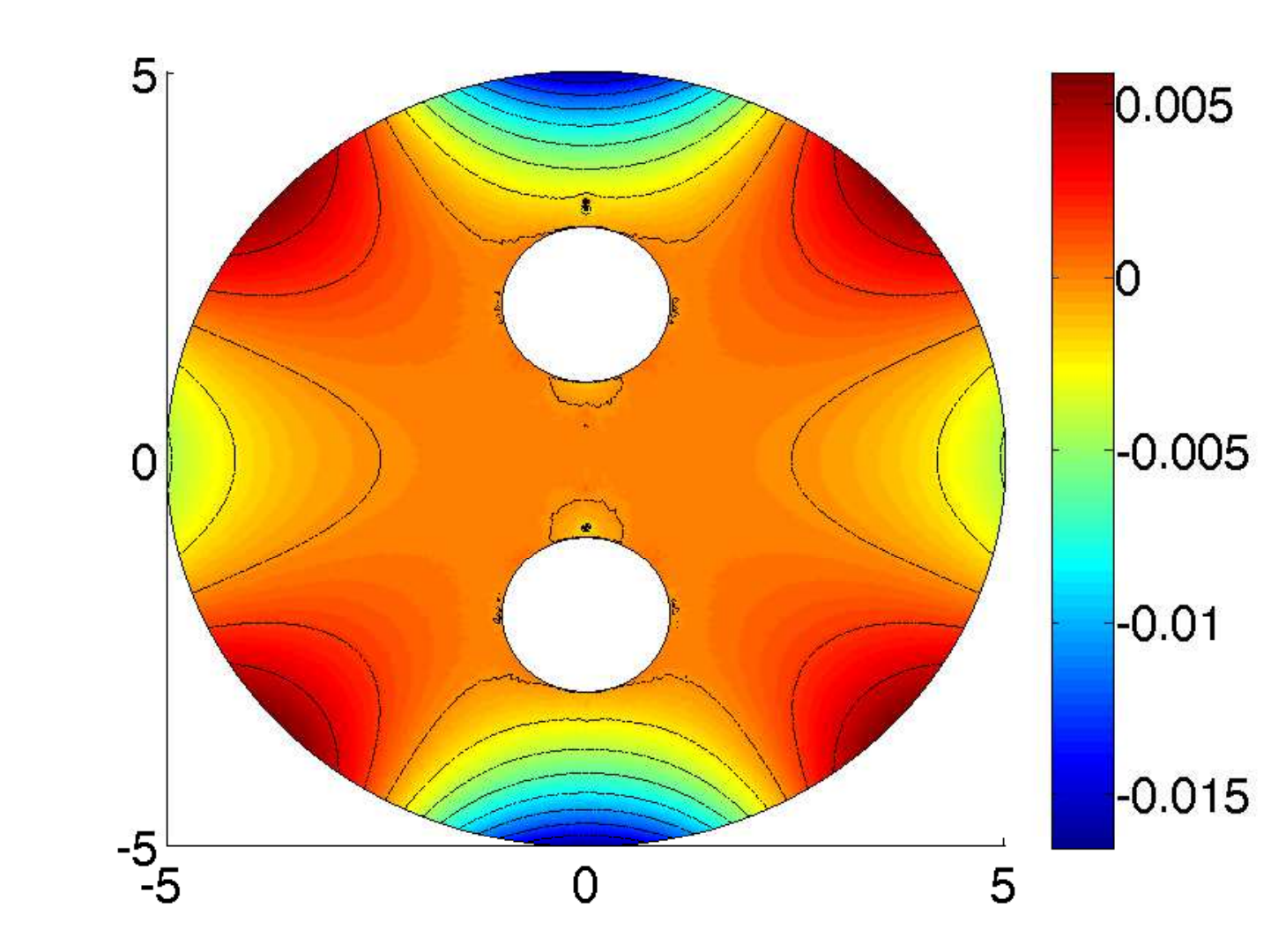} 
\includegraphics[width=32mm]{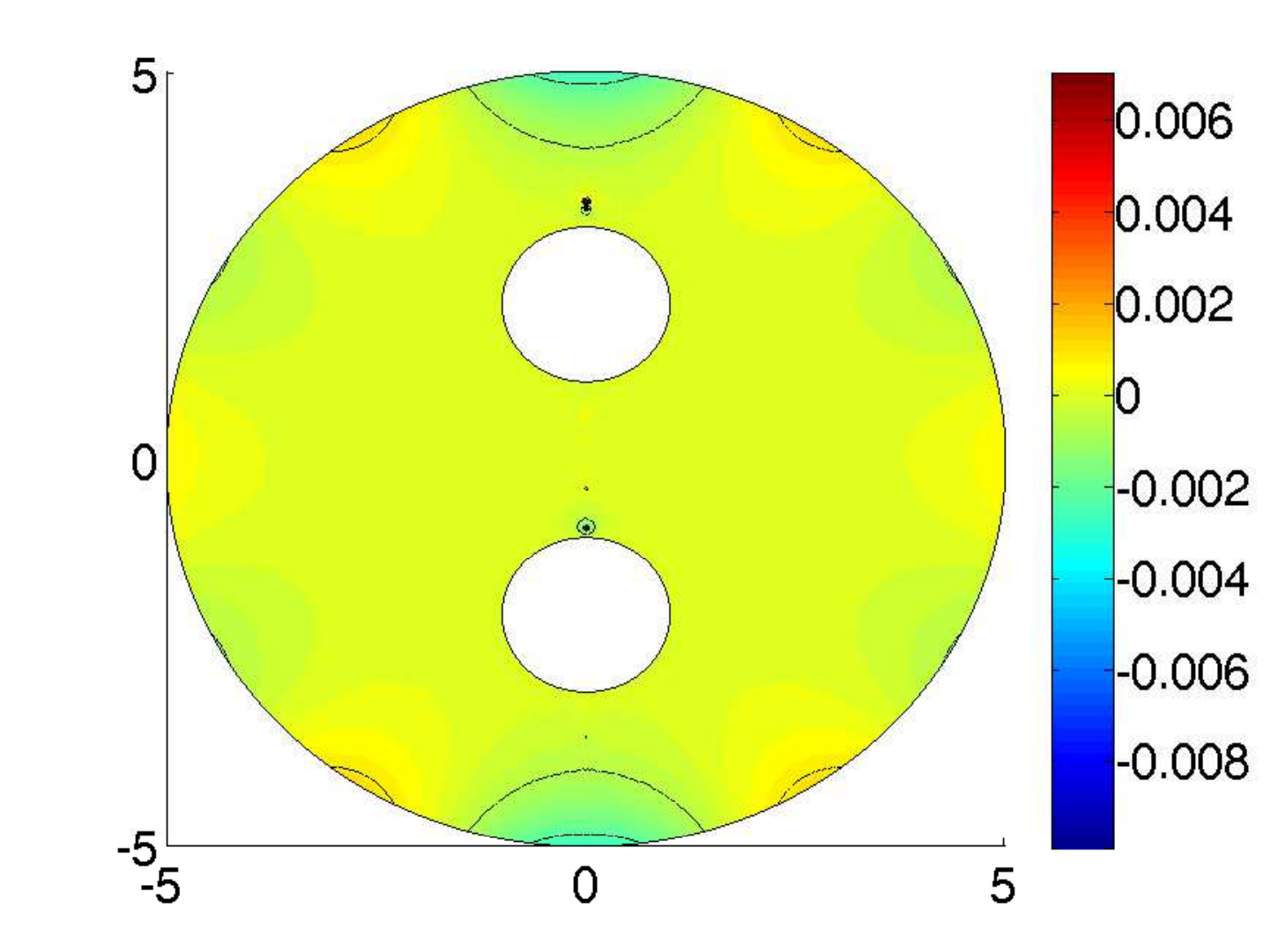} 
\includegraphics[width=32mm]{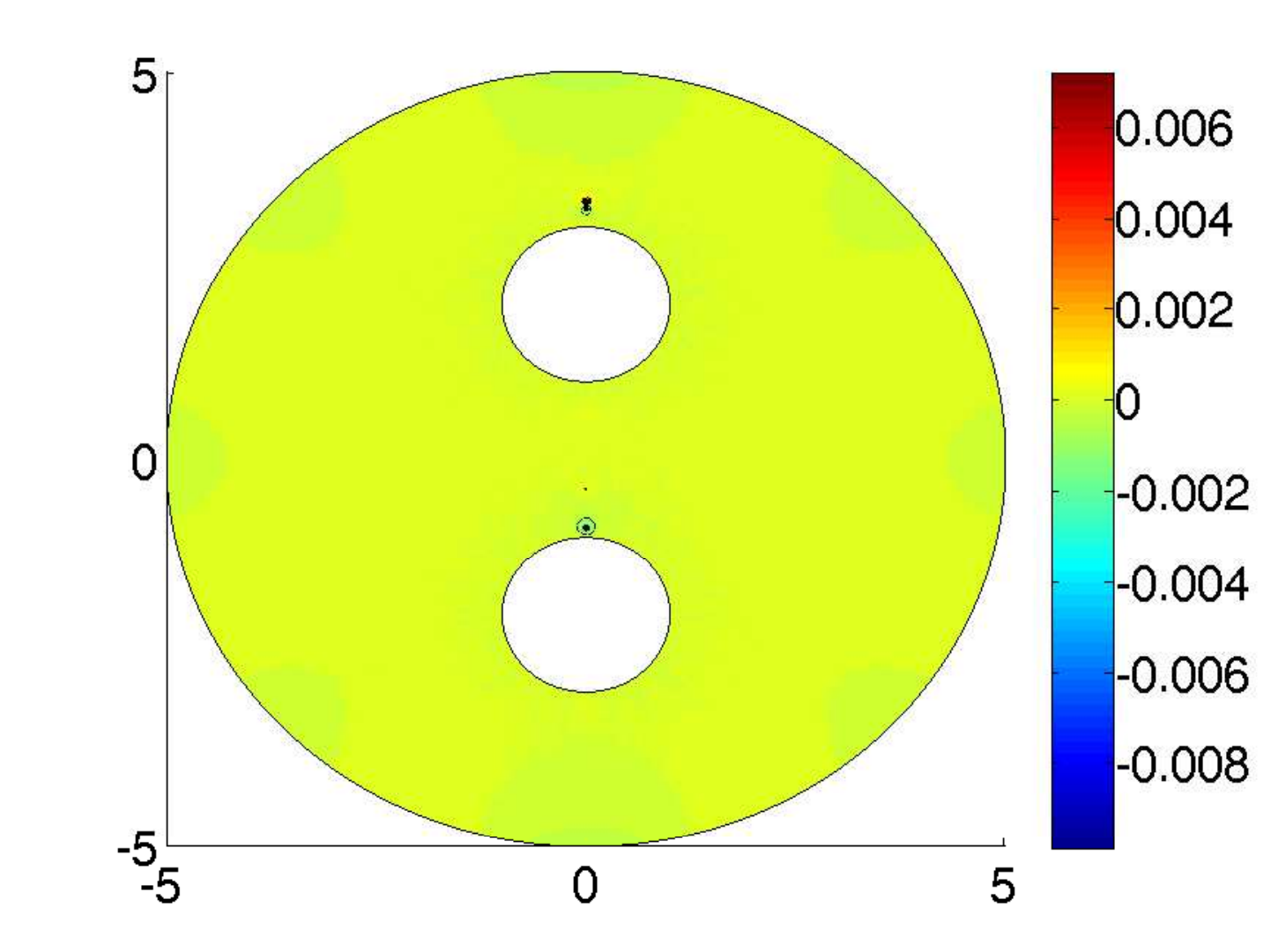} 
\includegraphics[width=32mm]{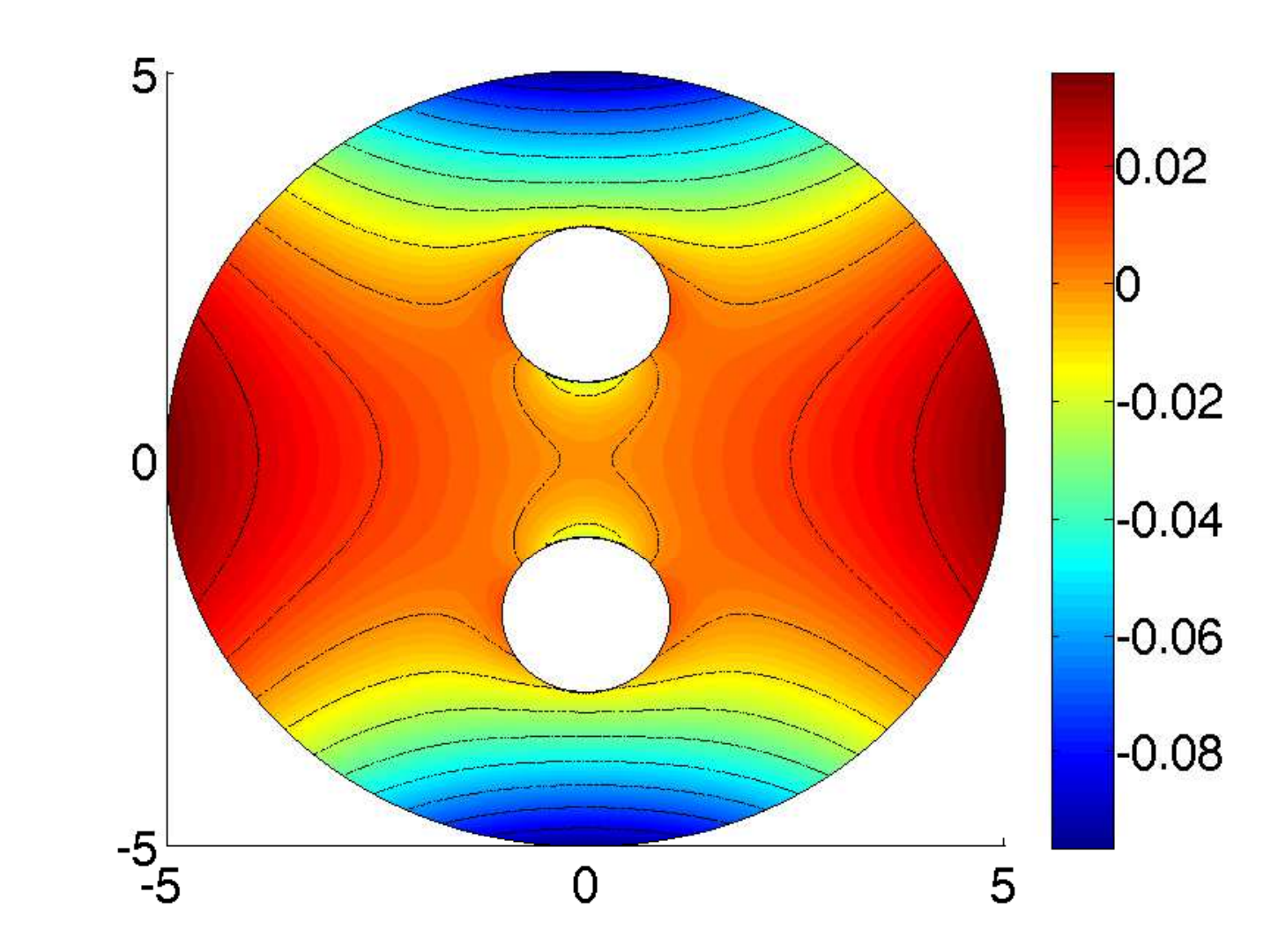} 
\includegraphics[width=32mm]{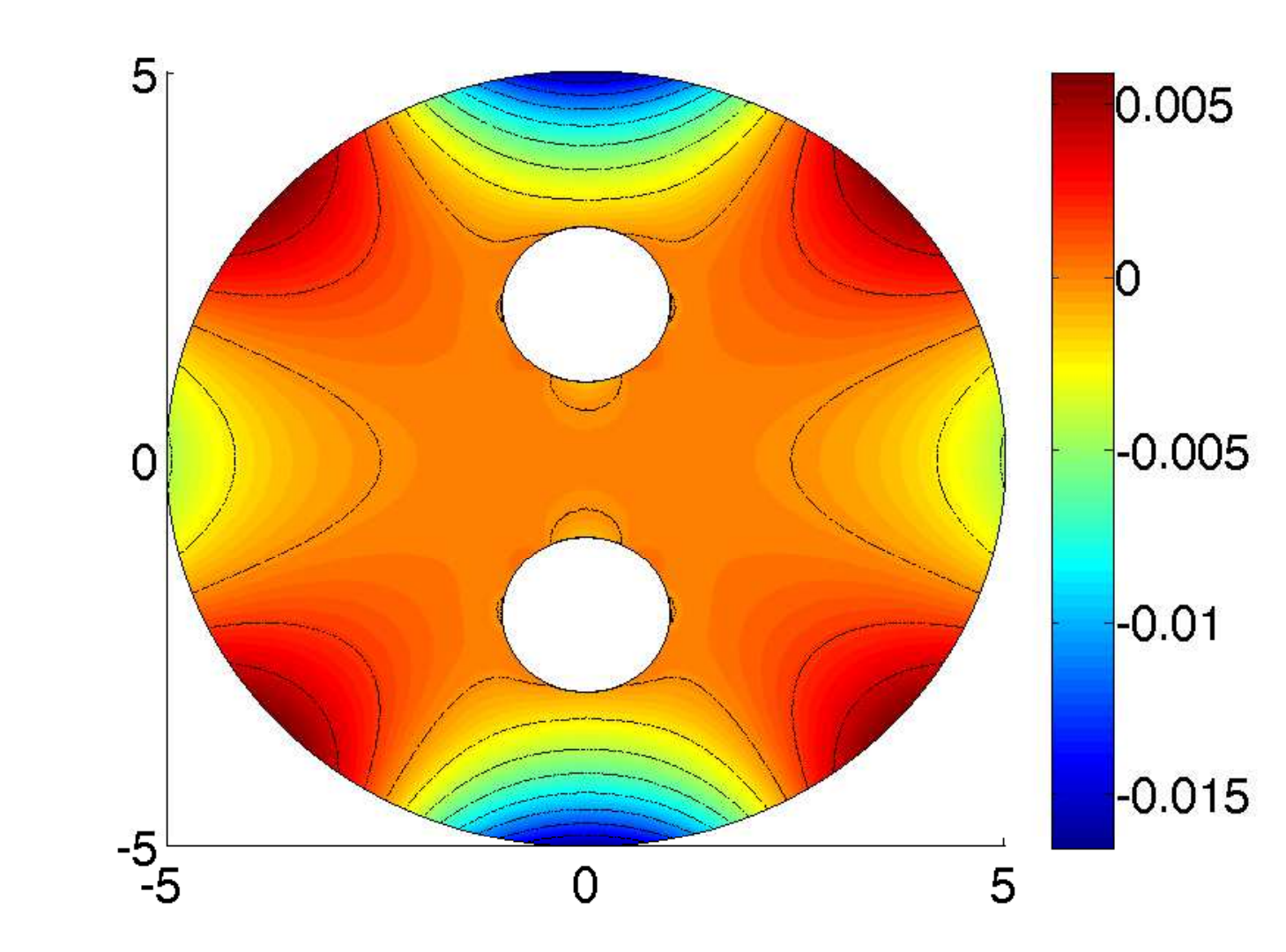} 
\includegraphics[width=32mm]{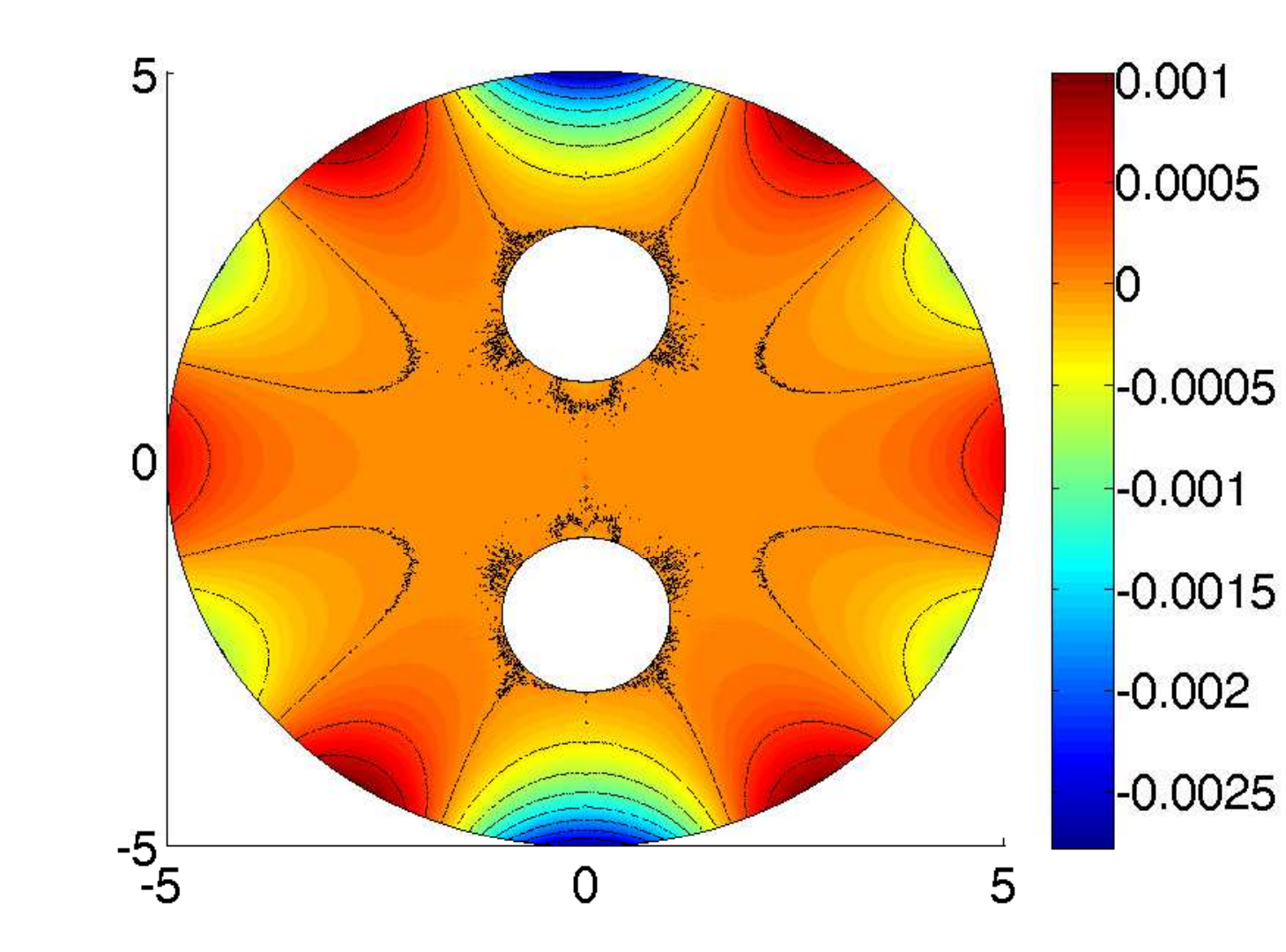} 
\includegraphics[width=32mm]{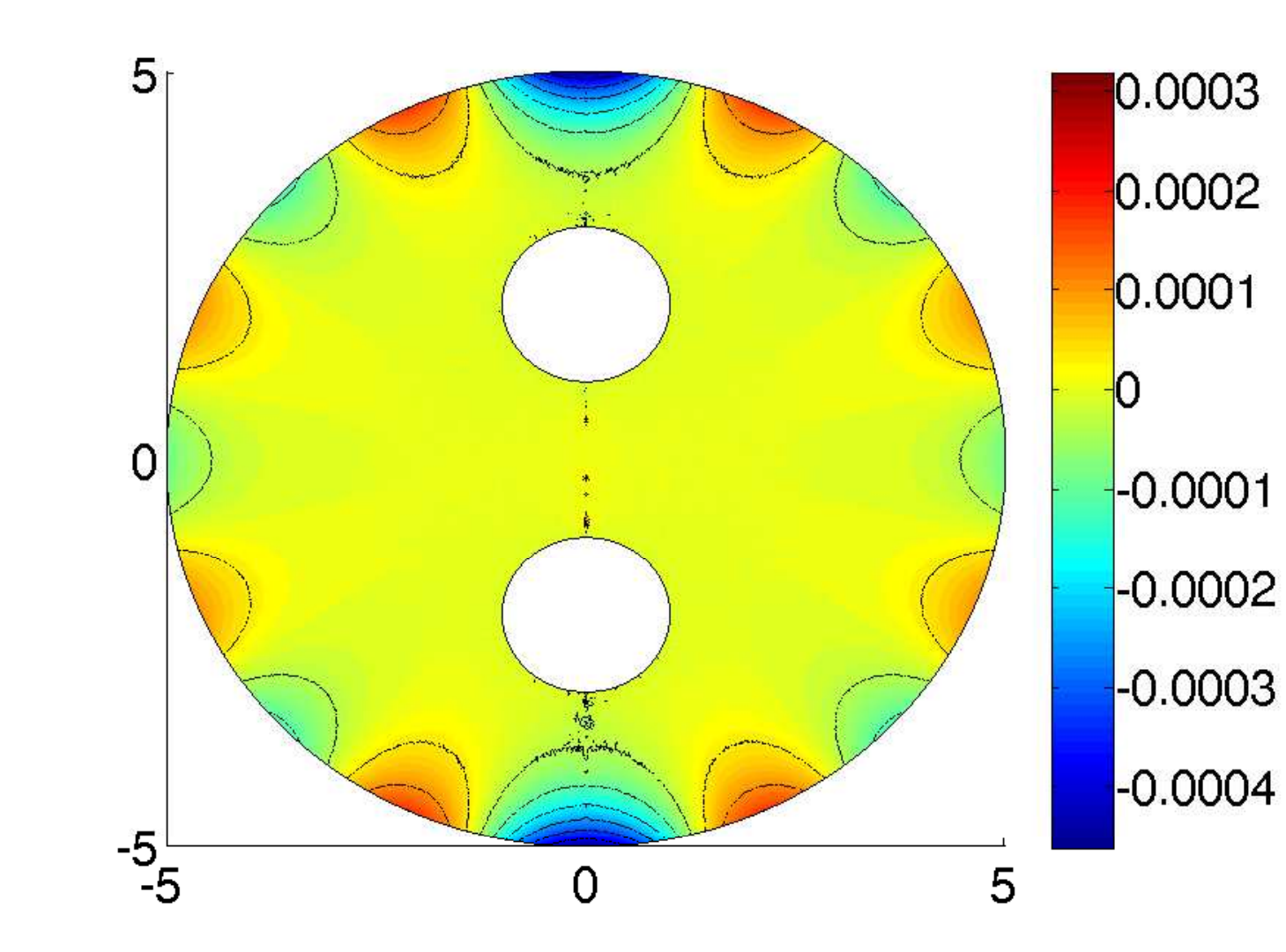} 
\end{center}
\caption{
Difference between the absorption probabilities $p_0(\xo)$ obtained by
the GMSV and by the FEM of Matlab PDE toolbox, for the domain composed
of two inner balls of radii $R_1 = R_2 = 1$ centered at $(0,0,\pm 2)$,
englobed by the outer source of radius $R_0 = 5$ centered at
$(0,0,0)$.  We set Dirichlet boundary conditions: $a_1 = a_2 = a_0 =
1$ and $b_1 = b_2 = b_0 = 0$.  Top/bottom rows correspond to two
maximal mesh sizes $h_{\rm max}$ of the FEM: $0.05$ (coarser) and
$0.02$ (finer).  Plots from left to right correspond to different
truncation degrees of the GMSV: $n_{\rm max} = 1, 3, 5, 7$.  The
maximal absolute errors are reported in Table \ref{tab:error}.  }
\label{fig:coaxial_error}
\end{figure}

\begin{figure}
\begin{center}
\includegraphics[width=32mm]{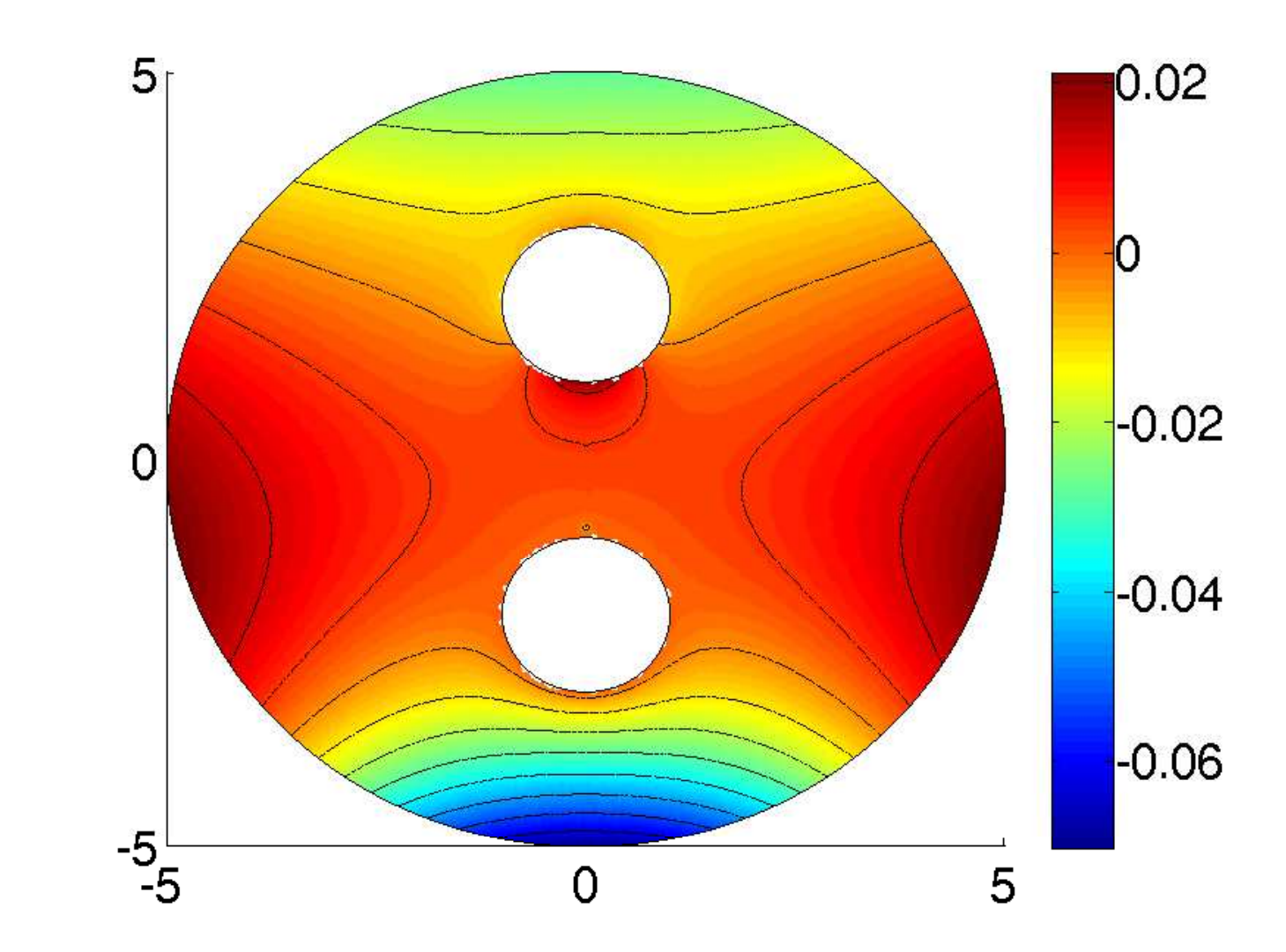} 
\includegraphics[width=32mm]{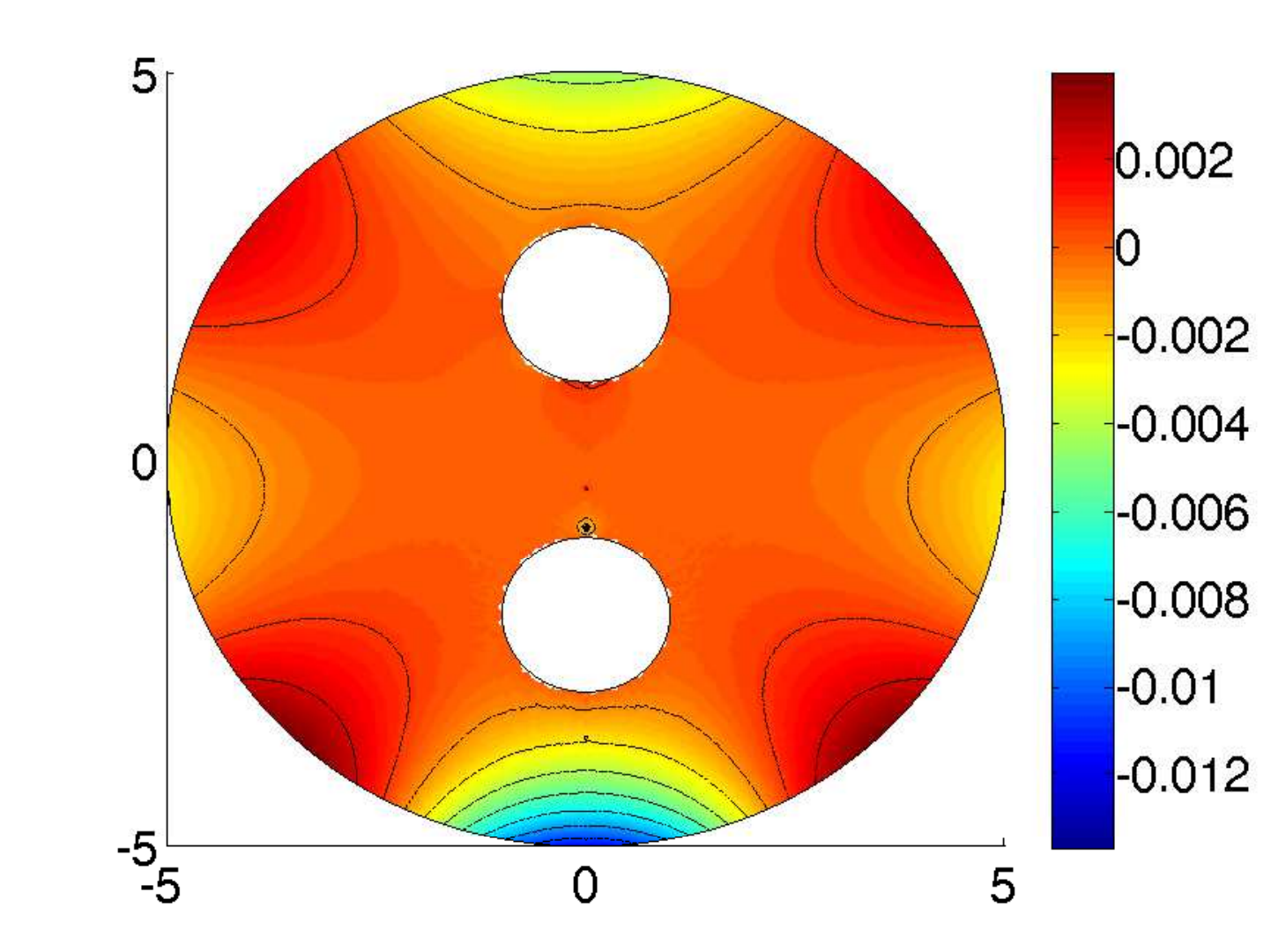} 
\includegraphics[width=32mm]{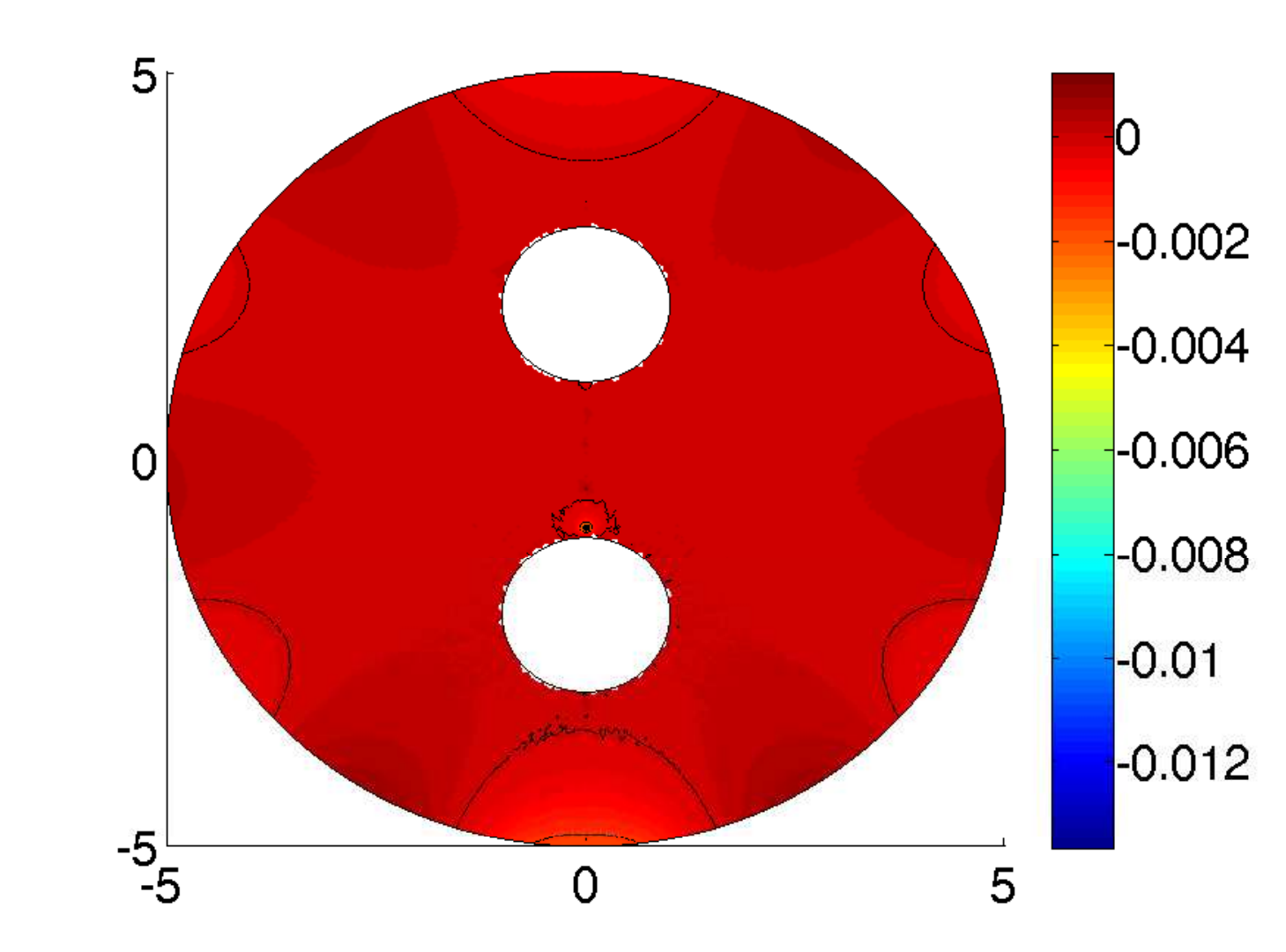} 
\includegraphics[width=32mm]{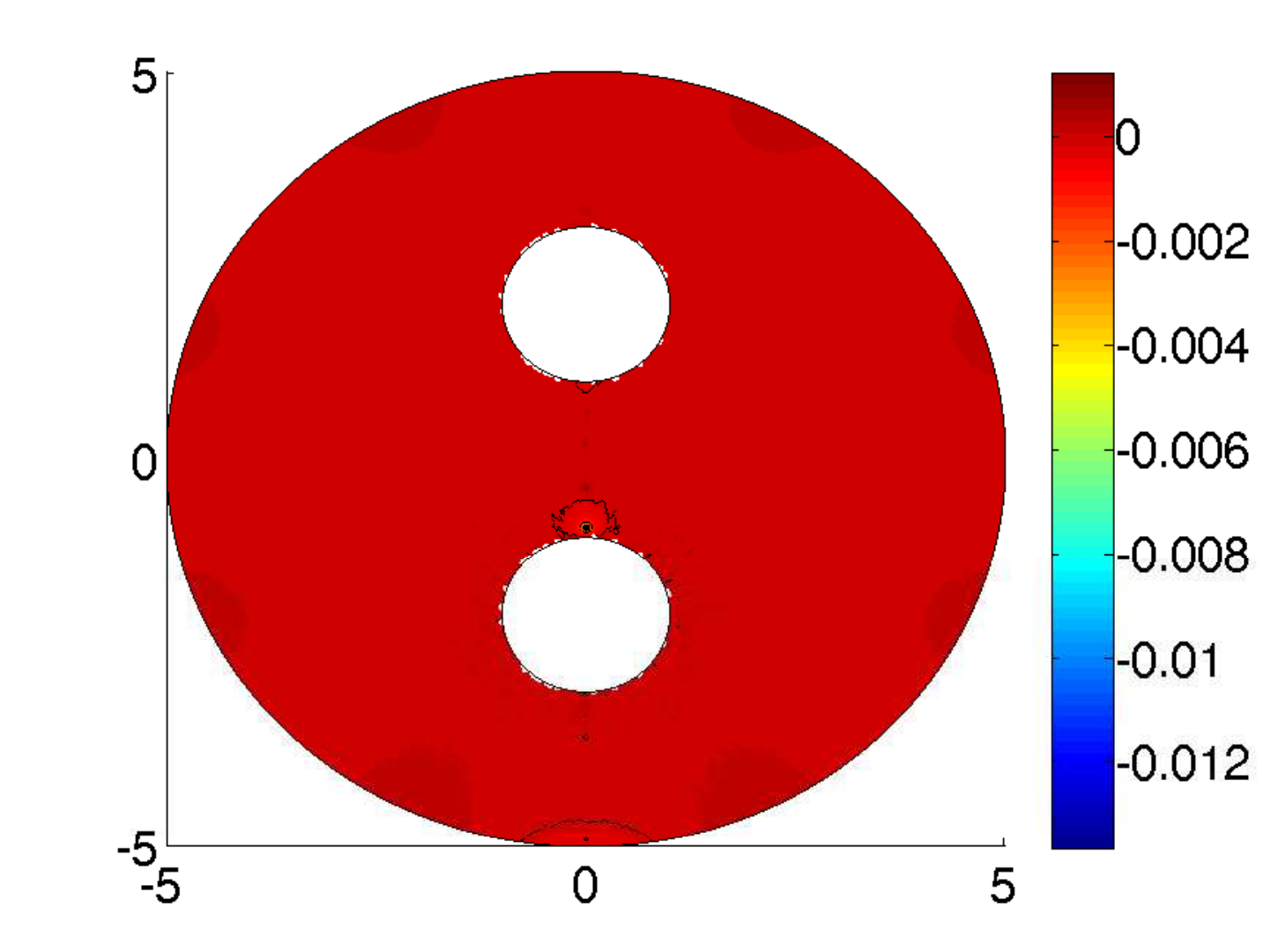} 
\includegraphics[width=32mm]{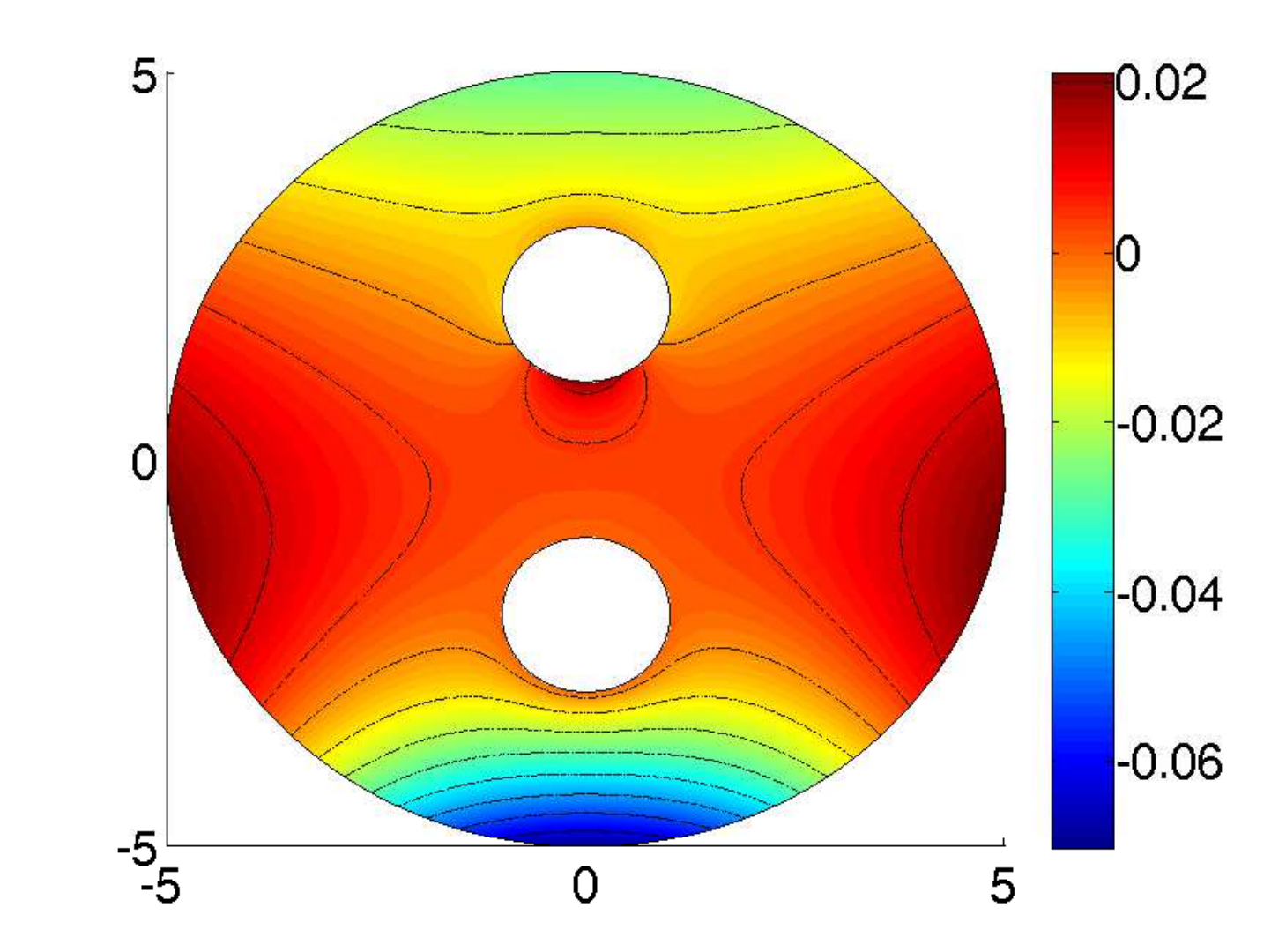} 
\includegraphics[width=32mm]{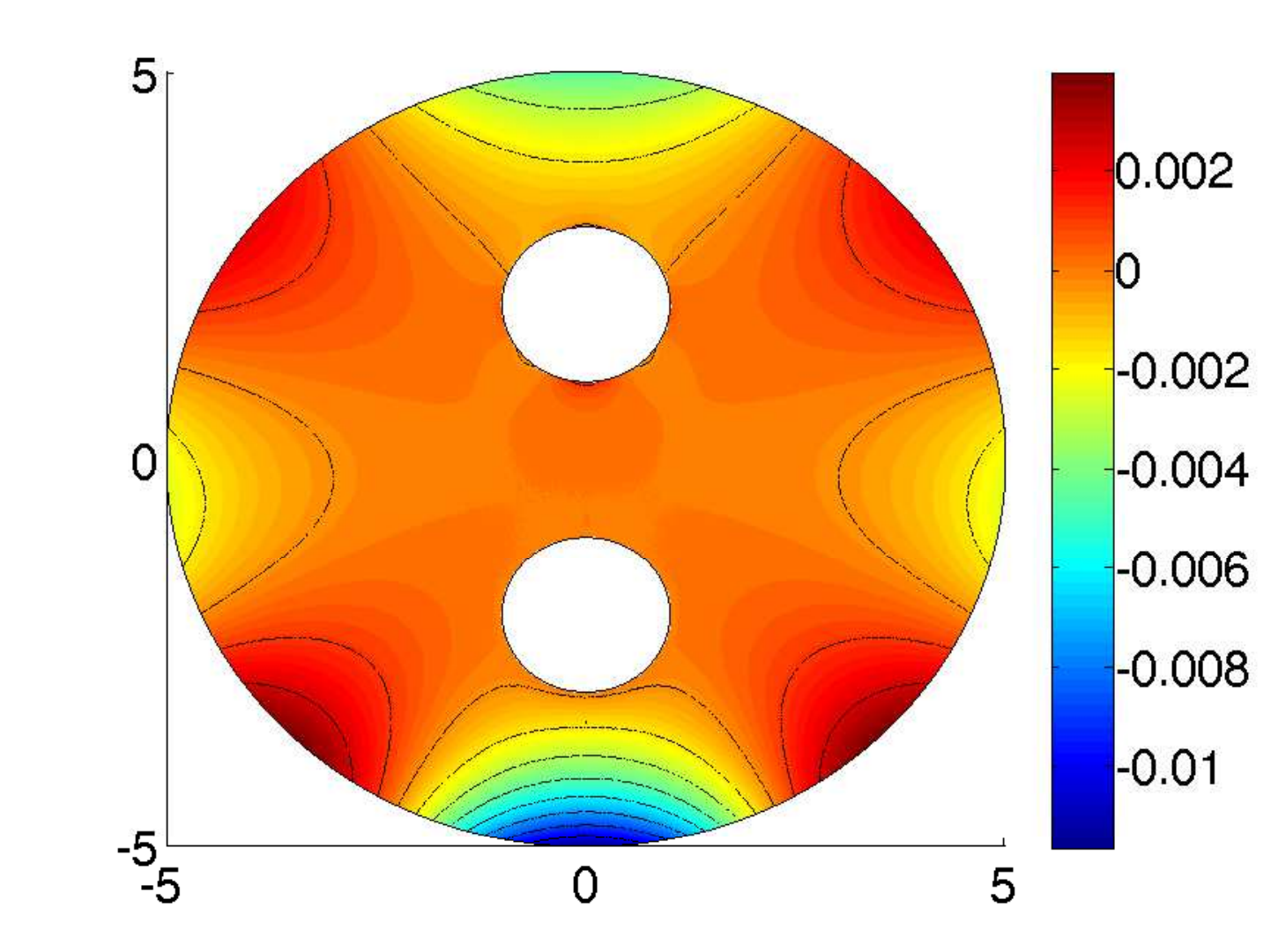} 
\includegraphics[width=32mm]{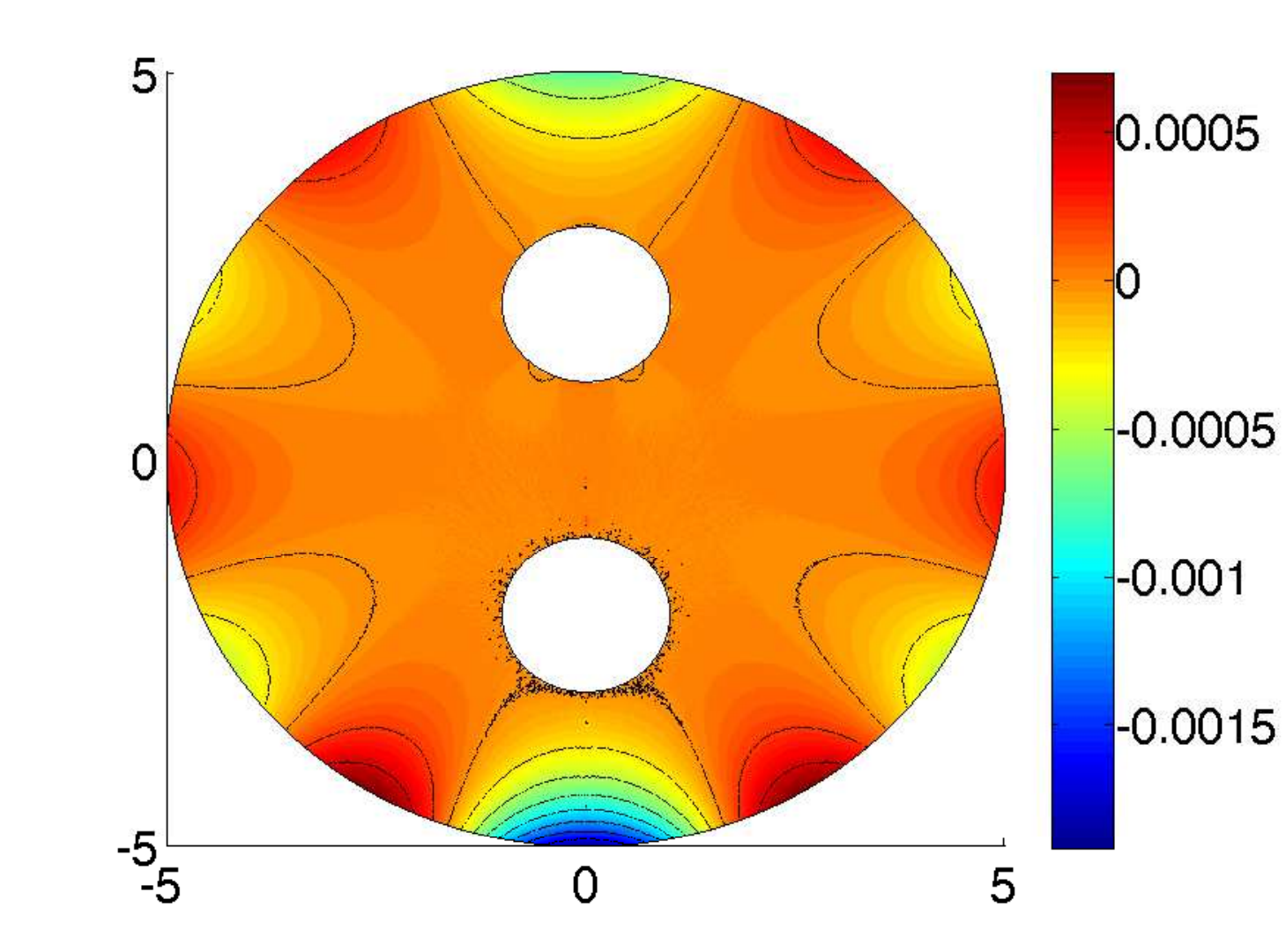} 
\includegraphics[width=32mm]{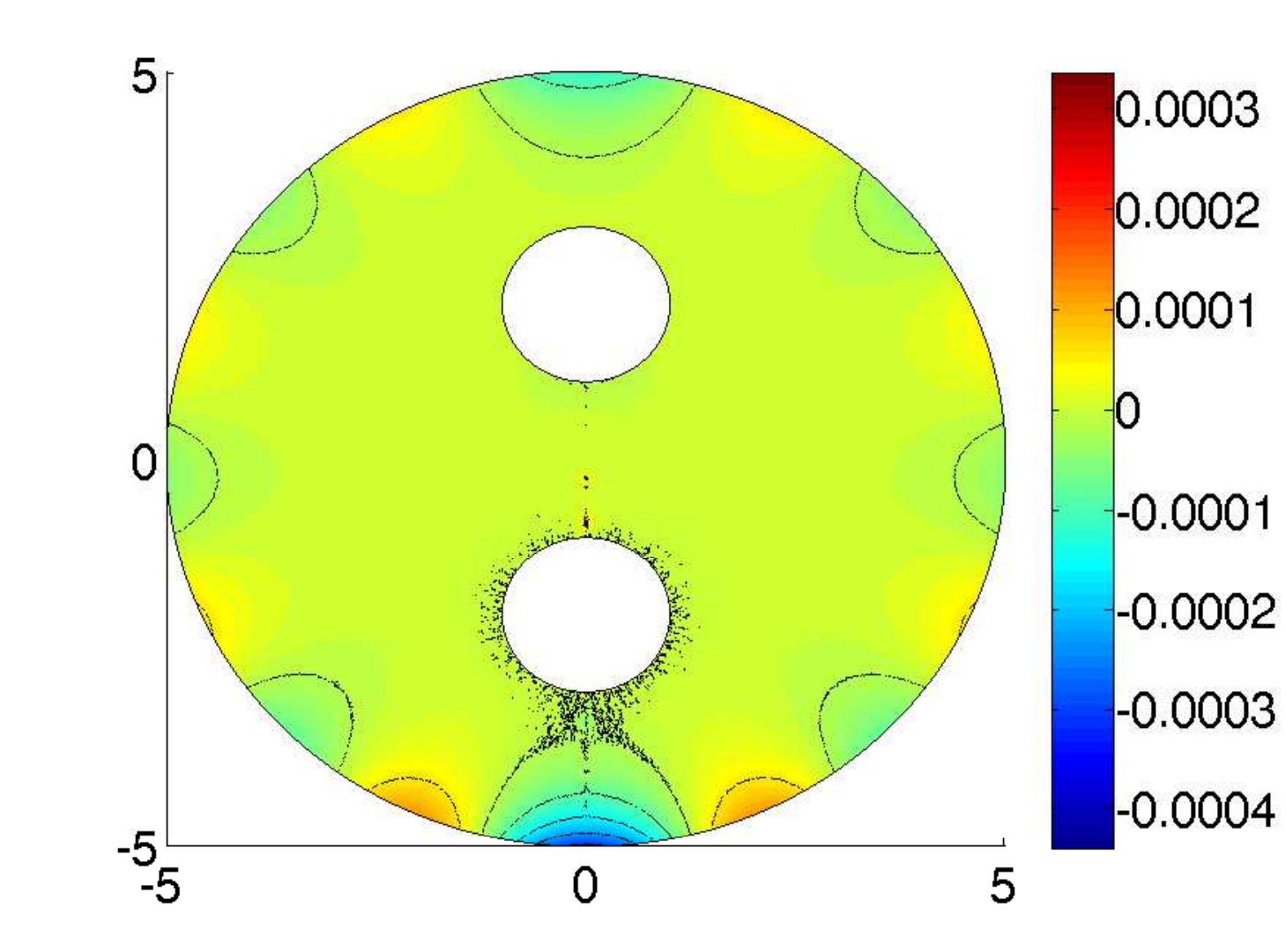} 
\end{center}
\caption{
Difference between the absorption probabilities $p_0(\xo)$ obtained by
the GMSV and by the FEM of Matlab PDE toolbox, for the domain composed
of two inner balls of radii $R_1 = R_2 = 1$ centered at $(0,0,\pm 2)$,
englobed by the outer source of radius $R_0 = 5$ centered at
$(0,0,0)$.  We set Dirichlet-Neumann boundary conditions at inner
balls: $a_1 = b_2 = a_0 = 1$ and $b_1 = a_2 = b_0 = 0$.  Top/bottom
rows correspond to two maximal mesh sizes $h_{\rm max}$ of the FEM:
$0.05$ (coarser) and $0.02$ (finer).  Plots from left to right
correspond to different truncation degrees of the GMSV: $n_{\rm max} =
1, 3, 5, 7$.  The maximal absolute errors are reported in Table
\ref{tab:error}.  }
\label{fig:coaxial_error_DN}
\end{figure}

Figure \ref{fig:coaxial_error} shows the difference between the
solutions obtained by the FEM and by the GMSV.  In the top panel, the
solution by the GMSV is compared to the coarser FEM solution with
$h_{\rm max} = 0.05$.  Increasing the truncation degree $n_{\rm max}$
from $1$ to $7$, one progressively improves the accuracy of the GMSV
solution.  The maximal absolute difference (i.e., the
$L_{\infty}$-norm) is reported in Table \ref{tab:error}.  One can see
that this difference stops to decrease for $n_{\rm max} \geq 5$.  This
reflects the fact that the difference is further determined by the
limited accuracy of the FEM solution.  In the bottom panel of
Fig. \ref{fig:coaxial_error}, a more accurate FEM solution with the
maximal mesh size $0.02$ is used for comparison.  In this case, the
maximal absolute difference progressively decreases for all considered
$n_{\rm max}$ up to $7$.  Similar behavior is observed for mixed
Dirichlet-Neumann boundary conditions set on two inner sinks
(Fig. \ref{fig:coaxial_error_DN}).

Qualitatively, the accuracy of the FEM solution with $h_{\rm max} =
0.05$ is comparable to that of the GMSV with $n_{\rm max} = 5$.
However, this FEM solution involves the triangulation of the planar
computational domain with $99\,494$ triangles and $K = 50\,190$
vertices and thus requires to solve numerically the system of $K$
linear algebraic equations.  In turn, finding the GMSV solution relies
on solving the system of $3(5+1)^2 = 108$ linear algebraic equations.
In addition to a $500$-fold reduction in the number of equations, the
GMSV provides the solution in an analytic form that can be easily
manipulated.  Moreover, we chose here the co-axial configuration of
sinks just to facilitate the use the FEM solution in a planar
computational domain.  In general, a three-dimensional computational
domain has to be discretized that would drastically increase the
number of linear equations in the FEM (to keep the same $h_{\rm max}$
and thus the same accuracy).  In contrast, the computational cost of
the GMSV does not depend on whether the configuration is co-axial or
not.  Finally, solving an exterior problem for two absorbing sinks
without an outer boundary by the GMSV involves even a smaller system
of linear algebraic equations, whereas the addition of an artificial
outer boundary is mandatory for the FEM.  We conclude that the GMSV
significantly outperforms the FEM for three-dimensional domains with
disconnected spherical boundaries, and is particularly valuable for
exterior problems.

\begin{table}
\begin{center}
\begin{tabular}{|c |c| c| c| c| c|}  \hline
   & $h_{\rm max}~ \backslash~ n_{\rm max}$ &     1    &    3    &     5   &    7    \\  \hline
\multirow{2}{5mm}[0mm]{DD} 
   & $0.05$                               & 0.0965   & 0.0166  & 0.0099  &  0.0099 \\
   & $0.02$                               & 0.0965   & 0.0166  & 0.0028  &  0.0005 \\  \hline 
\multirow{2}{5mm}[0mm]{DN} 
   & $0.05$                               & 0.0703   & 0.0137  & 0.0137  &  0.0137 \\
   & $0.02$                               & 0.0704   & 0.0117  & 0.0019  &  0.0004 \\  \hline
\end{tabular}
\end{center}
\caption{
Maximal absolute errors between the absorption probabilities
$p_0(\xo)$ obtained by the GMSV with the truncation degree $n_{\rm
max}$ and by the FEM of Matlab PDE toolbox with the maximal mesh size
$h_{\rm max}$ (see Figs. \ref{fig:coaxial_error},
\ref{fig:coaxial_error_DN}).}
\label{tab:error}
\end{table}

\section{Conclusion}
\label{sec:conclusion}

Using the classical translational addition theorems for solid
harmonics, we elaborated a general semi-analytical solution for
boundary value problems associated to the Laplace operator in
arbitrary configurations of non-overlapping balls in three dimensions.
We considered both exterior and interior problems with the most common
Dirichlet, Neumann, and Robin boundary conditions.  We also treated
the conjugate boundary value problems with diffusive exchange between
interior and exterior compartments.  In all cases, the solution is
based on the derived semi-analytical formula for the Green function
$G(\x,\xo)$, in which the dependence on points $\x$ and $\xo$ enters
\textit{analytically} through solid harmonics $\psi_{mn}^{\pm}$ while
the associated coefficients are obtained \textit{numerically} by
truncating and solving the established system of linear algebraic
equations.  In other words, although the solution is exact, its
practical implementation requires matrix truncation and inversion.
The desired accuracy of the solution is achieved by varying the
truncation degree.  The natural choice of solid harmonics as basis
functions that respect intrinsic symmetries of the domain, implies a
very rapid convergence of the numerical solution, as confirmed with
several examples.  Even the truncation to the zeroth degree, $n_{\rm
max} = 0$, known as the monopole approximation, can yield accurate
results, especially when the balls are small as compared to the
inter-ball distances.  Moreover, the computation does not involve
meshing of the domain that is often a limiting factor, especially in
three dimensions.  Once the coefficients in front of solid harmonics
are found, one can easily and rapidly evaluate the Green function at
any point of the domain.  Since irregular solid harmonics decay at
infinity, there is also no need for imposing an artificial outer
boundary to transform an exterior problem to an interior problem that
is needed for most other methods.  The long range character of the
fundamental solution $\G(\x,\xo)$ implies that the impact of such an
artificial boundary onto the solution can be significant even for
distant boundaries.  To reduce this impact in conventional methods,
one would need to put the outer boundary far away from balls that
would greatly increase the number of discrete elements and thus the
number of linear equations.  In contrast, the GMSV is even simpler for
exterior domains and provides superior computational efficiency.  To
summarize, the major advantages of the GMSV are: semi-analytical form
of the solution, mesh-free computation, very rapid convergence, and no
need for imposing artificial outer boundary to treat exterior
problems.

The Green function is also the key ingredient to access various
characteristics of stationary diffusion among partially reactive sinks
such as reaction rates, escape probability, harmonic measure,
residence time and mean first passage time, for which we provided
semi-analytical formulas.  Although our main focus was on applications
to diffusion-influenced chemical reactions, the proposed method is
also valuable in other fields in which the Laplace and Poisson
equations are relevant.  For instance, one can describe molecular
motion in biological tissues and heat transfer in heterogeneous media,
both modeled as non-overlapping balls (e.g., cells or tumors) immersed
in an exterior space.  The exchange between these compartments is
accounted via conjugate boundary conditions.  In electrostatics, the
Dirichlet Green function $G(\x,\xo)$ can be interpreted as the
electric potential created by a point charge at $\xo$ in the presence
of grounded balls.  In fluid dynamics, one can compute the velocity
potential of an incompressible flow in a pack of non-overlapping
spheres which is often used as a basic model of heterogeneous porous
media.

In spite of our focus on domains with disconnected spherical
boundaries, the GMSV is applicable to other canonical domains and
their combinations \cite{Traytak18}.  For instance, one can consider
spherical sinks englobed by a parallelepiped or by a cylindrical tube;
moreover, spherical sinks can be replaced by spheroids,
parallelepipeds, cylinders, or their combinations.  As a consequence,
such combinations of canonical domains provide a very flexible and
versatile tool for modeling structured or disordered media and the
related diffusion-reaction processes.  Future developments of the GMSV
for other canonical domains is a promising perspective of the present
work.

\section*{Acknowledgments}
DG acknowledges the support under Grant No. ANR-13-JSV5-0006-01 of the
French National Research Agency.

\appendix

\section{Technical derivations}
\label{sec:derivations}

\subsection{Newton's potential}
\label{sec:ANewton}

We use the Laplace expansion for the Newton's potential \cite{Epton},
\begin{equation}  \label{green7}
\frac{1}{\| \x - \xo\| } = \frac{1}{\| (\x - \x_i) - \L_{i} \|} 
= \sum\limits_{n=0}^\infty \sum\limits_{m=-n}^n (-1)^m \frac{r_{<}^n}{%
r_{>}^{n+1}} Y_{(-m)n}(\Theta_{i},\Phi_{i}) Y_{mn}(\theta_i,\phi_i) , 
\end{equation}
where $\L_{i} = \xo - \x_i$, $(L_{i}, \Theta_{i}, \Phi_{i})$ are the
spherical coordinates of $\L_{i}$, $r_{<} = \min( \|\x-\x_i\| ,
\|\L_{i}\|)$ and $r_{>} = \max( \|\x-\x_i\| , \|\L_{i}\|)$.  For $r_i
< L_{i}$, one has $r_{<} = r_i$ and $r_{>} = L_{i}$ so that
\begin{equation}  \label{green10a}
\frac{1}{\| \x - \xo\| } = \sum\limits_{n=0}^\infty
\sum\limits_{m=-n}^n (-1)^m \, \psi_{(-m)n}^{-}(L_{i},
\Theta_{i},\Phi_{i}) \, \psi_{mn}^{+}(r_i, \theta_i,\phi_i), 
\end{equation}
from which Eq. (\ref{green10b}) follows.  If $\x_i = 0
$, then this formula is reduced to
\begin{equation}
\G(\x,\xo) = \frac{1}{4\pi} \sum\limits_{n=0}^\infty P_n\left(\frac{(\x\cdot \xo)}{\|\x\|\, \|\xo\|}\right) 
\frac{\min\{\|\x\|,\|\xo\|\}^n}{\max\{\|\x\|,\|\xo\|\}^{n+1}} \,.
\end{equation}

In the opposite case $r_i > L_i$, one has $r_{>} = r_i$ and $r_{<} =
L_{i}$ so that
\begin{equation}  \label{green10a1}
\frac{1}{\| \x - \xo\| } = \sum\limits_{n=0}^\infty
\sum\limits_{m=-n}^n (-1)^m \, \psi_{(-m)n}^{+}(L_{i},
\Theta_{i},\Phi_{i}) \, \psi_{mn}^{-}(r_i, \theta_i,\phi_i), 
\end{equation}
from which Eq. (\ref{eq:Gfund_int}) follows.

\subsection{Derivation of the harmonic measure density}
\label{sec:HM_derivation}

Taking the derivative of Eq. (\ref{green10b}) with respect to $r_i$, one
finds 
\begin{equation}  \label{eq:dn_G0}
\left. \left( \frac{\partial \G(\x;\xo)}{\partial \n_\x} 
\right)\right|_{\x\in {\partial\Omega}_i} 
= - \sum\limits_{n,m} n V_{mn}^i \psi_{mn}^{-}(1,\theta_i,\phi_i) . 
\end{equation}
Similarly, the derivative of Eq. (\ref{eq:v_auxil2}) with respect to
$r_i$ yields
\begin{equation*}
\begin{split}
& \left. \left(\frac{\partial g(\x,\xo)}{\partial \n_\x} \right)\right|_{\pa_i} = 
\left. \left(\frac{\partial g_i(r_i,\theta_i,\phi_i;\xo)}{\partial \n_\x} \right)\right|_{\pa_i}
+ \sum\limits_{j=1, j\ne i}^N \left.\left(\frac{\partial}{\partial \n_\x} 
g_j(r_j,\theta_j,\phi_j; \xo)\right)\right|_{\pa_i} \\
& = \left.\frac{\partial}{\partial \n_\x} \sum\limits_{n,m} 
\left\{  A_{mn}^i \psi_{mn}^{-}(r_i,\theta_i,\phi_i) +
\left(\sum\limits_{j=1, j\ne i}^N \sum\limits_{l,k} A_{kl}^j U_{klmn}^{(-j,+i)} \right)
\psi_{mn}^{+}(r_i,\theta_i,\phi_i) \right\}\right|_{\pa_i} \\
& = \frac{1}{R_i}  \sum\limits_{n,m} \left\{ (n+1) A_{mn}^i \psi_{mn}^{-}(R_i,\theta_i,\phi_i) -
n \left(\sum\limits_{j=1, j\ne i}^N \sum\limits_{l,k} A_{kl}^j U_{klmn}^{(-j,+i)} \right) 
\psi_{mn}^{+}(R_i,\theta_i,\phi_i) \right\} \\
& = \frac{1}{R_i}  \sum\limits_{n,m} \left\{ (n+1) A_{mn}^i  -
n \left(\sum\limits_{j=1, j\ne i}^N \sum\limits_{l,k} \hat{U}_{mnkl}^{ij} A_{kl}^j \right) \right\}
\psi_{mn}^{-}(R_i,\theta_i,\phi_i) ,  \\
\end{split}
\end{equation*}
that can also be written as 
\begin{equation}  \label{eq:dg}
\left. \left(\frac{\partial g(\x,\xo)}{\partial \n_\x}
\right)\right|_{{\partial\Omega}_i} = \frac{1}{R_i} \sum\limits_{n,m}
\left\{ (2n+1) A_{mn}^i - n \bigl(\hat{\U} \A\bigr)_{mn}^i \right\} 
 \psi_{mn}^{-}(R_i,\theta_i,\phi_i) . 
\end{equation}
Recalling Eq. (\ref{eq:UA}), one gets a simpler form 
\begin{equation}
\left. \left(\frac{\partial g(\x,\xo)}{\partial \n_\x}
\right)\right|_{{\partial\Omega}_i} = \frac{1}{R_i} \sum\limits_{n,m}
\bigl[ (2n+1) A_{mn}^i - n \hat{V}_{mn}^i \bigr] 
 \psi_{mn}^{-}(R_i,\theta_i,\phi_i) . 
\end{equation}
Combining these results, we get Eq. (\ref{eq:HM}) for the harmonic
measure density.

\subsection{Computation of the flux}
\label{sec:Aflux}

The flux of particles onto the ball $\Omega_i$ is 
\begin{eqnarray}  \nonumber
J_i & := &\int\limits_{{\partial\Omega}_i} d\s \left . \left(-D 
\frac{\partial n}{\partial \n_{\xo}} \right)
 \right|_{\xo = \s} 
= - n_0 D \int\limits_{{\partial\Omega}_i} d\s \left . 
\left(\frac{\partial P_{\infty}}{\partial \n_{\xo}}
\right) \right|_{\xo =\s}  \\
& =& 4\pi n_0 D \sum\limits_{j=1}^N \int\limits_{{\partial\Omega}_i} 
d\s \left . \left(\frac{\partial A_{00}^j}
{\partial \n_{\xo}}\right) \right|_{\xo =\s} , 
\end{eqnarray}
where we used Eqs. (\ref{eq:Pinf}, \ref{eq:cPinf}).  According to
Eq. (\ref{eq:UA_Robin}), the derivative of $A_{00}^j$ can be expressed
as a linear combination of the derivatives of $\hat{V}_{mn}^k$.  We
show that
\begin{equation}
\label{eq:int_sphere_flux}
I_{mn}^{ij} := \int\limits_{{\partial\Omega}_i} d\s \, 
\left. \left(\frac{\partial \hat{V}_{mn}^j}{\partial \n_{\xo}} 
\right) \right|_{\xo = \s} = \delta_{n0} \, \delta_{m0} \, \delta_{ij} \, R_i ,
\end{equation}
from which Eq. (\ref{eq:Ji}) follows.  Indeed, for $j = i$, the
integral is
\begin{equation}
I_{mn}^{ii} = \int\limits_{{\partial\Omega}_i} d\s \, 
\frac{(-1)^m }{4\pi} R_i^{2n+1} \left. \left(- \frac{\partial
\psi_{(-m)n}^{-}(r_i,\theta_i,\phi_i)}{\partial r_i} \right)  \right|_{r_i = R_i}
 = \delta_{n0} \, \delta_{m0}\, R_i . 
\end{equation}
For $j \ne i$, we use the addition theorem (\ref{green7a}) to get
\begin{equation}
I_{mn}^{ij} = \int\limits_{{\partial\Omega}_i} d\s \, 
\frac{(-1)^m}{4\pi} R_j^{2n+1} 
 \sum\limits_{l,k} U_{(-m)nkl}^{(-j,+i)} \left(- \frac{\partial
\psi_{kl}^{+}(r_i,\theta_i,\phi_i)}{\partial r_i} \right) = 0 . 
\end{equation}

\subsection{Residence time}
\label{sec:Aresidence}

We use Eqs. (\ref{eq:G}, \ref{green10b}, \ref{green7a}) to write the
residence time ${\mathcal T}$ in a ball $\Omega_I$ of radius $R_I$
centered at $\x_I$ as
\begin{eqnarray}  \nonumber
{\mathcal T}(\xo) & = & \frac{1}{D} \int\limits_{\Omega_I} d\x \, G(\x, \xo) =
\frac{1}{D} \int\limits_{\Omega_I} d\x \biggl\{\sum\limits_{n,m} V_{mn}^I
\psi_{mn}^{+}(r_I,\theta_I,\phi_I) \\
\nonumber
& -& \sum\limits_{j=1}^N \sum\limits_{n,m} A_{mn}^j \sum\limits_{l,k}
U_{mnkl}^{(-j,+I)} \psi_{kl}^{+}(r_I,\theta_I,\phi_I) \biggr\} \\
& =& \frac{4\pi R_I^3}{3D} \biggl\{ \frac{1}{4\pi L_{I}} -
\sum\limits_{j=1}^N \sum\limits_{n,m} A_{mn}^j
\psi_{mn}^{-}(L_{Ij},\Theta_{Ij},\Phi_{Ij}) \biggr\}, 
\end{eqnarray}
where $\L_{Ij} = \x_j - \x_I$, $(L_{Ij}, \Theta_{Ij}, \Phi_{Ij})$ are
the spherical coordinates of $\L_{Ij}$, $L_{I} = \| \xo - \x_I\|$, and
$V_{mn}^I$ is given by Eq. (\ref{eq:V}) which is modified for the ball
$\Omega_I$.

\subsection{Integrals over balls}
\label{sec:Aint}

One can compute the integral of $\psi_{mn}^{-}(r_j,\theta_j,\phi_j)$
over any ball $\Omega_I$ (of radius $R_I$ and centered at $\x_I$),
which is not overlapping with the ball $\Omega_j$.  In fact, denoting
the local spherical coordinates associated to $\Omega_I$ as
$(r_I,\theta_I,\phi_I)$, one can use the I$\to$R addition theorem
(\ref{green7a}) for $r_I < L_{Ij}$ to write
\begin{eqnarray}
\nonumber
\int\limits_{\Omega_I} d\x \,\psi_{mn}^{-}(r_j,\theta_j,\phi_j) 
&=& \sum\limits_{l,k} U^{(-j,+I)}_{mnkl} \int\limits_{\Omega_I} d\x\,
\psi_{kl}^{+}(r_I,\theta_I,\phi_I) \\
\label{eq:psi_intI}
& =& \frac{4\pi R_I^3}{3} \, U^{(-j,+I)}_{mn00} = \frac{4\pi R_I^3}{3} \,
\psi_{mn}^{-}(L_{Ij},\Theta_{Ij},\Phi_{Ij}),
\end{eqnarray}
where $\L_{Ij} = \x_j - \x_I$, $(L_{Ij}, \Theta_{Ij}, \Phi_{Ij})$ are
the spherical coordinates of $\L_{Ij}$, and the mixed-basis elements
are given by Eq. (\ref{green8a}).
Similarly, the integral over the sphere ${\partial\Omega}_I$ reads
\begin{equation}
\int\limits_{{\partial\Omega}_I} d\s \,\psi_{mn}^{-}(r_j,\theta_j,%
\phi_j) = 4\pi R_I^2 \, \psi_{mn}^{-}(L_{Ij},\Theta_{Ij},\Phi_{Ij}).
\end{equation}

Now we consider a more complicated situation when $\Omega_j
\subset \Omega_I$.  We split the integration domain $\Omega_I$ into
two subsets, $\Omega_I^{<}$ and $\Omega_I^{>}$, such that
\begin{equation}
\begin{split}
\Omega_I^{<} & = \{ \x \in \Omega_I ~:~ \|\x - \x_I\| < L_{Ij} \},  \\
\Omega_I^{>} & = \{ \x \in \Omega_I ~:~ \|\x - \x_I\| > L_{Ij} \}. \\
\end{split}
\end{equation}
In each subset, we can use the appropriate addition theorem to compute
the integral.  Using Eq. (\ref{green7a}) for $r_I < L_{Ij}$ and
Eq. (\ref{ad2_II}) for $r_I > L_{Ij}$, we have
\begin{equation}
\int\limits_{\Omega_I^{<}} d\x \,\psi_{mn}^{-}(r_j,\theta_j,\phi_j) 
= \sum\limits_{l,k} U_{mnkl}^{(-j,+I)} \, 
\int\limits_{\Omega_I^{<}} d\x \,\psi_{kl}^{+}(r_I,\theta_I,\phi_I) 
= \frac{4\pi}{3} L_{Ij}^3  \, U_{mn00}^{(-j,+I)} 
\end{equation}
and
\begin{align}
\int\limits_{\Omega_I^{>}} d\x \,\psi_{mn}^{-}(r_j,\theta_j,\phi_j) 
& = \sum\limits_{l=n}^\infty \sum\limits_{k=n+m-l}^{m-n+l} U_{mnkl}^{(-j,-I)} \, 
\int\limits_{\Omega_I^{>}} d\x \,\psi_{kl}^{-}(r_I,\theta_I,\phi_I) \\
\nonumber
& = \sum\limits_{l=n}^\infty \sum\limits_{k=n+m-l}^{m-n+l} U_{mnkl}^{(-j,-I)} 
\, 2\pi \delta_{l0} \delta_{k0} (R_I^2 - L_{Ij}^2) 
= \delta_{n0} \delta_{m0} \, 2\pi (R_I^2 - L_{Ij}^2) , 
\end{align}
where we used $U_{0000}^{(-j,-I)} = 1$.

One may also need to compute the integral of
$\psi_{mn}^{-}(r_j,\theta_j,\phi_j)$ over $\Omega_I$ without any ball
$\Omega_i$:
\begin{equation}
\tilde{\Omega}_I = \Omega_I \backslash \bigcup\limits_{i=1}^N \Omega_i .
\end{equation}
We only consider the case when each ball $\Omega_i$ can be either
included into $\Omega_I$ (i.e., $\Omega_i \subset \Omega_I$), or lie
outside $\Omega_I$ (i.e., $\Omega_i \cap \Omega_I = \emptyset$).  In
other words, we do not allow the ball $\Omega_I$ to cut any ball
$\Omega_i$.  In this case, the integral over $\tilde{\Omega}_I$ is
simply the integral over $\Omega_I$ minus the integrals over each
$\Omega_i$.  First, we have
\begin{equation}
\int\limits_{\Omega_j} d\x \,\psi_{mn}^{-}(r_j,\theta_j,\phi_j) = \delta_{n0} \, \delta_{m0} \, 2\pi R_J^2 
\end{equation}
(although $\psi_{mn}^{-}$ is singular at $r_j = 0$, this singularity
is integrable for $n = 0$ due to the radial weight $r^2$, whereas the
symmetry of the integration domain $\Omega_j$ cancels the contribution
from other harmonics with $n > 0$).  Second, the integral of
$\psi_{mn}^{-}(r_j,\theta_j,\phi_j)$ over $\Omega_i$ (with $i \ne j$)
is given by Eq. (\ref{eq:psi_intI}).  Combining all these results, we
get
\begin{equation}
\begin{split}
\int\limits_{\tilde{\Omega}_I} & d\x \,\psi_{mn}^{-}(r_j,\theta_j,\phi_j) 
= 4\pi \biggl\{ \delta_{n0} \delta_{m0} \, \frac{R_I^2 - L_{Ij}^2 - R_J^2}{2} 
 + U_{mn00}^{(-j,+I)} \frac{L_{Ij}^3}{3} 
- \sum\limits_{i} \frac{R_i^3}{3} \, U_{mn00}^{(-j,+i)} \biggr\} , \\
\end{split}
\end{equation}
where the last sum is taken over the balls $\Omega_i$ (except
$\Omega_j$) which are included in $\Omega_I$.  This formula allows one
to integrate the solution over any ball $\Omega_I$ that does not cut
balls $\Omega_i$.

Using the addition theorem (\ref{ad2_II}), one can compute an integral
over a large sphere ${\partial\Omega}_I$ that englobes a ball
$\Omega_j$.  In fact, since $R_I > L_{Ij}$ because $\Omega_j \subset
\Omega_I$, one has
\begin{equation}  \label{eq:int_sphere}
\int\limits_{{\partial\Omega}_I} d\s \,
\psi_{mn}^{-}(r_j,\theta_j,\phi_j) = \sum\limits_{l=n}^\infty
\sum\limits_{k=n+m-l}^{m-n+l} U_{mnkl}^{(-j,-i)} 
\int\limits_{{\partial\Omega}_I} d\s \, \psi_{kl}^{-}(r_I,
\theta_I, \phi_I) = 4\pi R_I \, \delta_{n0} , 
\end{equation}
the last equality coming from the rotation symmetry of spherical
harmonics $Y_{kl}$ and from the identity $U_{0000}^{(-I,-i)} = 1$.
Note that this result depends neither on the location, nor on the
radius of the ball $\Omega_j$.

\section{Monopole approximation for interior problems}
\label{sec:MOA_int}

The monopole approximation for the interior problem of finding
chemical reaction rates was discussed in
\cite{Galanti16a,Galanti16b,TraytakPoster}.  Here, we briefly present
its extension for computing the Green function.

For the interior problem, one needs to modify the elements of
$\hat{\U}$ and $\hat{\V}$ corresponding to the outer boundary
$\partial\Omega_0$:
\begin{equation}
\hat{U}_{0000}^{i0} = R_i  \quad (i>0), \qquad \hat{U}_{0000}^{0j} = \frac{1}{R_0} \quad (j>0), 
\qquad \hat{V}_{00}^0 = \frac{1}{4\pi R_0}.
\end{equation}
With this modification, the boundary conditions read
\begin{subequations}
\label{eq:monop}
\begin{eqnarray}
(a_i+b_i) A_{00}^i + a_i R_i \sum\limits_{j(\ne i)=1}^{N} L_{ij}^{-1} A_{00}^j + a_i R_i A_{00}^0 & = & \frac{a_i R_i}{4\pi L_{i0}} 
\qquad (i = \overline{1,N}),\\
a_0 A_{00}^0 + \frac{a_0-b_0}{R_0} \sum\limits_{j=1}^{N} L_{ij}^{-1} A_{00}^j & = & \frac{a_0-b_0}{4\pi R_0} . 
\end{eqnarray}
\end{subequations}
If $a_0 \ne 0$, one can express $A_{00}^0$ from the last equation and
substitute it into the former ones that yields a closed system of
linear equations on $A_{00}^i$ for $i=\overline{1,N}$:
\begin{equation}
\left(\frac{a_i+b_i}{R_i} - c_0 \right) A_{00}^i + a_i 
\sum\limits_{j(\ne i)=1}^{N} \left(\frac{1}{L_{ij}} - c_0\right) 
A_{00}^j = \frac{a_i}{4\pi} \left(\frac{1}{L_{i0}} - c_0\right),
\end{equation}
with $c_0 = (a_0-b_0)/R_0$.  

Finally, if $a_0 = 0$ (i.e., the Neumann boundary condition at the
outer boundary), the last relation in Eq. (\ref{eq:monop}) is reduced
to
\begin{equation}
\sum\limits_{j=1}^{N} A_{00}^j = \frac{1}{4\pi} \,.
\end{equation}
In this case, Eqs. (\ref{eq:monop}) can be written as
\begin{equation}
A_{00}^i + c_i \sum\limits_{j(\ne i)=1}^{N} L_{ij}^{-1} A_{00}^j + c_i A_{00}^0 = \frac{c_i}{4\pi L_{i0}} \qquad (i = \overline{1,N}),
\end{equation}
with $c_i = a_i R_i/(a_i+b_i)$ (for $i = \overline{1,N}$).  Summing
these equations over $i$ from $1$ to $N$, one gets
\begin{equation}
\frac{1}{4\pi} + \sum\limits_{i=1}^{N} c_i \sum\limits_{j(\ne i)=1}^{N} L_{ij}^{-1} A_{00}^j + C A_{00}^0
= \sum\limits_{i=1}^{N} \frac{c_i}{4\pi L_{i0}} \,,
\end{equation}
where $C = c_1 + \ldots + c_{N}$.  Expressing $A_{00}^0$ from this
relation, one gets a closed system of linear equations on $A_{00}^i$
for $i = \overline{1,N}$:
\begin{equation}
A_{00}^i + c_i \sum\limits_{j(\ne i)=1}^{N} L_{ij}^{-1} A_{00}^j + 
\frac{c_i}{C} \left(\sum\limits_{k=1}^{N} \frac{c_k}{4\pi L_{k0}} - \frac{1}{4\pi}
- \sum\limits_{k=1}^{N} c_k \sum\limits_{j(\ne k)=1}^{N} L_{kj}^{-1} A_{00}^j \right)
= \frac{c_i}{4\pi L_{i0}} \,,
\end{equation}
or
\begin{equation}
A_{00}^i + c_i \sum\limits_{j(\ne i)=1}^{N} L_{ij}^{-1} A_{00}^j -
\frac{c_i}{C} \sum\limits_{j=1}^{N} A_{00}^j \sum\limits_{k(\ne j)=1}^{N} c_k L_{kj}^{-1} 
= \frac{c_i}{4\pi L_{i0}} - \frac{c_i}{C} \left(\sum\limits_{k=1}^{N} \frac{c_k}{4\pi L_{k0}} - \frac{1}{4\pi}\right),
\end{equation}
or
\begin{equation}
A_{00}^i\left(1 - \frac{c_i}{\ell_i} \right) + c_i \sum\limits_{j(\ne i)=1}^{N} A_{00}^j 
\left(L_{ij}^{-1} - \frac{c_i}{\ell_j} \right)
= \frac{c_i}{4\pi} \left(\frac{1}{L_{i0}} + 1 - \frac{1}{C} \sum\limits_{k=1}^{N} \frac{c_k}{L_{k0}} \right),
\end{equation}
where we denoted
\begin{equation}
\ell_j^{-1} = \frac{1}{C} \sum\limits_{k(\ne j)=1}^{N} c_k L_{kj}^{-1} .
\end{equation}


\end{document}